\documentclass[aps,preprint,pre,showpacs,nofootinbib,superscriptaddress,floatfix]{revtex4-1}
\usepackage{latexsym}
\usepackage{bm}
\usepackage{epsfig,amsmath, amsfonts, amssymb, xcolor,hyperref,wasysym}
\usepackage{stmaryrd}
\usepackage{soul}
\usepackage{textcomp}
\usepackage{graphicx}
\usepackage[T1]{fontenc}
\usepackage{color}
\newcommand{\sgn}[1][]{\ensuremath{\mathop{sgn}}}

\begin{document}

\title 
   {Collective behavior of colloids due to  critical Casimir interactions} 
\date{\today}
\author{A. Macio{\l}ek}
\email{amaciolek@ichf.edu.pl}
\affiliation{Institute of Physical Chemistry, Polish Academy of Sciences,
  Kasprzaka 44/52, PL-01-224 Warsaw, Poland}
  \affiliation{Max-Planck-Institut f{\"u}r Intelligente Systeme,
  Heisenbergstra{\ss}e 3, 70569 Stuttgart, Germany}
\affiliation{Institut f{\"u}r Theoretische Physik IV, Universit{\"a}t Stuttgart, 
  Pfaffenwaldring 57, 70569 Stuttgart, Germany}
\author{ 
S. Dietrich
}
\email{dietrich@is.mpg.de}
\affiliation{Max-Planck-Institut f{\"u}r Intelligente Systeme,
  Heisenbergstra{\ss}e 3, 70569 Stuttgart, Germany}
\affiliation{Institut f{\"u}r Theoretische Physik IV, Universit{\"a}t Stuttgart, 
  Pfaffenwaldring 57, 70569 Stuttgart, Germany}

\begin{abstract}
If colloidal solute particles are suspended in a solvent close to its critical point, they  act as cavities in a  fluctuating
medium  and thereby  restrict and modify the fluctuation spectrum in a way which 
depends on their relative configuration.
As a result effective, so-called
critical Casimir forces (CCFs) emerge  between the colloids.
The range and the amplitude of CCFs depend sensitively on the temperature and  the composition of the solvent as well as on the
 boundary conditions of the order parameter of the solvent at the particle surfaces. These remarkable, moreover universal features of the CCFs provide 
the  possibility for an active control  over the assembly of colloids. This has  triggered a recent
surge of  experimental
 and theoretical interest in these phenomena.
 We present an  overview of current research activities  in this area.
Various experiments demonstrate the occurrence of thermally reversible 
 self-assembly or aggregation or even  equilibrium phase transitions of colloids in the mixed
 phase below the lower consolute points of  binary solvents. We discuss the status
 of the theoretical description
 of these phenomena, in particular  the validity of a description  in terms of  effective, one-component colloidal 
 systems and the necessity of a full treatment of a ternary solvent-colloid mixture.
 We suggest  perspectives on the directions towards which future research 
 in this field might develop.

\end{abstract}
\maketitle

\section{Introduction}
\label{sec:intro}

Finite-size contributions to the free energy of a spatially confined fluid 
give rise to  an effective force per  
area acting on the confining surfaces~\cite{Evans:1990}. 
Fisher and de Gennes~\cite{Fisher-et:1978} made the crucial  observation that
this fluid-mediated interaction
acquires a universal, long-ranged contribution $f_C$ if the bulk 
critical point of the fluid is approached.
This is due to \emph{critical fluctuations}, and hence the notion `critical Casimir 
force', in analogy with quantum-mechanical Casimir forces which are due to quantum 
fluctuations of confined electromagnetic fields~\cite{Casimir:1948,Kardar-et:1999}. 
In  the case of colloidal suspensions with  near-critical suspending fluids (referred to as  solvents),  the typically micrometer-sized colloidal particles  act as  cavities inside the critical solvent.
At the colloid surfaces  these cavities impose boundary conditions for the fluctuating  order parameter 
of the solvent and perturb the order parameter field on the length scale of the bulk correlation length $\xi$.
Such modifications of the order parameter  and the restrictions of its fluctuation spectrum 
depend on the spatial configuration of the colloids. Following  the argument by Fisher and de Gennes~\cite{Fisher-et:1978}, 
 this gives rise to  CCFs  between the colloids, which is attractive for identical particles  and  has a range set by the bulk correlation length 
$\xi$ of the solvent. Since  $\xi(t=(T_c^{(s)}-T)/T_c^{(s)}\to 0)\sim |t|^{-\nu}$, where $\nu$ is a standard  bulk critical exponent, this range
diverges upon approaching the bulk critical temperature $T_c^{(s)}$ of the pure {\it s}olvent.

The collective behavior of colloids dissolved in the near-critical solvent 
 is determined by the interplay between the CCFs and other interactions acting between the constituents.
In general, in   colloidal suspensions the dissolved particles  interact  directly via van der Waals  interactions; these  are attractive and lead to  irreversible aggregation (called coagulation)  \cite{VarweyOverbeek:1948}.  
In  charge-stabilized suspensions, the colloids acquire surface charges due to dissociation of the surface groups in water
or due to chemical functionalization of  the surface of the particles.
This causes the formation of  electric  double layers  around the colloids and results in electrostatic repulsion between them.  
In  sterically-stabilized
suspensions, the short polymer chains 
grafted onto the surface of colloidal  particles give rise to a repulsive interaction which is  of entropic origin.
Additionally, the presence of
other smaller solute particles or macromolecular additives such as polymers, surfactants,  or micelles  \cite{Likos:2001},
induces effective entropic interactions between the colloids, called depletion forces, which
are predominantly attractive and short-ranged \cite{Asakura-et:1954,Vrij:76}.
If the CCFs between colloidal particles are  attractive and sufficiently strong to overcome the direct repulsive forces,
one may expect the occurrence of  a thermodynamically stable 
 colloid-rich liquid or solid phase - even in the absence of any direct attractive interactions. 
If the resulting attractive potential is sufficiently strong, the  condensation  transition from a colloid-poor ('gas') to a colloid-rich ('liquid') phase  may be preempted (on the characteristic time scales
of the observations) by the formation
of {\it non-equilibrium} aggregates in
which the colloidal particle stick together. 
In general, such aggregates may grow or shrink and their structure varies
from loose fractals through gels and glasses to crystals, 
depending  on the packing fraction of
the colloidal particles in the aggregates and
on the strength of the attraction among the colloidal particles.

Beysens and Est{\`e}ve \cite{Beysens-et:1985}  were the first to study experimentally    
aggregation phenomena for colloids suspended in  binary  solvents.
These authors studied silica spheres immersed in a  water-lutidine mixture
by using light scattering. They found the formation of  aggregates which sediment upon approaching the
bulk coexistence region of demixing from the one-phase region of the
binary liquid mixture at constant composition of the solvent.
Strikingly,  the observed aggregation was thermally reversible;
moving back the thermodynamic state deeply into the one-phase region
the sediments dissolved again. 
In the  decade following this pioneering work, quite a number of further  experiments 
were performed  leading to a similar behavior for various binary solvents and  a variety of  colloids.
The structure of the aggregates as well  as the the kinetics of aggregation and the reverse
process of fragmentation have been investigated.
Silica, quartz powder,  and polystyrene particles immersed in  water-lutidine mixtures were studied in 
Refs.~\cite{Gurfein-et:1989,gallagher:92,Broide-et:1993,narayanan:95,Kurnaz}, whereas 
in Refs.~\cite{narayanan:93,Jayalakshmi-et:1997,Kline-et:1994,Koehler-et:1997,
grull:97} other solvents were employed;
for corresponding reviews  see Refs.~\cite{Beysens-et:1994,Beysens-et:1999,Law:2001}. 
 These experiments, which were performed mostly in the one-phase region of a binary liquid mixture,
revealed  that reversible aggregation  (termed flocculation) is accompanied by a strong
adsorption phenomenon in the vicinity of the bulk two-phase coexistence curve.
Generically, colloidal particles have a preference for one of the
two components of the binary solvent. At the surface of the colloid
this preference gives rise to an effective surface field conjugate to 
the order parameter at the surface and thus leads to an adsorption 
layer rich in this preferred  component. The measurements   demonstrated
 that the temperature - composition $(T,c)$ region in which  colloidal aggregation appears is not symmetric
 about the  critical composition $c_c$ of the binary solvent. 
Strong aggregation occurs on that side of the  critical composition
 which is rich in the component not preferred by the colloids.

 Various mechanisms were put forward  for strong adsorption giving rise to attraction which in turn 
 could explain the occurrence of (non-equilibrium) flocculation.
For example,  capillary condensation and/or wetting
can occur  when  particles come close together via diffusion, even far off the critical composition of the solvent.
In this case a  liquid 'bridge' can  form which induces attractive solvation forces \cite{Bauer-et:2000}. 
Another possibility is that  the presence of an adsorption layer
 around the colloidal particles increases the strength of the direct attractive dispersion interactions.
However,  in the close vicinity of the bulk critical point of the solvent, in line with the predictions of Fisher and de Gennes,
effective attraction  induced by   critical fluctuations is expected be dominant. 
In their original paper, Beysens and Est{\`e}ve have
identified  an ``aggregation line'' in
the temperature-composition phase diagram of the solvent with a prewetting line.
However, they have observed that this  aggregation line extends to temperatures  below the lower demixing
critical point of a solvent. However, in general such an extent of prewetting lines  has not been found  up to now, neither for planar nor for  spherical substrates.
Actually, positive curvature even shortens prewetting lines \cite{Bieker-et:1998}. (Regrettably, the wetting behavior of this system in planar geometry has never been investigated.) 
The authors did not comment on this difference  nor did they admit the  role of critical fluctuations and Casimir forces for aggregation in the critical region.
They have presumed that the  aggregation process results from the attractive forces between the colloids
due to the presence  of the adsorption layer and have referred to de Gennes \cite{deGennes} as the one 
who had proposed such fluctuation induced interactions  at and near the bulk critical point of the solvent, but they have not put their results into the proper
context  of de Gennes's predictions. 
According to  another interpretation  of the experimental findings mentioned above, the observed phenomenon is regarded as a  precursor of 
a bona fide phase transition in the ternary mixture rather then a non-equilibrium flocculation of colloidal particles
\cite{Kline-et:1994,Jayalakshmi-et:1997,Koehler-et:1997}. A few theoretical \cite{Sluckin:1990} and
simulation \cite{Loewen:1995,Netz:1996} attempts have been concerned with such an interpretation.
The  status of knowledge about reversible aggregation of colloids 
 in  binary solvents up to the late 1990s
has been  reviewed by Beysens and Narayanan (see Ref.~\cite{Beysens-et:1999} and references  therein).
More recently, the scenario of 'bridging' transitions has again been studied theoretically 
in Refs.~\cite{Archer-et:2005,Okamoto-et:2011a,Okamoto-et:2013,Labbe-Laurent-et:2017}.

 In spite of the relevance of aggregation phenomena for the stability of colloidal suspensions,
the basic understanding of the collective behavior of colloids dissolved in a near-critical solvent has started to emerge only  recently.
This progress had to await  the advances made during the last decade concerning the statistical mechanical theory and computer simulations of CCFs.
The accumulated theoretical knowledge of two-body CCFs 
has triggered  also an increase  of experimental activities  in this field. This renewed  interest is driven by  application perspectives, in particular  concerning the buildup of  nanostructured materials of well defined structure by using self-assembly of colloidal particles. 
In order to achieve a desirable morphology of aggregates, one has to be able  to control  colloidal 
self-assembly and  to manipulate the  particles. The remarkable features of CCFs offer such possibilities.
The range and the strength of the CCFs, which depend sensitively on temperature via the bulk correlation length $\xi$,
 can be tuned reversibly and {\it continuously}  by moving the thermodynamic state of the solvent  around its critical point.
 The  sign of $f_C$ can be manipulated as well  by suitable surface treatments of the colloids
\cite{Hertlein-et:2008,Gambassi-et:2009,Nellen-et:2009}.
Additional interest is sparked by the potential relevance of CCFs for lipid  membranes. These are two-dimensional
(2d) liquids  consisting of two (or more) components, such as cholesterol 
and  saturated and unsaturated lipids, which can  undergo phase
separation into two liquid phases, one being  rich in the 
first two components and the other rich in the third \cite{membranes1}.
Lipid membranes serve as  model systems for 
cell plasma membranes \cite{LS}.  
Recent experiments suggest that cell membranes 
are tuned to the miscibility critical point of the  $2d$ Ising model \cite{membranes,Machta-et:2016}
so that CCFs may arise between
macromolecules embedded in the membrane \cite{sehtna,Machta-et:2012,Benet-et:2017}. 

 Other mechanisms, which {\texttwelveudash}  similar to the ones generating the critical Casimir effect {\texttwelveudash}  also  induce solvent-mediated  long-ranged interactions, occur 
inter alia in a chemical sol upon approaching its percolation transition \cite{Gnan-et:2014}, in a binary liquid mixture
subjected to a steady temperature gradient due to the concomitant nonequilibrium concentration fluctuations \cite{Kirkpatrick-et:2015}, 
in driven noncohesive granular media due to  hydrodynamic fluctuations \cite{Cattuto-et:2006}, or if  the solvent  
 comprises   active matter such as bacteria  or self-propelled colloidal particles \cite{Ray-et:2014,Ni-et:2015}.

Compared with other effective forces between colloid particles or macromolecules,   CCFs 
have two advantages. First, due to the concept of universality for critical phenomena, to
a large extent CCFs do not depend on  microscopic details of the system.
Second, whereas adding depletion agents or ions changes the resulting effective forces de facto
irreversibly, the tuning of $f_C$ via temperature is fully and easily reversible.

The present article  discusses current theoretical and numerical approaches towards the description of  
the static, equilibrium properties of colloidal suspensions with a near-critical binary 
solvent. The related  experimental body of research  is put into the corresponding context
 and  a number of intriguing  possible developments
are highlighted. Recent  developments concerning colloidal assembly due to  CCFs with a focus on the experimental observations
 is reviewed in Ref.~\cite{Schall_review}.
 
\section{Effective one-component approach}
\label{sec:eff}
A common approach to the statistical mechanics  description of   colloidal suspensions  follows the ideas 
developed for  multi-component molecular liquids, such as ionic solutions, by considering  the colloidal particles
as 'supramolecules' \cite{VNFA:78,Hansen:93,Likos:2001}. 
Within this approach, the degrees   
of freedom of the  solvent and the ions, in the case of charged-stabilized suspensions,  are  traced out 
in order to construct  an effective one-component system of colloidal particles interacting via  state- and 
 configuration-dependent forces.
For most cases, carrying out the integration over microscopic degrees  of freedom can be done only approximately, leading to
 additive pairwise  interactions between the colloidal particles (see the corresponding discussion below).
For any binary mixture with  pair interactions for which the volume integral is finite, a formal expression for an effective 
Hamiltonian, describing   particles of one species only but in  the presence of the particles of 
 the other species,  has been given in Ref.~\cite{Dijkstra-et:1999}.
This effective Hamiltonian consists of zero-body, one-body, two-body, three-body, and higher-body
interactions, which depend on the density of the second species and have to be determined one by one. 
For additive hard-sphere mixtures with a large size asymmetry, a comparison with direct simulations of true binary mixtures has shown that
the  pairwise (depletion) potential approximation of the effective Hamiltonian  between two large particles accounts remarkably well
 for the phase equilibria,  even in  limits for which one might expect that higher-body terms cannot be neglected.
This success encourages   one to use  an effective one-component approach (with the approximation of an  additive pairwise potential) to   colloids suspended in a near-critical solvent, despite the fact that
 the CCFs are inherently non additive. The critical Casimir interaction between two colloidal particles depends on  the
(instantaneous) spatial configuration of {\it all} colloids \cite{Mattos-et:2013,Mattos-et:2015,Hobrecht-et:2015,Volpe-et}.
Only for  dilute  suspensions or for temperatures sufficiently far away
from the bulk critical temperature $T_c^{(s)}$ of the pure solvent, such that
the range  of the critical Casimir interaction between the
colloids  is much smaller than
the mean distance between them, the assumption of pairwise additive CCFs is  expected
to be reliable. 

\subsection{Effective interactions}
\label{subsubsection:eff_inter}

In  most of the experimentally studied systems, the solvent is   a binary  mixture of molecular liquids
and the colloidal particles are micro-sized spheres of a radius $R$.
For such a sizewise highly asymmetric
multi-component system,  one can ignore the
discrete nature of the solvent  and use
a  simplified pair potential model for the background interaction potential between the colloids, which is
present also away from the critical temperature $T_c^{(s)}$ 
of the solvent. Such a model is supposed to    capture only 
the essential features of a stable suspension on the relevant, i.e., mesoscopic, 
length scale. Besides the van der Waals contribution, which will be discussed below, these features are the hard core repulsion for center-to-center 
distances $r<2R$ and a soft repulsive contribution, for which one can employ the Yukawa potential.
 This leads to
the screened Coulomb model of  suspensions which are charge-stabilized against flocculation \cite{Russel-et:1989,Hansen_Loewen:2000,Barrat-et:2003}:
\begin{equation}
\label{eq:1}
 V_{rep}(D)/(k_BT) = U_{rep}(D) = \left( U_0 /(\kappa D) \right)\exp(-\kappa D), \qquad D = r -2R >0,
\end{equation}
 where $D$ is the surface-to-surface distance and $k_B$ is the Boltzmann constant. The range $\kappa^{-1}$ of the repulsion is  the Debye screening length 
 $\kappa^{-1} = \sqrt{\epsilon\epsilon_0 k_B T/(e^2\sum_i\rho_i)}$ (see, e.g.,
Ref.~\cite{vdW}), where $e$ is the
elementary charge, $\epsilon$ the  permittivity of the solvent relative to the vaccum, $\epsilon_0  = 8.854\times 10^{-12}$ C$^2$/(Jm) is the vacuum permittivity, and $\{\rho_i\}$ the number densities of all ions (regardless
of the sign of their charges).
A simplified,  purely exponential form of  the repulsive 
pair potential,
\begin{equation}
\label{eq:2}
 U_{rep}(D) = A \exp(-\kappa D),
\end{equation}
is often used for 
 suspensions in which $\kappa^{-1 }\ll R$ for distances $2R> D > R+\kappa^{-1}$, for which  all  curvature effects associated
with the spherical geometry of the  colloidal particles  effectively drop out  \cite{Israelachvili:1998,Levin:2002}.
The corresponding condition  $\kappa^{-1 }\ll R$ is practically satisfied  for the  experimentally relevant 
systems for which the Debye
length is of the order of 10 nm and the  colloidal size of  the order
of 1$\mu m$. 
For the effective Coulomb interaction screened by
counterions the amplitude  $A$ is given by \cite{Russel-et:1989}
\begin{equation}
\label{eq:3}
	A= 2 \pi (\epsilon \epsilon_0 )^{-1} \Upsilon^{2}\kappa^{-2}R/(k_B T),
\end{equation}
where   $\Upsilon$ 
is the surface charge density of the colloid.
The purely exponential form of repulsion (Eq.~(\ref{eq:2})) can also describe  sterically stabilized suspensions beyond 
the hard-sphere model \cite{Israelachvili:1998}.
In that case the range $\kappa^{-1}$ of the repulsion is associated with the  length of the grafted polymers
and the strength $A$ of the repulsion depends on the surface coverage of grafted polymers.

Upon approaching the bulk critical point  $(T_c^{(s)}, c_c$) of the solvent,
CCFs between the particles emerge and the
corresponding pair  potential $V_C/(k_BT)=U_C$ adds to the background contribution.
In the well defined scaling limit of all length scales of the system being large on molecular scales, $U_C$ attains a scaling form \cite{Barber:1983,privman,Krech-et:1992}
in terms of suitable dimensionless scaling variables describing the distance between the colloids,
 the dependence on the thermodynamic state of the solvent, and
the shape of the colloidal particles.  
For example, for the spherical particles one has 
\begin{equation}
\label{eq:4}
\frac{V_C(D)}{k_BT}= U_C(D) \simeq \frac{R}{D} \Theta\left({\mathcal Y} = {\mathrm sgn}(t)\frac{D}{\xi_t},\; \Delta = \frac{D}{R},\; \Lambda = {\mathrm sgn}(h_b)\frac{D}{\xi_h} \right), \quad D = r -2R >0.
\end{equation} 
In Eq.~(\ref{eq:4}), $\Theta$  is a  universal scaling function. 
Here $\xi_t(t\gtrless 0)=\xi_{t,\pm}^{(0)}|t|^{-\nu}$, 
with  $t=\pm(T-T_c^{(s)})/T_c^{(s)}$  for an upper ($+$) 
and a lower ($-$) critical point, respectively,
is the true correlation length governing the exponential decay of the 
solvent  bulk two-point order parameter (OP) correlation function for $t\to0^{\pm}$ and $h_b=0$ where  $h_b$ is the bulk ordering field conjugate to the OP. 
The amplitudes $\xi_{t,\pm}^{(0)}$ (with $\pm$ referring to
the sign of $t$) are non-universal but their ratio $\xi_{t,+}^{(0)}/\xi_{t,-}^{(0)}$ 
is universal. The correlation length 
$\xi_h =\xi^{(0)}_h|h_b|^{-\nu/(\beta\delta)}$
governs the exponential decay of the solvent bulk two-point OP
correlation function for $t=0$ and $h_b\to0$, where $\xi^{(0)}_h$
is a non-universal amplitude related to $\xi_{t,\pm}^{(0)}$ via universal amplitude ratios; $\nu$, $\beta$, and $\delta$ are standard
bulk critical exponents \cite{Pelissetto-et:2002}.
For the demixing phase transition of 
a binary liquid mixture, the OP $\phi$ is proportional to the 
deviation of the concentration of species, say $a$, 
\begin{equation}
\label{eq:con}
 c_a = {\varrho}_{a}/({\varrho}_{a} + {\varrho}_{b})
\end{equation}
from its 
value $c_{a,c}$ at the critical point, i.e., $\phi \sim c_a - c_{a,c}$; 
here ${\varrho}_{\alpha}$, $\alpha\in\left\{a,b\right\}$,  are the number densities of the 
particles of species $a$ and $b$, respectively.  The 
bulk ordering field, conjugate to this order parameter, is  
proportional to the deviation of the difference $\Delta\mu=\mu_a-\mu_b$ 
of the chemical potentials $\mu_{\alpha}$, $\alpha\in\left\{a,b\right\}$, of the 
two species from its critical value, i.e., $h_b\sim \Delta\mu-\Delta\mu_c$. 
We note, that the actual scaling fields of fluids are 
linear combinations of $h_b$ and of the reduced temperature 
$t$. 

The  minimal model for a pair potential   describing the 
effects of a critical solvent on dissolved  colloids due to  CCFs, i.e., corresponding to
the sum of Eqs.~(\ref{eq:2})
and (\ref{eq:4}), has been used in
Refs.~\cite{Bonn-et:2009,Gambassi-et:2010,Mohry-et:2012a,Mohry-et:2012b,Mohry-et:2014,Nguyen-et:2013,Dang-et:2013}.
In Ref.~\cite{Zvyagolskaya-et:2011}, instead of a soft repulsive potential of the form as in Eqs.~(\ref{eq:2}) and (\ref{eq:3}), the electrostatic
repulsion has been modeled via a hard disc repulsion with an effective diameter. In some of the  
studies  cited above, a simplified functional form of the universal scaling function $\Theta$
has been employed ~\cite{Bonn-et:2009,Gambassi-et:2010,Nguyen-et:2013,Dang-et:2013}, such as using a form  valid only asymptotically 
for large values of the temperature scaling variable  neglecting the dependence on the other, also relevant  scaling variables. 
Since, however, the  shape of the total  pair potential depends sensitively on  details of the 
CCFs pair potential,  there is a need to discuss them (see Subsec.~\ref{subsection:CCFpot} below).

In order to be able to describe certain experimental systems, one has to   consider also
the interaction which  accounts for
effectively attractive dispersion forces. 
For two spheres, the (nonretarded) van der Waals forces contribute to the total potential 
 through a term (see, e.g., Refs.~\cite{Chen:1996,vdW})
\begin{equation}
\label{eq:5}
V_{\rm vdW}(\Delta)/(k_BT) = U_{\rm vdW}(\Delta) = - \frac{A_H/(k_BT)}{6} \left[\frac{2}{\Delta(\Delta + 4)} + \frac{2}{(2
  + \Delta)^2} + \ln\left( \frac{1+4/\Delta}{(1+2/\Delta)^2}\right)\right],
\end{equation}
where $A_H$ is the Hamaker constant. 
As $\Delta = D/R$ increases,
this term crosses over from the behavior $U_{\rm vdW}(D\ll R) \simeq
-(A_H/6) (R/D)^2$ to  $U_{\rm vdW}(D\gg R) \simeq
-(2A_H/3) (R/D)^3$. The Hamakar constant depends on the   dielectric properties of the 
materials involved in the experiment under consideration \cite{vdW}. 
By using index-of-refraction-matched colloidal suspensions \cite{Israelachvili:1998} its value can
be strongly reduced. This way 
 dispersion forces can be    effectively switched off. 
 (A detailed discussion of the Hamakar constant for a polystyrene
colloid near a  silica glass  substrate immersed in a mixture of water and 2,6-lutidine
can be found in Ref.~\cite{Gambassi-et:2009}.)

Additionally, in the presence of small co-solutes such as free polymer
coils or smaller colloids in  a stericly stabilized colloidal
suspension, one has to consider   depletion interactions which arise between large 
colloidal particles due to  entropic effects caused by the small solutes.
This  so-called depletion interaction  is mainly attractive and has a range 
 proportional to
the size of the depletant \cite{Asakura-et:1954,Vrij:76}. There are several theories and 
approximations for the depletion potential, which are summarized in Refs.~\cite{Goetzelmann-et:1998,Roth-et:2000}.
For example,  for a fluid of large hard spheres of  radius $R$ and a small spherical  depletant of diameter $\sigma$, which on its own behaves  as an ideal gas, the depletion potential is \cite{Vrij:76}
\begin{equation}
\label{eq:6}
  V_d(D)/(k_BT) =  U_d(D)=
    \begin{cases}
      -n_bV_{ov}(D), & 0 \le D \le \sigma \\
       0,  & D\ge \sigma,
    \end{cases}
\end{equation}
where $V_{ov}(D)= (\pi/6)(\sigma -D)^2(3R+\sigma+D/2)$.
In  Eq.~(\ref{eq:6}) $n_b$  is the bulk number density of the small  spheres. Note that $|V_d(D)|$ in Eq.~(\ref{eq:6})
 is equal to the pressure ($p=n_bk_BT$) times the overlap volume $V_{ov}$ between the excluded volumes denied to the centers of the small spheres around each big sphere.
Taking into account hard-core repulsion among  the depletants
produces  a repulsive contribution to the depletion interaction  as well as an oscillatory decay at  large distances \cite{Goetzelmann-et:1998,Roth-et:2000}.
For many actual colloidal suspensions  the depletion attraction is strong enough to induce colloidal aggregation and, despite of  
its rather short range,  a  gas-liquid-like phase separation, because even
a small degree of polydispersity  or non-sphericity of the
particles  causes the fluid phase to be not preempted by crystallization.

If  the depletant is only a co-solute in the suspension in which the solvent becomes critical, the resulting depletion potential simply 
adds to the background potential.  It may, however, occur that due to an effective depletant-depletant  interaction the depletant  itself  exhibits 
a phase transition with a critical point.   This means that the solvent, which is common to both the big spheres and the depletant,
does not display a critical point of its own as the one discussed above. A system which realizes this  interesting scenario was studied experimentally  in Refs.~\cite{Buzzaccaro-et:2010,Piazza-et:2011}
and via Monte Carlo simulations \cite{Gnan-et:2012a} 
(see Secs.~\ref{subsubsec:pp} and \ref{subsec:eff_dep}  below where we shall review the results of these studies).
In such a case one expects that  the depletant produces one unique  effective pair potential between the big particles. 
Far away from the critical point
of the depletant, this unique pair potential has the character of a depletion interaction, whereas close to the  critical region  of the depletant it 
should display the features of the  CCFs pair potential.
As has been shown in Refs.~\cite{Buzzaccaro-et:2010,Piazza-et:2011}, the framework of  density functional theory (DFT) \cite{Evans:1979}, which 
is commonly adopted in colloidal science, can provide both forms for the effective pair potential in the corresponding limits 
(see Sec.~\ref{subsection:CCFpot}  below for details).
In Ref.~\cite{Gnan-et:2012a},  the effective pair potential $V_{eff}(r)$ between two big hard-sphere colloids has been determined numerically  for two models of the depletant particles within
a wide range of state points, including the critical
region. 
In the first model, the spherical depletant particles interact via a pairwise
square-well potential (SW):
\begin{equation}
\label{eq:7}
  V^{SW}(r)=
    \begin{cases}
    \infty, & r < \sigma \\
      -d_w, & \sigma  \le r \le (1 + r_w \sigma) \\
       0,  & r \ge  (1 + r_w )\sigma,
    \end{cases}
\end{equation}
where $r$ is the  center-to-center distance between two depletant particles, $r_w$ is a dimensionless  well width,  and $d_w$ is the  well depth. The second
depletant model is  an anisotropic three-patches (3P) Kern-Frenkel system which consists of hard-sphere particles
 decorated with three attractive sites \cite{KF}. The critical packing fraction of these patchy particles  is very small - 
as in the experimental
system studied in Refs.~\cite{Buzzaccaro-et:2010,Piazza-et:2011}. 
For this kind of depletant, the numerically determined effective pair potential $V_{eff}(r)$ between two big spheres  has subsequently been  used in
 a grand canonical off-lattice MC simulation in order to analyze  the stability of the colloidal
 suspension  for a system of colloidal particles
  interacting via the pairwise additive interaction
 $ V_{HS} (r) + V_{eff}(r) $, where  $V_{HS} (r)$
  is a hard-sphere potential.
 The results of this study will be discussed below in Sec.~\ref{subsubsec:pp}.

\subsection{Critical Casimir pair potential}
\label{subsection:CCFpot}

A crucial ingredient for an effective one-component approach is to have an accurate
critical Casimir potential (CCP) between two colloidal particles.
This is not only relevant for the  investigation of aggregation or the bulk phase behavior of colloids, but it is 
also  of intrinsic scientific interest.
In recent years an experimental technique  (total internal reflection microscopy (TIRM) \cite{Walz-97,Prieve-99,h-th}) has been developed  which allows one to measure directly and with fN resolution
the  effective potential of the CCF between a colloidal particle, suspended in a near-critical binary mixture,
and a fixed object such as a planar wall \cite{Hertlein-et:2008,Nellen-et:2009,Gambassi-et:2009}.
Video microscopy has also been used in order to determine the  potential of the CCF between  two spherical
colloids \cite{Nguyen-et:2013,Dang-et:2013,Shelke-et:2013,Marcel-et}.
Although the resolution and the sophistication of such experiments increase, it is  difficult
to interpret the data  of these measurements mainly due to the inevitable, simultaneous presence of various contributions
to the effective pair  potential. 
Therefore  reliable  theoretical
results are required in order to improve the interpretations.

The solvent-mediated force between two spherical particles,  a  surface-to-surface distance $D$ apart, is defined
 as the negative derivative of the excess free energy ${\cal F}^{ex}={\cal F} -V f_b$,
\begin{align}
\label{eq:8}
f_s= -\frac{\partial {\cal F}^{ex}}{\partial D}
		 =-\frac{\partial( {\cal F} -V f_b)}{\partial D},
\end{align}
where $f_b$  is  the bulk free energy density of the solvent
and ${\cal F}$ is the free energy of the {\it s}olvent in
the macroscopically large volume $V$ excluding the volume of  two suspended
colloids. The CCF  $f_C$  is the long-ranged universal contribution to $f_s$ which emerges upon approaching the bulk critical point of the solvent.
The associated critical Casimir potential (CCP) is   
\begin{align}
\label{eq:9}
V_C(D,R,T,h_b) \equiv \int_D^\infty dz f_C(z,R,T,h_b)
\end{align}
so that the critical Casimir force (CCF) is given by $f_C(D,R,T,h_b)=-\frac{\partial}{\partial D}V_C(D,R,T,h_b)$.
According to finite-size  scaling theory \cite{Barber:1983,privman}, 
the  CCP exhibits  scaling
described by a universal  scaling function $\Theta$ as given by Eq.~(\ref{eq:4}). This scaling function 
is determined solely by the so-called universality class 
of the continuous phase transition occurring in the
bulk, the geometry of the setup, and the surface 
universality classes of the confining surfaces
\cite{Diehl:1986,Krech:1990:0,dantchev,gambassi:2009}.
The relevant bulk universality class for  colloidal suspensions is the Ising
universality class in spatial dimension $d=3$ or $d=2$. For the CCF one has
\begin{align}
\label{eq:9a}
f_C/(k_BT) \simeq \frac{R}{D^2} \vartheta\left({\mathcal Y}, \Delta, \Lambda \right)=\frac{R}{D^2}\left[ \Theta - {\mathcal Y}\frac{\partial}{\partial {\mathcal Y}}\Theta - \Delta\frac{\partial}{\partial\Delta}\Theta - \Lambda\frac{\partial}{\partial\Lambda}\Theta\right].
\end{align}

The main difficulty in determining theoretically the  scaling function of CCFs and their  potential 
lies in the   character of the critical fluctuations; an adequate treatment has to 
include  non-Gaussian fluctuations. Usually  colloidal particles
 exert a potential on the
surrounding fluid which  is infinitely repulsive at short distances and attractive at large distances. Such potentials give
rise to pronounced peaks in the density profile of the fluid near the surface of the  particle, i.e., there is strong  adsorption.
In terms of a field-theoretical description this corresponds to the presence of a strong (dimensionless) {\it s}urface field $h_s\gg 1$. For binary liquid mixtures one has $h_s \sim (\delta \Delta\mu_s)/(k_BT)$
so that there is a local increment  at the surface of the chemical potential difference between the two species. It  determines
which species of the solvent  is preferentially adsorbed at
the surface of the colloid. The preference for 
one component of a binary liquid mixture  may be so strong as to saturate the surface of the colloids
with the preferred component, which corresponds to $h_s = +\infty (-\infty)$. For two colloids this  gives rise to  symmetry-breaking boundary conditions (denoted by  $(+, +)$ or $(-, -)$) for the  solvent order parameter.
The resulting spatial variation of the order parameter poses a significant  complication
for obtaining analytic results. 
Moreover, non-planar geometries lower the symmetry of the problem.
Apart from a few exceptions and limiting cases, the presently available analytical results for  CCFs are of approximate character.

At first, for two spherical colloidal particles, 
the corresponding CCP  has been studied theoretically right at the bulk critical point.
De Gennes \cite{deGennes} has proposed the singular effective  interaction potential between two widely separated spheres by using a  free energy functional, which goes beyond mean-field in the sense 
that it incorporates the non-classical bulk critical exponents, but neglects the
critical exponent $\eta$; $\eta$ is the standard bulk critical exponent for the
two-point correlation function at criticality with $\eta \sim 0.04$ in $d=3$ and $\eta =1/4$ in $d=2$ \cite{Pelissetto-et:2002}. 
Within this approach he has found that for $d=3$ the energy of interaction in the limit $D\gg R$
is   $-21.3\times k_BT_c^{(s)}R/D$.  
For that interaction potential,  the  
so-called protein limit, which  corresponds to $D/R$, $\xi/D \gg 1$ and which is based on exact arguments using conformal invariance,
renders,  within a  small-sphere expansion \cite{Burkhardt-et:1995,Eisenriegler-et:95},
\begin{eqnarray}
 \label{eq:10}
U_C(D;T = T_c^{(s)},h_b=0,R)\sim R^{d-2+\eta}D^{-(d-2 +\eta)} & \stackrel{d=3}{=}& (R/D)^{-(1 +\eta)} \\
                                                               &\stackrel{d=2}{=}& (R/D)^{\eta}, \nonumber           
\end{eqnarray} 
which implies for the scaling function in Eq.~(\ref{eq:4}) $\Theta\left(\mathcal{Y}=0, \Delta=D/R \to \infty, \Lambda = 0\right) \sim \Delta^{-(d-3+\eta}) 
\stackrel{d=3}{=} \Delta^{-\eta} $. In Ref.~\cite{Burkhardt-et:1995} the amplitude of $U_C$  in $d=3$ has been  estimated to be slightly larger than $\sqrt{2}$, which is a factor of ca 15
smaller than  the above prediction given by de Gennes \cite{deGennes}.
In the  opposite, the so-called Derjaguin limit $D \ll R$, 
one has \cite{Burkhardt-et:1995,Eisenriegler-et:95} 
\begin{eqnarray}
 \label{eq:11}
  U_C(D;T=T_c^{(s)},h_b=0,R) \sim R^{(d-1)/2}D^{-(d-1)/2} &\stackrel{d=3}{=}& R D^{-1}=\Delta^{-1}  \\ 
                                                         &\stackrel{d=2}{=}& (R/D)^{1/2}=\Delta^{-1/2}, \nonumber                                                      
 \end{eqnarray}
which in turn implies for the scaling function in Eq.~(\ref{eq:4})
 $\Theta\left(\mathcal{Y}=0, \Delta = \frac{D}{R}\to 0, \Lambda = 0\right) = const$.
These  results  confirm that the  CCFs  can indeed successfully compete with direct dispersion
\cite{Dantchev-et:2007,Dantchev-et:2017} or electrostatic forces in determining the
stability and phase behavior of colloidal systems.

\subsubsection{Mean field theory}
\label{subsubsec:mf}

Concerning the full range of parameters, theoretical predictions for the universal scaling function of the CCP between spheres are
 available only within
mean field  theory \cite{Hanke-et:1998,Schlesener-et:2003}.
Within this Landau - Ginzburg - Wilson approach,
 the CCF is conveniently calculated using the stress tensor ${\cal T}(\phi({\bf r}))$ in terms of the mean field profile $\phi({\bf r})$
 \cite{Eisenriegler-et:1994}:
\begin{align}
    \label{eq:12}
      {\bf f}_C = k_BT \int_{\cal A}d^{d-1} r {\cal T}(\phi({\bf r})) \cdot {\bf n}  
\end{align}
where ${\cal A}$  is an arbitrary $(d-1)$-dimensional surface enclosing a colloid, ${\bf n}$ is  the outward normal  of this surface,
and   ${\bf f}_C = f_C {\bf e}$ is
the force between two colloids, where ${\bf e}$ is a unit vector along the line connecting 
their centers. The orientation of ${\bf e}$ is such that $f_C< 0 \;(>0)$ corresponds to
attraction (repulsion). In most cases the  mean-field profile $\phi({\bf r})$  is  determined by numerical minimization of 
the Landau-Ginzburg-Wilson  Hamiltonian  encompassing  suitable surface contributions from the colloid surfaces in order to account for  symmetry-breaking
boundary conditions  there. 
In   $d=3$ spatial dimensions such  a theory is approximate. In the spirit of a systematic expansion in terms of $\epsilon=4-d$, it is exact in $d=4$ 
for four-dimensional spheres.
Recently, the same approach has been employed in order to calculate the scaling function of the CCFs
 for three-dimensional spheres posing
as hypercylinders $(H_{d=4,d^*=3})$ in spatial dimension $d=4$ \footnote{In the present  notation $d$ and $d^*$ correspond to $D$  and  $d$ used in Ref.~\cite{Mohry-et:2014}, respectively.}
  ~\cite{Mohry-et:2014}, where  $H_{d,d^*}=\{\bf{r}=(\bf{r}_{\perp},\bf{r}_{||})\in \mathbb{R}^{d^*}\times \mathbb{R}^{d-d^*}\mid |\bf{r}_{\perp}|\le R \}$.
The obtained  results
differ from the ones for four-dimensional spherical particles $H_{d=4,d^*=4}$ in $d=4$. This raises the question whether $H_{d=4,d^*=3}$ or $H_{d=4,d^*=4}$ 
renders the better
mean-field approximation
for the physically relevant case of three-dimensional spheres $H_{d=3,d^*=3}$ in $d=3$. 
Due to this uncertainty more accurate theoretical approaches are highly desirable. Recent  MC simulations  of a sphere near a wall constitute a first step in this direction \cite{Hasenbusch}.

The  mean field Landau-Ginzburg-Wilson theory is  a versatile approach for calculating CCFs. 
In the case of the simple film geometry, this approximate approach  reproduces correctly the qualitative behavior of CCFs 
(i.e., their sign, functional form, and  structure)
for various combinations of surface universality classes.
Therefore it  has been used to calculate CCFs for various shapes of colloidal particles and for other geometries.
Non-spherical, i.e.,  highly ellipsoidal or  sphero-cylindrical colloids,  or elongated particles  such as cylindrical micelles~\cite{cyl_mic}, 
block copolymers~\cite{block_copol},
the mosaic tobacco virus~\cite{mos_tob_vir}, and carbon nanotubes~\cite{carbon_nan} are experimentally available and widely used in the corresponding current research efforts, with 
application perspectives towards new materials in mind. 
The  orientation dependent CCP for ellipsoidal particles near a planar wall
at  vanishing bulk field $h_b = 0$ has been studied in Ref.~\cite{Kondrat-et:2009}. In this case,  due to the anisotropy of the particles,
there is not only a force  but  also a torque acting on the particle. 
This may lead to additional interesting  effects such as the orientational ordering of nonspherical colloids in a critical solvent.
The behavior of hypercylinders $H_{d=4,d^*=3}$ and $H_{d=4,d^*=2}$ near  
planar, chemically structured substrates  have been studied in Ref.~\cite{Troendle-et:2009-10}.
In Ref.~\cite{Mattos-et:2013}, the scaling function associated with CCFs for a system consisting of two spherical particles 
facing a planar, homogeneous substrate has been calculated. This  allows one to determine
 the change of the lateral CCF between two colloids upon approaching a wall which acts like a large third body.
 Within the applied mean field theory, this many body contribution can reach up to 25$\%$ of the pure pair interaction. As one would expect, the
 many-body effects were found to be  more pronounced for small   distances, 
as well as for temperatures close to criticality. This trend has been confirmed by studying three parallel hypercylinders within mean field theory \cite{Mattos-et:2015}
and three discs in $d=2$ within MC simulations \cite{Hobrecht-et:2015}. Three-body interactions in $d=3$ for spherical colloids have been determined experimentally \cite{Volpe-et}.
However, at the present stage these data cannot yet be compared quantitatively with theoretical results.
In Ref.~\cite{Labbe-Laurent} the CCP   for and the ensuing alignment of cylindrical colloids near chemically
patterned substrates  has been determined within mean field theory. The case in which the  particles 
exhibit spatially inhomogeneous surface properties, forming so-called  Janus particles
which carry two opposing boundary conditions, has  also been considered \cite{Labbe-Laurent-et:2016}. The experimental fabrication of such particles is of
research interest in itself \cite{fabrication,Labbe-Laurent-et:2016} as is the theoretical understanding of the interactions between spherical \cite{spherical_Janus,Labbe-Laurent-et:2016} or 
non-spherical Janus particles \cite{Labbe-Laurent-et:2016,nonspherical_Janus}, because they are considered
to be promising building-blocks for self-assembling materials~\cite{materials,Iwashita-et:2013,Iwashita-et:2014}.

\subsubsection{Beyond mean field theory}
\label{subsubsec:bmf}

Beyond mean field theory, theoretical predictions for  the CCP
 for two  spheres immersed in a critical fluid are available from conformal field methods,
which, however,  are restricted to the bulk critical point of the solvent  and to $d=2$.
As mentioned in  passing, the  exact analytical results  for the limiting behavior of spheres
which nearly touch (Eq.~(\ref{eq:11})) and spheres which are widely separated (Eq.~(\ref{eq:10}))  have 
been obtained by using conformal invariance of the free energy  
and taking  the Derjaguin limit in the first case and applying the small-sphere expansion in  the second case, respectively~\cite{Burkhardt-et:1995}.
 For the $2d$ Ising universality class, the CCP has been calculated numerically  \cite{Burkhardt-et:1995} within  the full range of distances $D$  between two spheres.  
 More recently  this was achieved
 also analytically \cite{Machta-et:2012}
 via the partition function of the  critical Ising model
on a cylinder, using a conformal mapping onto an
annulus.
For non-spherical colloidal particles with their dummbell or lens shapes being small compared to the correlation length and to the interparticle distances,
exact results for the orientation-dependent CCFs have been  obtained by using a small-particle operator expansion and by exploiting  conformal invariance
for $d \lesssim 4$  and $d=2$ \cite{Eisenriegler:2004}.
 In fact, in $d=2$  conformal field theory provides a general scheme for critical Casimir
interactions between two (or more) objects of arbitrary
 shape \cite{Bimonte-et:2013}. This can be  achieved by using the local conformal mappings
of the exterior region of two such objects  onto a circular annulus for which the stress tensor is known.
Assuming the availability of  a
simple transformation law for the stress tensor under any  such (local) conformal mappings,
the CCFs are  obtained from the contour integral of the transformed  stress tensor 
along a contour surrounding either one of the two objects  \cite{Bimonte-et:2013}.

Another nonperturbative approach, which allows one to calculate CCFs directly at a fixed specific spatial dimension
 \textthreequartersemdash  \; potentially  an advantage
over field-theoretic approaches based on  a systematic expansion in terms of  $\epsilon = 4-d$ \textthreequartersemdash \;  is the use of semi-empirical free energy functionals for 
 critical inhomogeneous fluids and Ising-like systems. 
They have been developed by Fisher and 
Upton~\cite{Fisher-et:1990} in order to extend the original de 
Gennes-Fisher critical-point ansatz~\cite{Fisher-et:1978,Fisher-et:1980}.
Upon construction, these functionals  fulfill the necessary analytic properties as a 
function of $T$ and a proper scaling behavior for arbitrary $d$.  The only  input  
  needed is the bulk  Helmholtz free energy and the values of the critical exponents. (However, the available functional is valid only for symmetry breaking boundary conditions.)  The  predictions  of this functional 
for films with $(+,+)$ boundary  conditions are in  very good agreement with previous results
obtained $\epsilon$-expansion and  conformal invariance for the scaling function of the order parameter  and for the 
critical Casimir amplitudes \cite{Borjan-et:1998}. 
Also the predictions for the full scaling functions of the  CCFs
at $h_b =0$ \cite{Borjan-et:2008} are in  good agreement with results from Monte Carlo
simulations. These  predictions have been obtained from a linear parametric 
model, which  in the neighborhood of a critical point
provides a simple scaled  representation    of the Helmholtz free energy in terms of  \hspace{1pt} ``polar'' coordinates
$(r,\theta)$ centered at the critical point $(t,h_b,\phi)=(0,0,0)$, where 
$r$ is a measure of the  distance from the critical point,  and which assumes a linear relationship between
the OP $\phi$ and $\theta$. 
A similar local-functional approach proposed by Okamoto and Onuki~\cite{Okamoto-et:2012} 
uses a  form of the bulk Helmholtz free energy which differs from
 the one employed in Ref.~\cite{Borjan-et:2008}, in that it is 
a field-theoretic expression for the free energy with  (in the sense of  renormalization group theory) renormalized coefficients.
Such a version does not seem to produce more accurate results for the Casimir amplitudes ~\cite{Okamoto-et:2012}.
  Within this renormalized local-functional approach, the  CCFs  between  two spherical particles immersed in a near-critical binary 
mixture (consisting of $a$ and $b$ particles) has been   calculated  as a function of  both scaling fields, $T$ and $h_b$ \cite{Okamoto-et:2013}. 
 The focus of the study  is the situation that the  $b$-rich phase  forms on
the colloid surfaces in thermal equilibrium with the subcritical $a$-rich 
bulk solvent ($T <T_c^{(s)}$  and $h_b < 0$). This  gives rise to a bridging transition between
two spherical particles and the resulting effective forces are of different nature than the CCFs. 
We note that the validity of  the extended 
de Gennes-Fisher or renormalized local-functional approach 
in the presence of bulk ordering fields has not yet been tested, not even for the simple film geometry.

\subsubsection{Computer simulations}
\label{subsubsec:cs}
In cases in which one cannot obtain analytical results, Monte Carlo simulations offer a highly welcome tool in order to overcome 
the  shortcomings of approximate theoretical approaches 
 and   to study  CCPs within the whole
temperature range and also in the presence of the bulk ordering field $h_b$. In the case that the solvent  is a simple  fluid and in the spirit
of the universality concept of critical phenomena,  one can 
study the simplest representative of the corresponding universality class of the critical solvent, e.g.,  spin models such as the Ising lattice gas model.
In the case of a lattice, the derivative in Eq.~(\ref{eq:8}) is replaced  by a finite difference $\Delta {\cal F}^{ex}$ associated with a  single lattice spacing. 
This also requires to introduce a lattice version of a spherical  particle.
In general, Monte Carlo methods are  not efficient to determine
quantities, such as the free energy, which cannot be expressed in terms of  ensemble
averages. Nevertheless, free energy differences  can be cast into such a form via, e.g., 
the so-called ``coupling parameter approach'' (see, e.g., Ref.~\cite{Mon}). 
This approach can be applied  for     systems characterized by two  distinct Hamiltonians ${\cal H}_0$ and ${\cal H}_1$
 but the same configurational space.  In such a case, one can introduce the {\it cr}ossover Hamiltonian ${\cal H}_{cr}(\lambda) = (1-\lambda){\cal H}_0 + \lambda {\cal H}_1$,
 which  interpolates between ${\cal H}_0$ and ${\cal H}_1$  as the  crossover parameter $\lambda \in [0,1]$  increases from 0 to 1.
The difference between the free energies of two systems characterized by these different Hamiltonians  can be conveniently expressed
as an integral  with respect to $\lambda$ over canonical ensemble averages of ${\cal H}_1-{\cal H}_0$ (with the 
averages taken by using the corresponding  crossover Hamiltonian for a given
value of the coupling parameter).
 Alternatively, the free energy difference can be determined  by 
integrating the corresponding difference of internal 
energies over the inverse temperature. 
 The drawback of both methods is that they usually require  
 knowledge of the corresponding {\it bulk} free energy density (as in the case of CCFs for a slab geometry). The accurate computation of the {\it bulk}
free energy density poses a numerical challenge by itself and extracting it from finite-size data requires a very
accurate analysis. Moreover, the  internal energy differences, when determined by using  standard MC algorithms,  are affected by
 huge variances,  especially  for non-planar geometries  such as the sphere - planar wall or the sphere - sphere geometry. In such geometries,
the differences between local energies of the two systems with different (by  one lattice spacing) sphere - wall or sphere - sphere 
distances  are large only close to the sphere or the wall. This implies that 
the variance is dominated by the remaining part of the system, for which the local energy difference is very small. 
Recently a more sophisticated algorithm \cite{Hasenbusch} and new 
approaches ~\cite{Hobrecht_Hucht:14} have been developed in order to
overcome these  problems.
The numerical  method proposed in Ref.~\cite{Hobrecht_Hucht:14} is analogous  to the experimental one used  by Hertlein {\it et al.}~\cite{Hertlein-et:2008}
according to which the CCP  is  inferred directly  
from the Boltzmann distribution function of the  positions of the two interacting  objects.
(In this experiment a sphere performs Brownian motion near a planar substrate.)
So far, the improved algorithm of Ref.~\cite{Hasenbusch} and the dynamic method of Ref.~\cite{Hobrecht_Hucht:14}
have been applied only to the slab  geometry and to the geometry of a single sphere (or disc) near a planar wall. 
The only available simulation data  for  two quasi-spheres in the $d=3$  Ising model have been   
obtained   by using  a method based on the integration of the local magnetization over the applied local magnetic field \cite{Vasilyev:2014}.
In that study, the CCP  has been  calculated at fixed distances between the two spheres  as a function of the temperature scaling variable (related to 
${\mathrm sgn}(t)(D/\xi_t)$)
for a few values of $h_b$,  or as a function of the bulk field scaling variable (related to ${\mathrm sgn}(h_b)(D/\xi_h) $) 
for several temperatures. Results of this calculation have been  obtained only  for   
small separations $D\le 2R$,  where 
$R=3.5$ (in  units of the lattice spacing $a$)   is the radius of the particle, because  the strength  of the  potential decreases rapidly upon increasing  $D$.
The accuracy of these data deteriorates at low temperatures.

Alternative ways of computing the sphere - sphere CCP via MC simulations have been used for  two-dimensional lattice models. 
One of them \cite{Machta-et:2012} uses Bennet's method \cite{Bennet:76},
according to which one can efficiently   estimate the free energy difference between two 
canonical ensembles, characterized by two different energies $E_0$ and $E_1$ with the same configuration space, provided 
that these two ensembles exhibit a significant overlap of common configurations.
This method has been employed for   the
 $2d$ Ising model,  with frozen spins forming two quasi-spheres with an effective radius $R$; in the reference ensemble 
 (with  energy $E_0$)
  the quasi-spheres are separated by a distance  
 $D$ whereas in the second ensemble  (with  energy $E_1$), they are separated by a distance  $D+a$ ~\cite{Machta-et:2012}.
In order to estimate the corresponding free energy difference,  within this approach
 one considers a trial move which keeps the  configuration space the same but switches
 the energy from  $E_0$ to $E_1$, e.g., by mimicking the move of  one of the particles from a distance $D$  to a distance $D+a$ by
 suitably changing and exchanging spins.
The estimate of  the  free energy difference is given by $F_1 - F_0  = -(1/k_BT)\ln\langle \exp(E_0-E_1)/(k_BT)\rangle_0$, where the canonical ensemble average $\langle \; \cdot \; \rangle_0$ is 
 taken with respect to the ``reference'' ensemble with the energy $E_0$. Integrating this free energy difference up to infinity, one obtains 
 the CCP. Reference~\cite{Machta-et:2012} provides the results for the CCP
 as a function of  distance $D/R \gtrsim 25$ for  four values of the temperature
 in the disordered phase of the solvent, i.e.,  $T\ge T_c^{(s)}$ and $h_b=0$. At the critical temperature, 
 these MC simulation
 results agree well with the analytical ones obtained from  conformal field theory.
Still another route towards  determining  the CCP from  MC simulations has been followed in Refs.~\cite{bob-et:2014,Tasios-et:2016} for 
two quasi-discs immersed in  a $2d$ lattice model of a binary liquid mixture  \cite{Rabani-et:2003}.
Here, the pair potential $V_C(x,y)$ has been obtained from the numerically determined
probability $P(x,y)$ of finding one colloid at position $(x,y)$
provided that another one is fixed at the origin:   $V_C(x,y) = -k_BT \ln(P(x,y)/P(\infty,\infty))$. 
In order to determine  $P(x,y)$ accurately at fixed values of $h_b\sim \Delta\mu-\Delta\mu_c$ and $T$,  the  so-called Transition Matrix Monte Carlo technique~\cite{Errington:2003} has been employed.
This technique relies on monitoring the attempted transitions
between macrostates, as defined, for example, by  specific positions of the two quasi-discs, and on using this knowledge in order to infer their
relative probability $P(Y)/P(Y')$, where $Y$ and $Y'$ are two distinct states of the system;  once sufficient transition data are
 collected,  the entire probability distribution $P(Y)$ can be constructed. 
The studies in Refs.~\cite{bob-et:2014,Tasios-et:2016} have been focused on the case that the  b-rich phase  adsorbs on
the colloid surfaces in thermal equilibrium  with a supercritical a-rich 
bulk solvent ($T \ge T_c^{(s)}$  and $h_b \le 0$). It has been found that data for $U_C = V_C/(k_BT)$ obtained for  $h_b=0$ and   $D \ll R$ 
collapse on a common master curve if multiplied by $(D/R)^{1/2}$ and plotted as a function of $D/\xi_t$, i.e., 
$(D/R)^{1/2}U_C = \Theta\left(\mathcal{Y}= {\mathrm sgn}(t)D/\xi_t, \Delta = D/R =0,\Lambda = {\mathrm sgn}(h_b)D/\xi_h = 0 \right)$. 
The scaling exponent 1/2 for the prefactor agrees with the prediction in Eq.~(\ref{eq:11}) for $d=2$, which is obtained based on the conformal
invariance in the  Derjaguin limit $D\ll R$, and which implies $\Theta(y \to 0, \Delta \to 0, \Lambda \to 0) = const$.

The standard algorithms used for the MC simulation studies described above are based on    trial moves  generating a trial configuration, which are local (Metropolis-type flips of spins)
and hence  become very slow near the critical point. This critical slowing down effect can be weakened by using  so-called cluster algorithms such as the Swendsen and Wang \cite{Swendsen-et:1987} 
 or  the  Wolff algorithm \cite{Wolff:1989}  in which   instead of a single spin a whole cluster of spins is flipped simultaneously. The Wolff algorithm constructs clusters which  consist of spins that are aligned and  connected by bonds. The proof that the Wolff algorithm obeys detailed balance, and hence  generates
 the Boltzmann distribution, hinges on the spin inversion symmetry of the Hamiltonian. (The Wolff algorithm can be generalized to  systems which contain bulk or surface fields \cite{Binder_Landau}.)
 One can exploit also  other symmetries in order to to develop a cluster method, for example, by using geometric operations
 on the spin positions such as, e.g., point reflection or rotation with respect to a randomly chosen  ``pivot''.
Hobrecht and Hucht~\cite{Hobrecht-et:2015} have 
extended  the geometric cluster algorithm (GCA), introduced by Heringa and Bl\"ote for bulk Ising models in the absence of external fields \cite{Heringa-et:98}, 
 to the case of  Ising systems containing areas of spins with a fixed orientation, facilitated by infinitely strong bonds, which mimic a colloidal suspension. The GCA makes use of the invariance of the  Hamiltonian 
  with respect to  a point inflection in order to construct
 two symmetric clusters of spins which are then exchanged.  Contrary  to the  Wolff cluster algorithm, the GCA conserves the order parameter.
The modification due to  Hobrecht and Hucht consists of including  into the clusters not only spins but also the  bond configuration  between them.
This  way, the particles \texttwelveudash encoded into the bond configuration \texttwelveudash can be moved and the configuration of a solvent  represented by the Ising spins can be updated within one cluster step.
In the cluster exchange the neighboring lattice sites $i$ and $j$ as well as the connecting bond $\langle ij\rangle$ are mapped via
point reflection with respect to a pivot onto the sites $i'$ and $j'$ and the bond $\langle i'j'\rangle$, respectively. 
Using this apparently very efficient MC cluster algorithm, the authors of Ref.~\cite{Hobrecht-et:2015}
have studied two-dimensional systems with a fixed number of identical, disclike particles defined as  regions of fixed positive spins, which  thus effectively impose  symmetry-breaking (+) BCs onto the surrounding free spins. Within this scheme, they have
calculated the two- and three-body CCP at the bulk critical point of the Ising model.   
The authors report strong  
 finite-size effects: for   periodic simulation boxes with a fluctuating total magnetization, the presence of  a nonzero number density of colloidal particles with a non-neutral surface preference 
 for up and down spins 
 shifts the system away from the critical point.  As a result, their MC results for the  CCP do not exhibit   the form expected  to hold at the critical point for a single pair of particles in solution  (Eqs.~(\ref{eq:10}) and (\ref{eq:11}). In order to suppress this effect,
the authors of Ref.~\cite{Hobrecht-et:2015} have proposed to use a fixed total magnetization $M=0$ or to insert in addition  the same number of 
particles but with the opposite surface preference.

\subsubsection{Derjaguin approximation}
\label{subsubsec:DA}

Within the so-called  Derjaguin approximation  curved, smooth surfaces 
are approximated by surfaces which are a steplike sequence of parallel planar pieces~\cite{Derjaguin:1934}.
Between two vis-\`{a}-vis,   flat pieces of the opposing surfaces  partitioned this way, locally
a force  acts like in the slab geometry and the total force is taken
to be the sum of the forces between each individual pair with the appropriate areal weight. 
The   Derjaguin approximation is widely used to estimate CCFs forces between colloidal particles,  because
the forces between parallel surfaces  are  much easier to calculate. Indeed, for  
the slab geometry the CCFs  are known even beyond mean field theory
(see below).
Therefore,  by using this  approximation one can  account  even for non-Gaussian critical fluctuations, albeit  at the expense of  
not fully considering  the shape of the particles. The Derjaguin  approximation  is valid  for temperatures
which  correspond to $\xi_t \lesssim R$, because under
this condition  the CCFs between the colloids act only at
surface-to-surface distances $D$ which are small compared with  $R$.
In many cases this approximation is surprisingly reliable even
for $D\lesssim R$ \cite{Gambassi-et:2009,Troendle-et:2009-10}.  
In the absence of a  bulk ordering field, i.e., for $h_b=0$
and  strongly adsorbing  $(+,+)$ confining surfaces 
the results for the critical Casimir interactions in the film geometry have been  provided by field-theoretical studies \cite{Krech:1997},
the extended de Gennes - Fisher local functional
method ~\cite{Borjan-et:2008,Okamoto-et:2012,Mohry-et:2014},  
and Monte  Carlo simulations of the Ising model 
\cite{Vasilyev-et:2007,Vasilyev-et:2009,Hasenbusch:2012}  or   improved
models which offer the benefit that the amplitude of
the leading bulk correction to scaling vanishes \cite{Hasenbusch:2012,Hasenbusch:2010,Francesco-et:2013,Hasenbusch:2015}.
Within  MC simulations  also the case of weakly  adsorbing surfaces  has been considered \cite{Vasilyev-et:2011,Hasenbusch:2011}. 
In this latter case  the corresponding surface field $h_s$  might be so small that upon approaching the critical point one effectively observes
a crossover of the type of boundary condition imposed on the order parameter from symmetry preserving to symmetry breaking boundary conditions.
In this case  there appears to be no effective enhancement of the order parameter upon approaching the confining wall.
The CCP  reflects such crossover behaviors;
depending on the film thickness, the CCP  can even change sign  \cite{Vasilyev-et:2011,Hasenbusch:2011,Mohry-et:2010}.
 On the basis of scaling arguments one expects that for moderate adsorption preferences the scaling function of the CCP in film geometry
 additionally depends on the dimensionless scaling variables $y^{(i)}_{s} = a_ih^{(i)}_{s}L^{\Delta_1/\nu}, i= 1 , 2$, where $h^{(1)}_{s}$
 and $h^{(2)}_{s}$ are the effective surface fields at the two confining surfaces, $a_1$ and $a_2$  are  nonuniversal amplitudes,  $\Delta_1 =0.45672(5)$ \cite{Hasenbusch:2011}
 is the surface crossover exponent at the so-called ordinary surface phase transition.
 
\begin{figure}
	\includegraphics[width=0.8\linewidth]{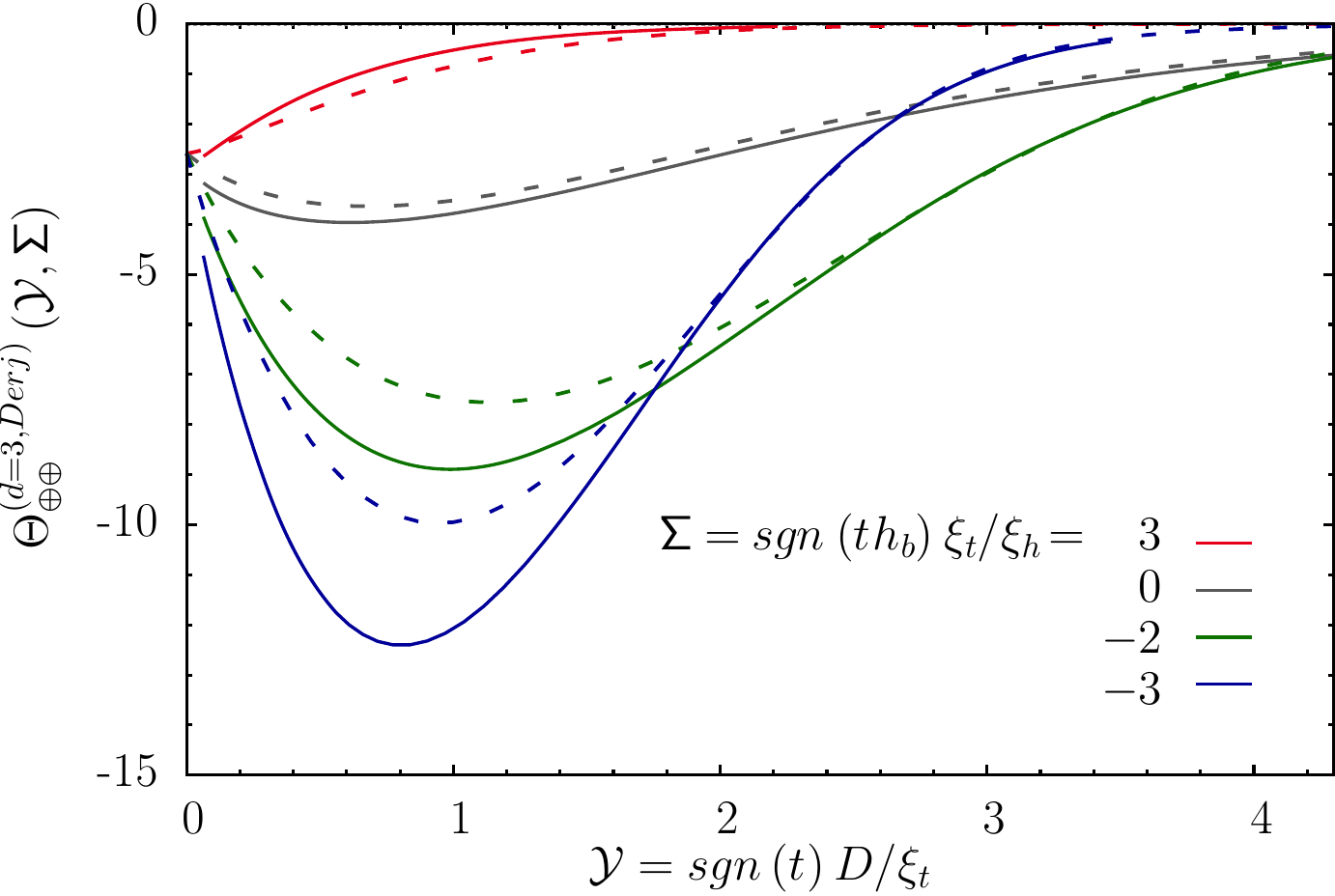}
	\caption{Scaling function $\Theta_{\oplus\oplus}^{(d=3,Derj)}(\mathcal{Y},\mathsf{\Sigma}) \simeq (D/R) U_c(D;t,h_b,R)$ (see Eq.~(\ref{eq:4}) and the main text) of 
	the sphere - sphere CCP, as obtained within the  Derjaguin approximation  by using the extended 
         de Gennes-Fisher functional (solid lines) and  the ``dimensional'' approximation (dashed lines)  as a function
         of the surface-to-surface distance $D$ (in units of $\xi_t$) for several values of the scaling variable $\mathsf{\mathsf{\Sigma}} = \Lambda/\mathcal{Y}= \mathrm{sgn}(th_b)\xi_t/\xi_h$
         related to the bulk ordering field $h_b$. At fixed temperature in the one-phase region of the solvent ($t>0$), 
         the CCP  is shorter-ranged  and much weaker for   $h_b>0$ (i.e., $\Sigma > 0$), which favors the same $(+)$ phase of the solvent 
         as the one preferred by the colloid surfaces, whereas $h_b < 0$ (i.e., $\Sigma < 0$) favors the other phase. }
	\label{fig:1}
\end{figure}

The knowledge of  the dependence of  CCPs  on the bulk ordering field $h_b$ is rather limited, although it is   crucial for understanding the aggregation of colloids near the 
bulk critical point of their solvent. The presently available MC simulations for  Ising films provide  such
results only along the critical
isotherm $T=T_c^{(s)}$ \cite{Vasilyev-et:2013}.
In Refs.~\cite{Schlesener-et:2003,Mohry-et:2012a,Mohry-et:2012b,Mohry-et:2014}
the variation of CCFs  with  $h_b$ has  been  approximated by adopting the functional  {\it 
form}  obtained within   mean field theory (MFT, $d=4$)  by using a field-theoretical approach within the framework
of the Landau - Ginzburg theory, but by keeping the actual critical exponents in $d=3$ for the scaling variables.
Within  this ``dimensional'' approximation, the scaling function $\vartheta^{(d=3)}_{||} = f_C D^3/(k_BT\mathcal{S})$ of the CCF for the film geometry $(||)$ with macroscopically 
large surface area $\mathcal{S}$ of one wall (see the text below Eq.~(\ref{eq:9})) is taken  to be
$\vartheta^{(d=3)}_{||}(\mathcal{Y} = \mathrm{sgn}(t)D/\xi_t, \mathsf{\Sigma}=\Lambda/\mathcal{Y}= \mathrm{sgn}(th_b )\xi_t/\xi_h )\simeq \vartheta^{(d=3)}_{||}(\mathcal{Y},\mathsf{\Sigma}=0)\vartheta^{(d=4)}_{||}(\mathcal{Y},\mathsf{\Sigma})/\vartheta^{(d=4)}_{||}(\mathcal{Y},\mathsf{\Sigma}=0)$,
where $\vartheta^{(d=3)}_{||}(\mathcal{Y},\mathsf{\Sigma}=0)$ has been adopted from MC simulation data ~\cite{Vasilyev-et:2007,Vasilyev-et:2009}.
(The normalization by $\vartheta^{(d=4)}_{||}(\mathcal{Y},\mathsf{\Sigma}=0)$ eliminates  a nonuniversal prefactor carried by the Landau - Ginzburg expression for the scaling function of CCFs.)  
In  Ref.~\cite{Mohry-et:2014} the scaling functions $\vartheta^{(d=3)}_{||}$ resulting from the ``dimensional'' approximation have been compared with those obtained 
within the extended de Gennes - Fisher local functional. This allows us to compare in Fig.~\ref{fig:1}  (unpublished) results for the scaling function $\Theta_{\oplus\oplus}^{(d=3,Derj)}(\mathcal{Y},\mathsf{\Sigma})$ of the sphere - sphere CCP (see Eq.~(\ref{eq:4})) within
the Derjaguin approximation, i.e., $\Theta_{\oplus\oplus}^{(d=3,Derj)}(\mathcal{Y},\mathsf{\Sigma})\simeq (D/R) U_c =  \pi \int_1^{\infty} \mathrm{d}x(x^{-2}-x^{-3})\vartheta^{(d=3)}_{||}(x\mathcal{Y},\mathsf{\Sigma})$ \cite{Mohry-et:2014}, where $\vartheta^{(d=3)}_{||}$ is obtained by using these two aforementioned approaches.
Only for weak bulk fields the curves in Fig.~\ref{fig:1} for 
both approaches compare well.
Otherwise there are clear quantitative discrepancies, the origin of which is  not clear. The ``dimensional'' approximation might deteriorate upon increasing  $h_b$. On the other hand, as already  mentioned in passing, the reliability of the local functional
approach for  $h_b \ne 0$ has not yet been tested systematically. For example,  the result of the local functional
approach for the scaling function of the  CCFs along the critical isotherm differs substantially from the one obtained from  MC simulations \cite{Mohry-et:2014}.
 In Refs.~\cite{Buzzaccaro-et:2010,Piazza-et:2011} results of a long wavelength analysis of density functional theory for the scaling function of CCF  have been reported (see also Sec.~\ref{subsubsec:pp} below ); they are in  qualitative agreement with the results
obtained using the extended de Gennes - Fisher local functional \cite{Mohry-et:2014}.
The first microscopic off-lattice results for the scaling function of CCF, both along the critical isochore and in the off-critical regime, have been worked out in Ref.~\cite{Anzini-et:2016}.
The authors of Ref.~\cite{Anzini-et:2016} have studied a hard core Yukawa model of a fluid by using density functional theory (DFT) within a specific weighted density approximation  \cite{Leidl-et:1993}, coupled with  hierarchical reference theory \cite{Parola-et:1995}.
This kind of  DFT weighted density approximation, related to the one proposed early on by Tarazona and, independently, by Curtin and Ashcroft \cite{Tarazona:1985,Curtin-et:1985}, captures the short-ranged correlations of the underlying hard sphere fluid rather accurately, whereas the hierarchical reference theory is able to account also for the critical properties of a homogeneous fluid.
In Ref.~\cite{Anzini-et:2016}, this technique has been applied  in order to determine  the effective interaction between two hard walls immersed  in that fluid.
This approach facilitates to investigate the crossover between  depletion-like roots of such effective forces at high temperatures and the critical Casimir effect upon approaching the critical point of the fluid.
It appears that for hard core Yukawa fluids the universal features of CCF  emerge only in close neighborhood of the critical point. 
The predictions obtained in Ref.~\cite{Anzini-et:2016} for the scaling function of the CCF for various temperatures along the critical isochore ($h_b=0$)
differ significantly from the Ising model MC simulation results in Ref.~\cite{Vasilyev-et:2009}.
Moreover, the data along various isotherms do not collapse as  expected from scaling theory. The authors of Ref.~\cite{Anzini-et:2016}
interpret these deviations as an indication of  strong corrections to scaling occurring in the hard core Yukawa fluid.

As far as   experimental data for the {\it film} geometry are concerned, they are available  only indirectly from  measurements of
 the thickness of wetting films; they correspond to opposing $(+,-)$
boundary conditions \cite{Fukuto-et:2005,Rafai-et:2007}. The  scaling function
of the CCF determined  by Fukuto {\it et al.} \cite{Fukuto-et:2005} compare well with the theoretical predictions.
The measurements performed by  Rafai {\it et al.} \cite{Rafai-et:2007} have provided  
data which show stronger deviations  from the theoretical curves, in particular around the maximum
of the CCF.
The scaling function of CCFs for wetting films
corresponding to $(+,+)$
boundary conditions could not be determined due to the collapse of the incomplete wetting film 
in the course of the measurements \cite{Rafai-et:2007}. 
Based on  exact results  in $d=2$ \cite{Evans_Stecki, Abraham_Maciolek:2010,Abraham_Maciolek:2013}  and in  $d=4$ \cite{Krech:1997},  and based  on the predictions  from the local-functional approach in $d=3$ \cite{Borjan-et:2008},
 it is expected that the CCF $f_C(D)$  between two planar walls  with symmetry-breaking boundary conditions at separation $D$
 varies as  $\exp(-D/\xi_t)$  for $D\gg \xi_t$ and  $t >0$, i.e., in the one-phase region of the solvent.
This implies that the asymptotic behavior of  the scaling function is given by $\vartheta^{(d=3)}_{||}(\mathcal{Y}= \mathrm{sgn}(t)D/\xi_t\gg 1,\mathsf{\Sigma} = \mathrm{sgn}(th_b )\xi_t/\xi_h=0)=\mathcal{A}_+\mathcal{Y}^d\exp(-\mathcal{Y})$, where $\mathcal{A}_+=1.2-1.5$ is a universal number valid for $(+,+)$ boundary conditions~\cite{Gambassi-et:2009}.  
By applying the Derjaguin approximation to this functional form of the CCFs, one obtains the following  asymptotic behavior of the scaling function 
 $\Theta_{\oplus\oplus}^{(d=3,Derj)}(\mathcal{Y},\mathsf{\Sigma}=0)$ for the sphere - sphere CCP : 
\begin{equation}
\label{eq:13}
 \Theta_{\oplus\oplus}^{(d=3,Derj)}(\mathcal{Y}\to\infty,\mathsf{\Sigma}=0) = \pi \mathcal{A}_+\mathcal{Y}e^{-\mathcal{Y}}, \qquad D\gg \xi_t, \qquad t>0,
\end{equation}
 so that in this limit $U_C^{(d=3,Derj)}=\pi\mathcal{A}_+\frac{R}{\xi_t}e^{-D/\xi_t}$ (Eq.~(\ref{eq:4})).
 
\subsubsection{Pair potential in the presence of depletants}
\label{subsubsec:pp}

In order to calculate the effective pair potential between two large hard spheres  immersed into a fluid
of depletants close to their  gas-liquid critical
point,  the authors of  Refs.~\cite{Buzzaccaro-et:2010,Piazza-et:2011} have used   density functional theory  (DFT),  which is a powerful tool for describing  equilibrium properties of colloidal suspensions  (see the last two paragraphs in Sec.~\ref{subsubsection:eff_inter}).  
The hard spheres  have been analyzed within  the Derjaguin approximation, which  is   inherent in recent approaches to 
depletion forces acting  in hard-sphere mixtures. In this context, the Derjaguin approximation relates 
the force between the two big objects  to the integral of the solvation force $f_{solv}(D) = -(1/\mathcal{S}) (\partial\Omega/\partial D)_{\mu,T,\mathcal{S}} -p)$
 of the small particles (i.e., the depletant agents) confined between two parallel planar walls  with cross-sectional area $\mathcal{S}$.  (The film solvation force per surface area
 $\mathcal{S}$, $f_{solv}$, is an excess pressure over the bulk value $p$ of the confined fluid  described by the grand potential
$\Omega$.) 
In the limiting case of hard walls exposed to an ideal gas of depletants, this relation  reproduces the well known Asakura-Oosawa result for depletion forces \cite{Asakura-et:1954}. This scheme can be
fruitfully applied also to   interacting systems.
In fact such a relation is equivalent to the general formula obtained by Derjaguin \cite{Derjaguin:1934} relating the force between two convex
bodies to the  free energy, in excess of its bulk value, of a fluid confined between planar walls. Within an actual DFT approach this formula holds even for approximations of the excess part
of an intrinsic Helmholtz free energy functional of a fluid.  
The authors of  Refs.~\cite{Buzzaccaro-et:2010,Piazza-et:2011} have employed   the square gradient
local density approximation for the intrinsic Helmholtz free energy functional entering into DFT, which is valid for spatially slowly varying depletant number  densities $\rho$ such that $\nabla\rho/\rho \ll 1/\xi_t$.
They have simplified  the Derjaguin approximation by replacing the two big spheres by two planar parallel walls.
Within the square gradient local density approximation, the effective (solvation) force per area $f_{solv}$ between two parallel walls is expressed solely in terms 
of the bulk free energy density $f(\rho)$ of the  depletant host fluid and in terms of the value of the equilibrium number density profile $\rho(z=z_{mid})$
of such a fluid at  the midpoint $z=z_{mid}$ between the two walls.  The solution of the  extremum condition for the free energy functional provides
the second equation relating, albeit in an implicit way, $f_{solv}$ and $\rho(z=z_{mid})$ to the density at the wall $\rho(z=0)$, which in turn depends on the wall-fluid interactions. 
In order to incorporate the non-Gaussian behavior near the gas-liquid critical point of the depletant, in the spirit of the Fisk - Widom or local-functional approach \cite{Fisk:1969,Fisher-et:1990,Okamoto-et:2012}, the authors of Refs.~\cite{Buzzaccaro-et:2010,Piazza-et:2011} have used the scaling form for the singular part of the bulk free energy density of the fluid
$\left(f(\rho)-f(\rho_c)\right)/(k_BT) = a_{11}t^{2-\alpha} \Psi(x)$ with the scaling variable $x=b_1\phi t^{-\beta}$, where $\phi = \rho -\rho_c$ is the bulk order parameter OP, $a_{11}$ and $b_1$ are non-universal, dimensional constants, and $\beta$ and $\alpha$ are bulk critical  exponents~\cite{Pelissetto-et:2002}.
In the scaling limit, the general expression for the effective force between two parallel walls 
 can be expressed in terms of  $\Psi(x)$, the rescaled midpoint OP $x_0=b_1\left(\rho(z_{mid})-\rho_c\right)t^{-\beta}$, and  the universal amplitude ratio
 $g_4^+$ \cite{Pelissetto-et:2002}. This expression takes the scaling form characteristic of  CCFs.
By using a  parametric expression for $\Psi$ \cite{Pelissetto-et:2002} and the critical exponents of $3d$ Ising universality class,   the authors have obtained 
 very good agreement between their analytic theory and  the scaling function of CCFs obtained  from  MC simulations \cite{Vasilyev-et:2009} at $h_b=0$.
This is not surprising because, within  the square gradient local-density approximation,   DFT   reduces to the  local-functional approach
which proved to  capture correctly critical fluctuations \cite{Fisher-et:1990,Borjan-et:1998,Borjan-et:2008}, at least for $h_b=0$.

This theoretical  approach  has been followed up  by  MC simulation studies \cite{Gnan-et:2012a} aiming at the computation of
the CCP  between   two hard-sphere colloids suspended
in an implicit solvent  in the presence of interacting depletant particles.  
In this  off-lattice MC simulations of fluctuation induced forces,  the effective potential  between two hard spheres
has been determined upon  approaching the  gas-liquid critical point of the depletant for two different depletant models, one for SW and one for 3P particles (see Eq.~(\ref{eq:7}) and thereafter).
Given the computational limitations, the authors have considered the  size ratio $q$ between 
the hard-sphere depletant  and the hard-sphere colloid within the range  $ 0.05 < q < 0.2$.
The resulting effective colloid-colloid force has been evaluated by using canonical Monte Carlo simulations for various  fixed values 
of the depletant  concentration  in the reservoir.  The method consists of  performing virtual displacements of each colloid
from its fixed position  
and of computing the probability  of encountering at least one collision  with  a depletant particle. 
The effective colloid-colloid potential follows from integrating the corresponding force. These  MC results show 
that upon cooling the effective potential between two colloidal particles   gradually  looses its high-temperature, pure hard-sphere depletion  character 
 of exhibiting oscillations and transforms  into
a completely attractive potential with a progressive and significant increase of its  range,  signaling the onset of critical Casimir forces.
For large distances between the surfaces of the two colloids, the MC data for the CCP fit well to the asymptotic form given by Eq.~(\ref{eq:13}).
 This numerical study has been extended to the case in which colloids interact with  SW depletant particles; this interaction has been  continuously
modified from hard-core repulsion to strong attraction, thus changing from $(-,-)$ boundary conditions (i.e.,  preference for the gas phase)  to $(+,+)$ boundary conditions (i.e., preference for the liquid phase)
\cite{Gnan-et:2012c}. For strong  colloid-depletant attraction, the effective colloid-colloid potential exhibits oscillations, as they occur for
  the high-temperature depletion potential, modulating  its exponentially decaying attractive tail.
The variation of the colloid-colloid effective potential upon crossing over from $(-,-)$ to $(+,-)$  and from $(+,+)$ to $(+,-)$ 
boundary conditions has been determined, too. In the asymptotic spatial  range these two crossovers are the same.  However, in the 
  numerical study of Ref.~\cite{Gnan-et:2012c} only the behavior at  short distances has been probed, where the effective potential 
  is dominated by  non-universal aspects of the solvent-colloid interaction  so that the aforementioned pairs of boundary conditions are no longer equivalent.

\subsubsection{ Main properties of CCFs}
\label{subsubsec:sumCCP}

\begin{figure}
	\includegraphics{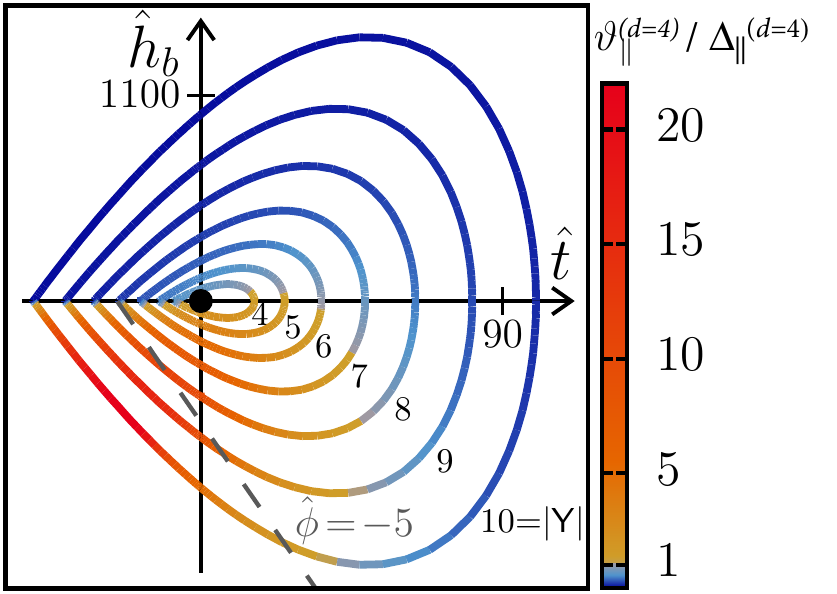}
	\caption{ Normalized  mean-field CCF  scaling 
	function $\vartheta^{(d=4)}_{||}(\mathsf{Y},\mathsf{\Sigma}) = D^4f_{C}^{||}/(k_BT{\cal S})$, where $ \mathsf{Y}=\mathrm{sgn}(t)D/\xi(t,h_b)$ and $\mathsf{\Sigma}=\mathrm{sgn}(th_b)\xi_t/\xi_h$, for films
	(of thickness $D$ and macroscopically large cross-sectional area  ${\cal S}$) along isolines of constant scaling variable  
	$\mathsf{Y}=4,5,\ldots, 10$  (from the inner to the outermost ring) 
	 in the thermodynamic  state space   of the  solvent spanned by 
	 $\hat t = (D/\xi_{t,+}^{(0)})^{1/\nu}t$ and 
	$\hat{h}_b = (D/\xi_h^{(0)})^{\beta\delta/\nu}h_b$ (see Fig. 1 in Ref.~\cite{Mohry-et:2014});
	$\xi(t,h_b)$ is the bulk correlation length of the solvent  with $\xi_t=\xi(t,h_b=0)$ and $\xi_h=\xi(t=0,h_b)$.
	(Note that depending on the particular thermodynamic path under consideration, 
        representations of the scaling function of the critical Casimir force can be more convenient  in terms of other
        scaling variables, such as in Fig.~\ref{fig:1} where  $\mathcal{Y}=\mathrm{sgn}(t)D/\xi_t$ is chosen.
	The color along the lines of constant $\mathsf{Y}$  indicates the 
	absolute value $|\vartheta^{(d=4)}_{||}/\Delta^{(d=4)}_{||}|$.  
	 The \emph{bulk} critical point of the solvent $(\hat t,\hat{h}_b)=(0,0)$ is 
         indicated by $\bullet$. The region  shown here lies {\it above}
	 the capillary transition critical point,
	 where the \emph{film} coexistence line ends.
	 For $(+,+)$ boundary conditions as considered here, the  capillary condensation transition occurs 
	 for  $\hat t<0$ and $\hat{h}_b<0$.	 %
	The  dashed line indicates the path of constant order parameter $\phi$ of the solvent
        $\hat\phi=\frac{D}{\xi_{t,+}^{(0)}}\phi/\mathcal{B}=-5$, where $\mathcal{B}$ is the   non-universal amplitude 
        of the bulk OP $\phi=\mathcal{B}t^{\beta}$.  Within mean-field theory
	$\nu=\beta=1/2$ and $\nu/(\beta\delta)=1/3$. 
	$\Delta_{||}^{(d=4)}=\vartheta^{(d=4)}_{||}(\mathsf{Y}=0,\mathsf{\Sigma}=0)$.}
	\label{fig:2}
\end{figure}

We  close  this  section by listing  the main properties of  critical Casimir interactions between
two colloidal particles in very dilute suspensions. 
The  results for spheres in $d = 3$, obtained from  the  variety of  methods described  above,  tell that they share
the same qualitative features of the CCFs with ``spheres'' in $d=2$ and $d=4$ as well as with  planar walls.
This refers to the property that  for like boundary conditions CCFs are attractive 
and it refers to the position  of a force maximum along various thermodynamic paths.
 Around the consolute point of a binary solvent the main features   of the CCFs  between two planar walls 
are  summarized in Fig.~\ref{fig:2}
in terms of  the  force  scaling function $\vartheta^{(d=4)}_{||}(\mathsf{Y},\mathsf{\Sigma})$ obtained 
from  Landau theory (see Fig.~1 in Ref.~\cite{Mohry-et:2014}). 
The main message conveyed by Fig.~\ref{fig:2}  is  the  
asymmetry  of  the CCFs around the critical point
of the solvent with  the  maximum strength  occurring at $h_b<0$.
This asymmetry is due to the presence of  surface fields, which in films  lead to capillary condensation
\cite{Evans:1990}, whereas between spherical colloids  to  bridging
\cite{Bauer-et:2000,Archer-et:2005,Okamoto-et:2013}. Near these phase transitions, 
the effective force acting between the confining 
surfaces is attractive, exhibits (within Landau theory) a  cusp, and is very strong: the depth of the corresponding
CCPs can reach a few hundred $k_BT$. 
The concomitant strong increase of the absolute value of the force is  reflected 
 by its corresponding universal scaling function and extends  
to the thermodynamic region above the capillary condensation  or bridging 
critical point, even to temperatures  $T >T_c^{(s)}$. 

\subsection{Stability}
\label{subsec:stab}

Knowing the effective pair potential for the colloids one can investigate the stability of colloidal suspensions and the aggregation of colloids.
These phenomena  are related to
kinetic processes (see Ref.~\cite{Russel-et:1989} and references therein), which 
 are based on the diffusion of single particles in the presence
of other particles of the same kind, interacting with them via
interaction potentials which contain both attractive and repulsive contributions.
Aggregation occurs if 
the attractive interactions of the particles  dominate over  their thermal kinetic energy, which is responsible for  the Brownian motion of the particles.
Hydrodynamic interactions may also  play a role, e.g., by slowing down the aggregation process for solvents
with  high viscosity.

In order to quantify the behavior of  interacting particles, which irreversibly stick together 
once their surfaces touch each other, Fuchs has introduced 
the concept of a  stability ratio \cite{Fuchs:1934}. It is defined as the ratio $W=J_0/J$
between   the Brownian motion induced pair formation rate $J_0$ in the absence  of other than excluded volume interactions 
between the particles,   and the corresponding formation rate $J$ of  particle pairs in the presence of
such interactions. 
$W$ can be calculated by extending   Smoluchowski's diffusion equation for the
radially symmetric relative motion of  two coagulating spherical 
particles of radius $R$ in order  to account for their interaction potential \cite{Fuchs:1934}:
\begin{equation}
  \label{eq:15}
	W = 2R \int_{2R}^{\infty}\frac{\exp\left\{U(r)\right\}}{r^2} \text{ d}r.
\end{equation}
The analysis, which   leads to this expression, is valid only in the early stages of coagulation before triplets etc. are formed.
 However, it neither  deals with  the very beginning of the coagulation
process but considers only the steady-state situation, which is  established quickly. 
In the analysis based on  Smoluchowski's equation   hydrodynamic
interactions are neglected.
From the definition of  $W$ it follows that 
for hard spheres $W=1$, while for $W>1$ ($W<1$) the repulsive (attractive) 
part of the pair  potential $(k_BT)U$ dominates. In the case  of  a potential 
barrier, i.e., if $U(r)\gg 1$ for a certain range of distances $r>2R$, 
which leads to $W>1$, one can expect that on intermediate time scales the suspension will 
be in  a (meta)stable homogeneous state. The cluster formation will set in   only  
on a very large time scale. This time scale is proportional 
to the ratio between the characteristic times for diffusion of a single particle, 
$t_{diff}$, and for the formation of a pair of particles, $t_{pair}$ \cite{Russel-et:1989}:
\begin{equation}
\label{eq:16}
   t_{diff}/t_{pair}=3 \eta/W
\end{equation}
with  the packing fraction $\eta=(4\pi/3)R^{3}\rho$ where  $\rho$ is the 
number density of the colloidal particles.
For a near-critical solvent, the pair  potential can be 
taken as (see Eqs.~(\ref{eq:2}) and (\ref{eq:4}); $\mathsf{\Sigma} = \Lambda/\mathcal{Y}$)
  \begin{widetext}
\begin{equation}
\label{eq:17}
   U(r)=
    \begin{cases}
      \infty, & D<0 \\
      U_{rep}+U_C^{(d=3)}=
      A\exp(-\kappa D)+(1/\Delta)\Theta^{(d=3)}(\mathcal{Y},\Delta, \mathsf{\Sigma}),  & D>0.
    \end{cases}
\end{equation}
\end{widetext}
For the mixed phase of the solvent the ratio $W$ has been calculated in Ref.~\cite{Mohry-et:2012b}.
The results of this calculation show that the CCF
can lead to a rapid coagulation,  setting  in within a
narrow temperature interval, and that the extent of the coagulation region reflects the fact that  the CCFs
are stronger for compositions of the solvent slightly poor in the component preferred by the colloids.
The effective pair potential given by Eq.~(\ref{eq:17}) is applicable only for  sufficiently large
distances $D\gtrsim\kappa^{-1}$, because it takes into account only the electrostatic repulsive interactions
 and neglects possible short-ranged
contributions to the effective  van-der-Waals interactions.
Furthermore, the CCP attains its universal form
as given by the scaling function $\Theta^{(d=3)}$   only in the scaling limit, i.e., for distances $D$ 
which are sufficiently large compared with the correlation length amplitude 
$\xi_{t,+}^{(0)}\approx 0.25nm$.  Analogously, also $\xi_t$ and $\xi_h$
must be sufficiently large compared with microscopic scales.
From the behavior of the scaling function $\Theta^{(d=3)}$  (discussed  in Ref.~\cite{Mohry-et:2012b}  within the Derjaguin
approximation)  it follows  that along the typical thermodynamic paths realized experimentally, 
the {\it range} of  attraction due to the CCFs grows steadily upon
increasing the bulk correlation length $\xi_t$, but the {\it amplitude} of the CCFs
is a  nonmonotonic function of $\xi_t$ with its maximal strength attained for an intermediate value of
$\xi_t$. Depending on the values of $A$ in Eq.~(\ref{eq:17}) and of $\xi_t$, the CCFs compensate the repulsion for all
values of $D$ or only within certain ranges of $D$, i.e.,  a
secondary attractive minimum of $U(r)$  can occur  at a certain distance $D_{min}$ (in
addition to the primary, global minimum at $D \to 0^+$) while for small distances $D < D_{min}$  and for $D<0$ the
potential maintains a repulsive part. The presence of 
the repulsive barrier and of the attractive secondary minimum in the effective potential, and thus the occurrence of coagulation, depends on temperature.
But  also in the case,
in which for {\it all} values of $\xi_t$ a repulsive barrier
remains, coagulation can appear, due to a deep secondary minimum.
The nonmonotonic dependence of the maximal strength of
the CCF on temperature results in a  nonmonotonic
behavior of the ratio $W$, which is  shown  in Figs.~2 and 3 of Ref.~\cite{Mohry-et:2012b}.

In Ref.~\cite{Mohry-et:2012b}, the analysis of the stability ratio for the diversity of possible shapes of the effective pair potential given 
in Eq.~(\ref{eq:17}) has been complemented  by the analysis of the bulk structure of  colloidal suspension.
To this end the  radial distribution function $g(r)$ has been calculated  within  the integral equation approach 
by using  the hypernetted-chain and the Percus-Yevick closure \cite{Hansen-et:1976}. The results for both types of closure are almost
the same.
(The applicability and reliability of this integral equation approach 
is discussed in detail in Ref.~\cite{Caccamo:1996}.)
 For temperatures
far away from the critical temperature of the solvent, the colloids behave  effectively as hard spheres with
an effective diameter $\sigma > 2R$ due to the soft repulsive background contribution $U_{rep}$. 
Accordingly, for such values of $\xi_t$,
$g(r)$ has the corresponding characteristics of a fluid of hard
spheres, such as the rather broad first peak for small values of $D$. Due to the emerging attractive CCFs, for increasing $\xi_t$
the radial distribution function $g(r)$ is enhanced close to the surfaces of the colloids. This implies an enhanced short-ranged order
and that the formation
of colloidal dimers is favored. The way in which the  shape of the radial distribution function $g(r)$ changes upon increasing the 
temperature reveals whether the effective potential exhibits a repulsive
barrier at small values of $D$ and is attractive throughout large
distances or whether  an attractive minimum develops  at
 intermediate values of $D$ upon increasing temperature while repulsion
 remains at small and large values of $D$.

\begin{figure}
	 \includegraphics{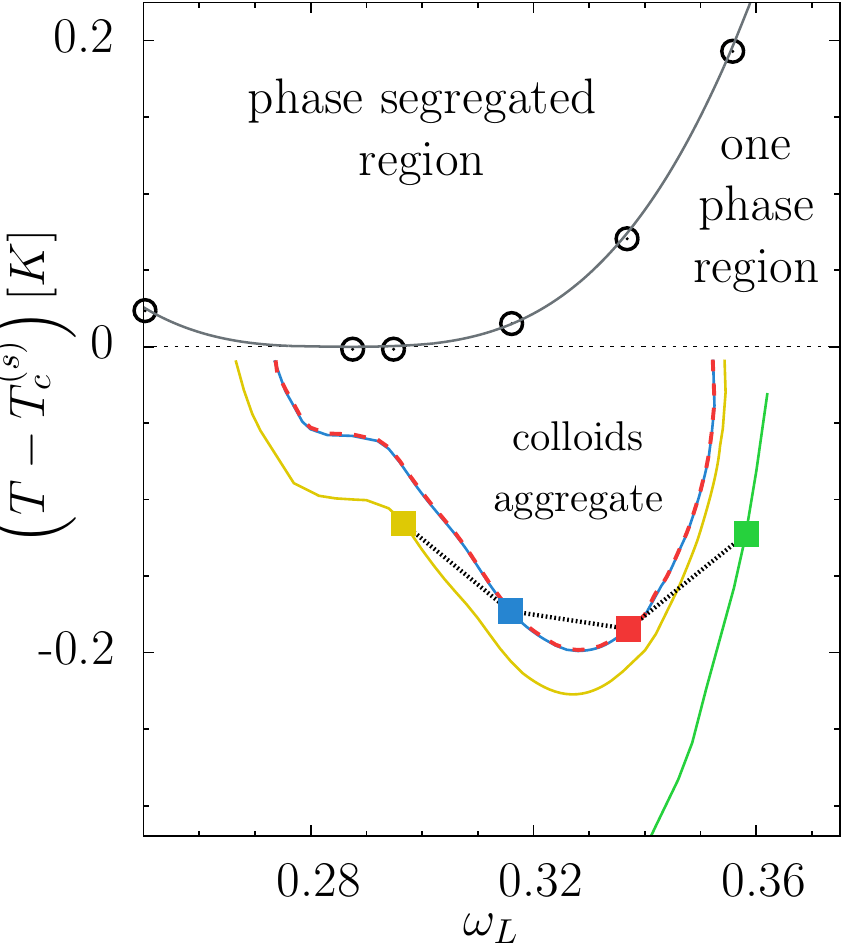}
	 \caption{ 
	 The experimentally determined coexistence points  [$\odot$] of the binary liquid mixture lutidine-water  \cite{gallagher:92}
	 which exhibits a lower, continuous  demixing phase transition. $T_c^{(s)}$ is the critical
	 temperature of this demixing transition and $\omega_L$ is the mass fraction of lutidine. These coexistence points on the binodal of phase segregation  agree well with the
        relation $|\omega_L-\omega_{c,L}|=B_{\omega}|t|^{\beta}$ , where $\beta = 0.3265$ and $B_{\omega} = 0.765$ (dark gray line).
	 Squares denote the  experimentally obtained state points of the onset of aggregation 
         (the straight black dotted lines in between are a guide to the eye)  taken from the middle of Fig.~1 in 
         Ref.~[11(b)]. 
         Each isoline of constant $B_2$  (full, colored lines; for visibility of the blue line
         the red one is dashed, both lines nearly coincide) corresponding to one of
         the state points (squares),  is calculated by using the effective potential given by Eq.~(\ref{eq:17}); their
         values are $B_2/(\frac{4\pi}{3}R^3) = -67$ (red square), -65 (blue square), -23 (yellow square), and 5.4 (green square)  (Fig.~8 in Ref.~\cite{Mohry-et:2012b}). 
          Each $B_2$-isoline can capture some qualitative trends of
	 the possible shape of the line of  onset of aggregation. The blue and the red lines 
	 reveal agreement, but the yellow and the green one not.
	 }
	\label{fig:3}
\end{figure}

One can address the issue whether a relationship
can be established between the onset of aggregation and the behavior of a certain quantity, which is accessible both
theoretically and experimentally, such as the second virial coefficient $B_2$. 
For dilute suspensions, the second virial coefficient \cite{Hansen-et:1976} 
 provides information about the strength
of the radially symmetric attraction between spherical particles:
\begin{equation}
\label{eq:18}
 B_2 = \frac{1}{2}\int \text{d}^3r \left(1-e^{-U(r)}\right) = 2\pi \int_{0}^{\infty}\text{d}r\hspace*{1pt} r^{2} \left(1-e^{-U(r)}\right) =  2\pi \int_{0}^{\infty}\text{d}r\hspace*{1pt} r^{2} \left(1-g(r)\right).
\end{equation}
Beyond the ideal gas contribution it determines the leading non-trivial
term in the expansion of the pressure
$p(\rho)/(k_BT\rho)=1+B_2\rho+\ldots$ in terms of powers of
the number density $\rho$. In Ref.~\cite{Mohry-et:2012b}  $B_2$ has been calculated for the potential given by Eq.~(\ref{eq:17})
at the state points for which, according to the  experiments  reported in Ref.~\cite{gallagher:92},  aggregation sets in. It turns out that
at these states of aggregation onset the values of $B_2$
are close to each other  and that to  a certain extent those $B_2$-isolines, which emerge by belonging to these experimental data points,  agree with each other and with the possible
shape of the aggregation onset line (see Fig.~\ref{fig:3}). However, based on this analysis one cannot
state definitely  that
$B_2$ can   serve as a quantitative indicator for the onset of
aggregation. Further efforts, both theoretically and experimentally,   are needed to provide more
precise values of the relevant quantities.

\subsection{Phase behavior}
\label{subsec:therm}

In   suspensions with sufficiently large packing fractions of colloids,  
the attraction among them due to CCFs can induce a so-called ``liquid-gas'' phase separation  of the colloids, i.e., the separation 
of two phases which differ with respect to their colloidal number density.
Adopting the  effective one-component
approach allows one to use standard liquid state theory in order to determine the onset of  phase separation.
Within this effective approach
 feedback mechanisms of the colloids,  acting on the solvent and  changing its critical
behavior, are neglected. Therefore this approximation does not allow one to describe reliably all details of the full
many-component system.
However, one can identify certain regions  of the thermodynamic phase
space for which this approach is applicable.
One expects the effective one-component model to work
well for temperatures corresponding to the one-phase region
of the pure solvent and for an intermediate range of values of the colloid   number density $\rho$.  The latter 
  should be large enough so that the competition between the configurational entropy and the potential
energy due to the effective forces can induce  a phase separation, but  small enough so that the
approximation of using an effective pair potential between the
colloids is valid and the influence of the colloids on the phase
behavior of the solvent is  secondary. 

\begin{figure}
\begin{center}
 	\includegraphics[width=\textwidth]{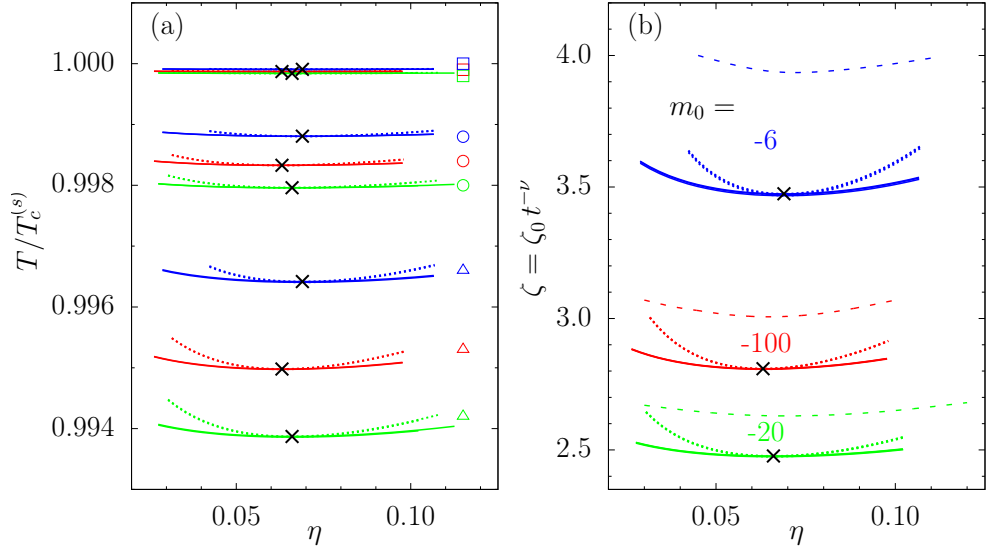}
\end{center}
\vspace{-15cm}
	\caption{In the plane spanned by temperature and colloidal 
	packing fraction $\eta$:
	(a) colloidal gas-liquid phase coexistence curves 
	$T^{(\mathrm{eff})}_{cx}(\eta=(4/3)\pi R^3\rho\,\vert\,c_a)$ (full lines, see, c.f., Fig.~\ref{fig:5}), spinodals 
	(i.e., loci of mean-field divergence of the isothermal compressibility $\chi_T$, dotted lines), and the
	critical points $T^{(\mathrm{eff})}_{c}$ (crosses) of 
	an effective, one-component system of large colloidal particles  as obtained by {\it density functional theory} (Fig. 2 in Ref.~\cite{Mohry-et:2012a}). 
        These particles of radius $R$ interact via an effective 
	potential given by Eq.~(\ref{eq:17}) with  parameters $\kappa R=10$ and $A=1000$ \cite{Mohry-et:2012a}  taken at $T^{(s)}_{c}$). The curves correspond to a {\it{}s}olvent with 
	a lower critical temperature $T_c^{(s)}$ and with various fixed solvent 
	compositions $c_a$ (Eq.~(\ref{eq:con})) represented by the variable $m_0=\mathrm{sgn}(\phi)(\zeta_0)^{1/\nu}|\mathcal{B}/\phi|^{1/\beta} = -100$ (red), $-20$ (green), 
	$-6$ (blue).  $\mathcal{B}$ is defined via the shape of the solvent binodal $\phi=\mathcal{B}t^{\beta}$ and 
	$\zeta_0=\kappa\xi^{(0)}_{t,+}$  (see below). The bulk critical exponents used here are $\nu=1/2$ and $\beta=1/2$.
	Close to the phase separation of the solvent the dominant temperature 
	dependence within the effective approach described  by Eq.~(\ref{eq:18})  is that of the critical Casimir forces 
	(CCFs),  encoded in
	$\zeta(t=1-T/T_c^{(s)})=\mathrm{sgn}(t)\kappa\xi_t(t)=\mathrm{sgn}(t)\zeta_0|t|^{-\nu}$. 
	Therefore, if the temperature (see panel (a)) is expressed in terms of $\zeta$ (see
        panel (b)) the members of each set of  curves with equal color in (a), corresponding to
        various  values of $\zeta_0$ ($\zeta_0=0.01\; (\Box)$, $0.05\; (\circ)$, and 
        $0.1\; (\vartriangle)$) fall de facto on top of each other and are characterized by $m_0$.
	The dashed lines in (b) correspond to the spinodals  determined within the {\it integral equation approach}. 
	(Due to  $m_0\sim(c_a-c^{(s)}_{a,c})^{-1}$, the quantities $m_0= \pm \infty$ correspond to the
	critical composition $c_{a,c}^{(s)}$). For solvent compositions  which                            
	are somewhat poor in the component preferred by the colloids, 
	i.e., for intermediate negative values such as $m_0 \simeq -20 $, 
	the critical Casimir forces are strongly attractive. Therefore, for them short
	correlation lengths suffice to bring about phase separation;
	accordingly the binodals occur at small values of 
	$\zeta$. Here 
	only thermodynamic states of the solvent which 
	are in the one-phase region, i.e., for $t>0$, are considered.
      }
      \label{fig:4}
\end{figure}

In Ref.~\cite{Mohry-et:2012a}, for particles interacting via the effective pair potential given in Eq.~(\ref{eq:17}), 
 the phase coexistence curve $T^{(\mathrm{eff})}_{cx}(\eta=(4/3)\pi R^3\rho\,\vert\, c_a)$ with the critical point temperature $T_c^{(eff)}$ and  the spinodal (i.e., the loci where, within mean-field theory, the isothermal compressibility 
 $\chi_T$ diverges) have been calculated using  density functional theory within the so called random phase approximation \cite{Evans:1979} and the integral 
 equation  approach \cite{Hansen-et:1976}. The results obtained from these two approaches, which are presented
 in  Fig.~\ref{fig:4} (Fig.~2 in Ref.~\cite{Mohry-et:2012a}),
  differ only slightly.
 The loci of the phase separation depends sensitively on the strength $A$ of the repulsive part of the effective pair potential (Eq.~(\ref{eq:3})) and the
 solvent compositions.  For  solvent 
 compositions, which are somewhat poor in the component preferred by the colloids, even short 
 correlation lengths suffice to bring about phase separation. The critical value $\eta_c$ of the colloidal packing fraction   $\eta=(4\pi/3)R^3\rho$ (in terms of number density $\rho$)
 is rather small,
i.e., $\eta_c\approx 0.07$, because 
the effective hard sphere diameter $\sigma$, which results from the 
soft repulsion $U_{rep}$, is larger than $2R$.
Furthermore, the binodals shown in Fig.~\ref{fig:4} are
rather flat compared with, e.g., the ones for hard
spheres interacting via a short-ranged, attractive, and temperature
independent potential.
In the present system, the  deviation of $T$ from the critical temperature 
$T_c^{(s)}$ which  corresponds to a range of $\eta$
for the coexisting phases as large as  the one shown in Fig.~\ref{fig:4},
is about $1$\textperthousand, whereas for a system of hard spheres with an  additional
attractive  interaction  the corresponding temperature deviation  is a few percent.
The critical temperature $T_c^{(eff)}$ obtained  within DFT agrees  with the
simple prediction $B^*_2 = B^*_{2,c}$, acting as an implicit equation for  $T_c$,
 as suggested by Vliegenthart and Lekkerkerker \cite{Vliegenthart-et:2000}
 and Noro and Frenkel \cite{Noro-et:2000} (VLNF), where $B_2^*(T) \equiv B_2(T)/B_2^{(HS)}$ is the reduced second virial coefficient and $B^*_{2,c}$ 
 is the critical value of  $B_2^*$ for Baxter's model of adhesive hard spheres ~\cite{Baxter:1968}.
$B_2^{(HS)} =\frac{2\pi}{3}\sigma^3$ is the second virial coefficient
of a suitable reference system of hard spheres (HS)
with diameter $\sigma$.  The effective hard-sphere diameter can be taken as $\sigma=\int_0^{r_0}(1-\exp (-U(r))\mathrm{d}r $, with $U(r=r_0)=0$.  
One can  adopt also other  definitions of  $\sigma$ (for a  discussion see Ref.~\cite{Andersen-et:1971}).
According to VLNF, $B_2^*$ is 
a useful indicator of the occurrence of a phase separation into
a colloidal-rich (``liquid'') and a colloidal-poor (``gas'') phase. An  extended law of corresponding states proposed by VLNF predicts that
the value of the reduced second virial coefficient
$B_2^*$
at the critical point is the same for all systems 
composed of particles with short-ranged attractive interactions, regardless of the 
details of these interactions.
This (approximate) empirical rule is supported by experimental data
\cite{Vliegenthart-et:2000} and  theoretical results \cite{thN}. 

\section{Multicomponent mixture}
\label{sec:ter}

\subsection{General discussion}
\label{subsec:gd}

\begin{figure}
	\includegraphics[scale=0.9]{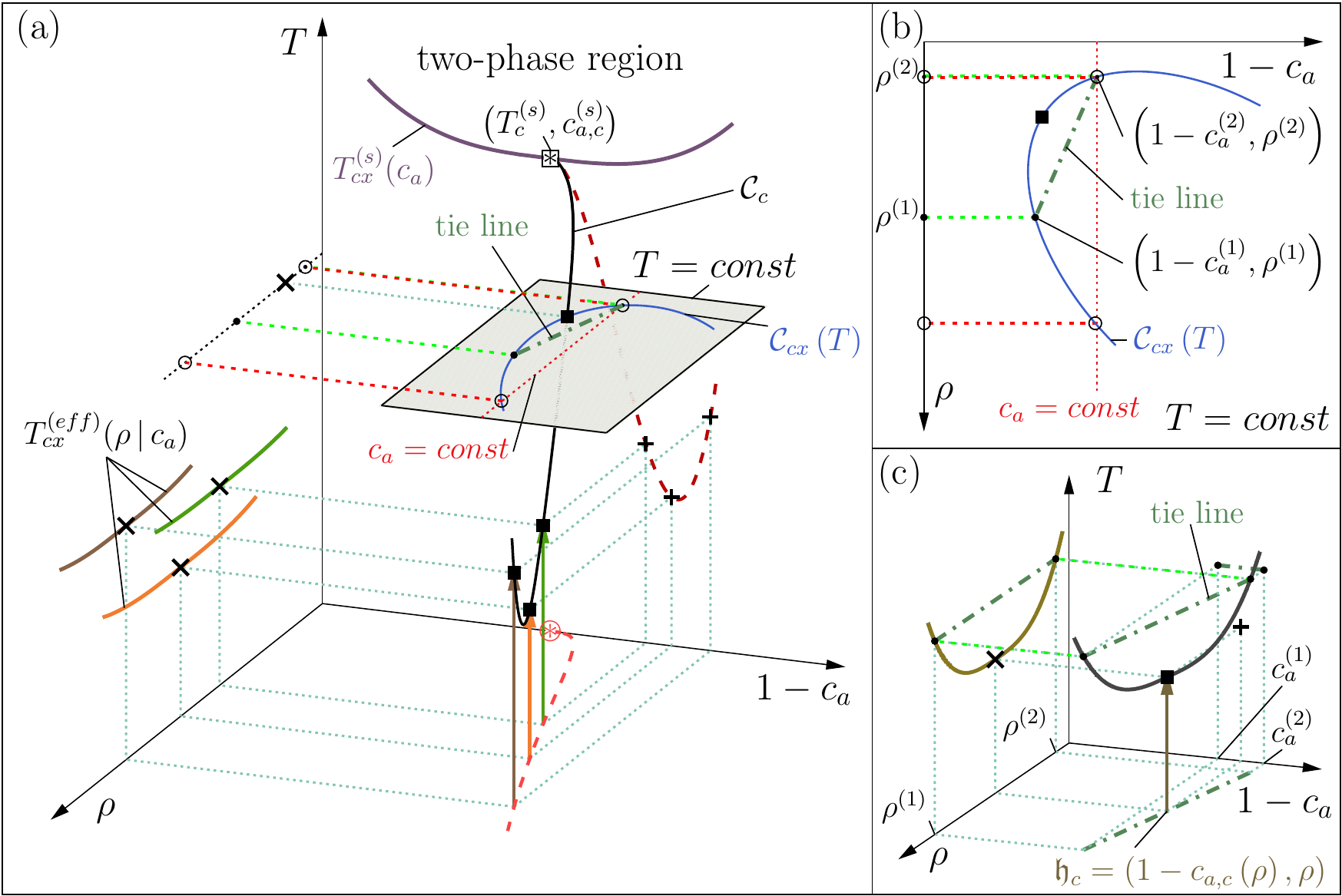}
	\caption{
	Sketch of the phase diagram for colloids immersed in a binary   
	liquid mixture at fixed pressure corresponding to a liquid state of the
        system  (Fig.~1 in Ref.~\cite{Mohry-et:2012a}).
	Upon adding colloids  the  phase separation curve 
        $T^{(s)}_{cx}(c_a)$ of a pure solvent  in the $(T,c_a,\rho=0)$ plane 
        extends to a  tube-like two-phase region $T_{cx}(c_a,\rho)$
	in the $3d$ thermodynamic  (td)  space spanned by the temperature $T$, the concentration $c_a$,
	and  the colloidal number density $\rho$.
        The lower critical point  $\boxast$ ($T_c^{(s)},c_{a,c}^{(s)},\rho=0$) of a pure solvent
	extends to a line  $\mathcal{C}_c$ (black curve) of critical points 
	(some of which are shown as black squares). Its shape reflects the fact that
	the $2d$ manifold $T_{cx}(c_a,\rho)$ of coexisting states is not straight 
	but bent and twisted due to the specific properties of the critical Casimir potential.
 	The red dashed curves denote the projections of $\mathcal{C}_c$  onto the planes $(\rho,c_a)$ and $(T,c_a)$.	 
	Within the effective one-component approach the coexistence curves  are explicit
	functions of $\rho$ only and depend parametrically on the overall
	concentration $c_a$.
	All three panels show that in general for $T=const$ 
        the coexisting phases (i.e., the points connected by a tie-line) 
        differ with respect to both  $\rho$  and $c_a$. Thus the 
	effective one-component approach has a limited applicability for determining 
	the phase diagram. Experimentally or within suitable, sufficiently rich  models, upon increasing 
        temperature (along thermodynamic paths  indicated by vertical arrows)  one is able to determine a  coexistence curve (black line in (c)) in 
	the $3d$ td space. For a description of further details see the caption of Fig.~1 in Ref.~\cite{Mohry-et:2012a}.
    } 
    \label{fig:5}
\end{figure}

For the kind of systems considered here, the determination 
of phase equilibria  is rather subtle because \texttwelveudash{} due to the concomitant adsorption phenomena, which  are state
dependent  \texttwelveudash{} the effective potential between the colloids
depends on the thermodynamic state itself.
Experiments  have revealed \cite{gallagher:92,grull:97} 
that for suspensions very dilute in colloids and
with a phase separated solvent, basically all colloidal particles are 
populating the phase rich in the component, say $a$, preferred by the colloids
(this phase has concentration $c_a^{(1)}$).
This implies that in coexisting distinct phases the effective potential acting between 
the particles is different. This is not captured by the effective potential approach discussed  above. 
Besides the colloid-solvent also  the solvent-solvent interactions can influence the effective 
potential and, accordingly,  the phase behavior of the effective 
colloidal system. This  has been demonstrated  by recent
MC studies in which various kinds of model solvents have
been used \cite{Gnan-et:2012a}. On the other hand, 
the presence of colloidal particles may alter the phase behavior of the solvent.
In the case of molecular fluids, it is well established  that the phase diagrams of 
ternary mixtures, as they emerge from those of binary mixtures by adding a third component, are   distorted and  become more complex
relative to the underlying original binary ones \cite{Andon-et:1952,Prafulla-et:1992}.  
Similar distortions and complex features of phase diagrams are observed  experimentally  
\cite{Kline-et:1994,Jayalakshmi-et:1997,Koehler-et:1997} upon  adding 
colloidal particles to  binary solvents. In particular, one finds a decrease of the lower critical temperature.
These observations  tell  that for  reliably determining such phase diagrams the full, many component mixture  has to be considered. 
 The importance of considering the colloidal suspension with a binary solvent
as a  truly ternary mixture has been already pointed out
in Ref.~\cite{Sluckin:1990}. 
Taking into account the  features of the CCP one can try to predict the 
expected  topology of the phase diagrams for  ternary solvent-solvent-colloid mixtures \cite{Mohry-et:2012a}.
Upon adding colloids, the two-phase  region of the demixed phases of the pure solvent extends 
into the three-dimensional thermodynamic space of the actual colloidal
suspension. At fixed pressure this space is  spanned  by $T$, the concentration $c_a$ of the component
$a$ of the binary solvent, and, e.g., 
by the colloidal number density $\rho$. 
The actual shape  of the two-phase region forming a tube-like manifold  is expected to depend sensitively on all
interactions present in the ternary mixture
(see Fig.~\ref{fig:5}). 
The foremost difficulty in describing such mixtures theoretically or in dealing with them via computer simulations consists of   the 
fact that the sizes of their constituents differ by, a few, 
orders of magnitude. This property distinguishes them significantly 
from mixtures of  hard spheres, needles, and polymers  \cite{Schmidt-et:2002,Schmidt:2011}, for which
the constituents are of comparable size, i.e., 
none of them is larger by a factor of ten or more than the others.
 In contrast to molecular ternary mixtures as modeled,  e.g., in terms of a  lattice gas  
\cite{terMixTheory}, in colloidal suspensions the colloidal particles influence 
the other two components not only by direct interactions but also 
via strong entropic effects. These occur because  their surfaces 
act as confinements to fluctuations of the concentration of the 
solvent and they also generate a sizeable excluded volume for the solvent 
 particles. 
\subsection{Monte Carlo studies}
\label{subsec:mcm}

 MC simulations for a lattice model of ternary mixtures, suitably mimicking  colloidal particles suspended in a near-critical binary
solvent, offer the   advantage over  other available approaches  that they account for  both 
fluctuations and  nonadditivity of the emerging CCFs. 
Due to the large size difference  between the
colloid and solvent particles and due to the critical slowing down of the
ternary mixture upon approaching its critical point, studying the corresponding
three-dimensional systems by MC simulations is computationally challenging.
As a computationally cheaper substitute, two-dimensional lattice models have been treated by this method in Refs.~\cite{bob-et:2014,Tasios-et:2016,Hobrecht-et:2015}.
In a first of such a MC simulation study \cite{bob-et:2014}, the colloids 
have been
modeled as  discretized hard discs of radius  $R$ (in units of the lattice spacing) occupying 
a fraction $\eta$ of sites on the square lattice. The remaining lattice sites have  been taken to be 
occupied by a solvent molecule of either species $a$ or $b$ with no empty sites left.
A similar  model of 
solvent-solvent-colloid mixtures has been considered earlier by  Rabani {\it et al.}  \cite{Rabani-et:2003}
in order to study drying-mediated self-assembly of nanoparticles.  
In Refs.~\cite{bob-et:2014,Tasios-et:2016},  only  the nearest
neighbor repulsive interactions between the components of a binary solvent have been considered.
This  drives  phase segregation, which in the absence of colloids has a critical point belonging to 
the $2d$ Ising universality class. In order to mimic the preference of the colloids for 
 component $b$ of the binary solvent, 
a nearest-neighbor attractive interaction between the colloid and component $b$  has  been taken into account. It has been assumed that there is 
 only a  nearest-neighbor repulsion between the colloid and component $a$.
In the limit $\Delta\mu_s = \mu_a - \mu_b \to  \infty$ of the chemical potential difference between the components $a$ and $b$,
  the ternary  mixture reduces to  a binary mixture of  $2d$  hard-discs and solvent  component $a$, which exhibits  coexistence
  between a fluid and a solid phase.
In the study by Edison {\it et al.} \cite{bob-et:2014}, the preclusion of  solvent-mediated colloidal
aggregation, arising from complete wetting and capillary
condensation,  has been implemented by focusing on colloids, which are 
immersed in a supercritical mixed phase of components  $a$ and $b$, such that  $b$-rich critical adsorption
layers  (for critical adsorption profiles at  spheres and cylinders see Refs.~\cite{Hanke-et:1999,Yubanaka-et:2017} and references therein) form on the colloid surfaces in equilibrium  with a supercritical $a$-rich solvent in the bulk. 
The careful simulation study in Ref.~\cite{Tasios-et:2016}, in which the solvent has been treated grand canonically at constant pressure,
has revealed the existence of 
gas-liquid and fluid-solid transitions, occurring in a region of the thermodynamic variables   $(\eta, t=(T-T_c^{(s)})/T_c^{(s)}, \Delta\mu_s = \mu_a - \mu_b)$ of ternary mixtures which  extends well away  from 
the critical point  of the solvent reservoir (especially concerning the fluid-solid transitions). It has been found that in  all  phases  
 the local solvent composition is strongly correlated
with the local colloid density. The coexisting colloidal liquid and solid phases
are poor in  component $a$, whereas in the gas phases 
coexisting with the liquid or the solid  phases  the solvent composition is very close to the composition of the solvent reservoir,  which, however, is far from its critical composition.
These features agree
with experimental observations \cite{gallagher:92,grull:97}
and cannot be captured by an effective one-component approach as discussed in the previous section.
As  expected \cite{Evans-et:1994}, all pair correlation functions decay exponentially on the scale of  the same correlation length.
Strikingly, the correlation length found 
in the homogeneous supercritical state of the ternary mixture was  much larger than the colloid
radius ($R=6$), which in turn clearly exceeded  the correlation length  of the solvent reservoir.
Upon adding a small volume fraction  of colloids, the gas-liquid critical point of a ternary mixture shifts continuously from that of the
colloid-free solvent to the negative values  of $\Delta\mu_s$, which decrease upon increasing $t$.
In addition,  the  MC simulation data indicate the existence of a second, lower metastable gas-liquid critical point  located at larger
values of $\eta$.  The   gas-liquid-solid triple point is also expected to occur  near the fluid-solid coexistence of pure hard discs shown as  the vertical dashed lines in Fig.~\ref{fig:6}.
The authors of Ref.~\cite{Tasios-et:2016} have checked to which extent a description in terms of  an effective pair
potential  can account for the phase behavior  observed  in their MC simulations. They have concluded that 
those approaches, which exclusively employ effective pair potentials, as
obtained from, e.g., planar slit studies combined with  the Derjaguin
approximation, overestimate the extent of gas-liquid coexistence
and underestimate how much the critical point of the ternary
mixture is shifted relative to  that of the solvent reservoir.

In Ref.~\cite{Hobrecht-et:2015},  the two-dimensional Ising model, with embedded colloids represented as 
disclike clusters of spins with fixed orientation, 
 has been treated by the geometric cluster algorithm (see Sec.~\ref{subsubsec:cs}). However, the focus of this study has not been the phase  behavior
of a ternary mixture but rather  the two- and three-body CCP. 
Employing the geometric cluster algorithm  facilitated the study of the  same lattice  model as in Refs.~\cite{bob-et:2014,Tasios-et:2016}  but in $d=3$~\cite{Tasios-et:2017}. As anticipated in Refs.~\cite{bob-et:2014,Tasios-et:2016}, the phase diagram
 displays the same qualitative features as in the two-dimensional case.

\subsection{Mean field theory}
\label{subsec:mftm}

\begin{figure}
%
         \includegraphics[width=0.43\textwidth]{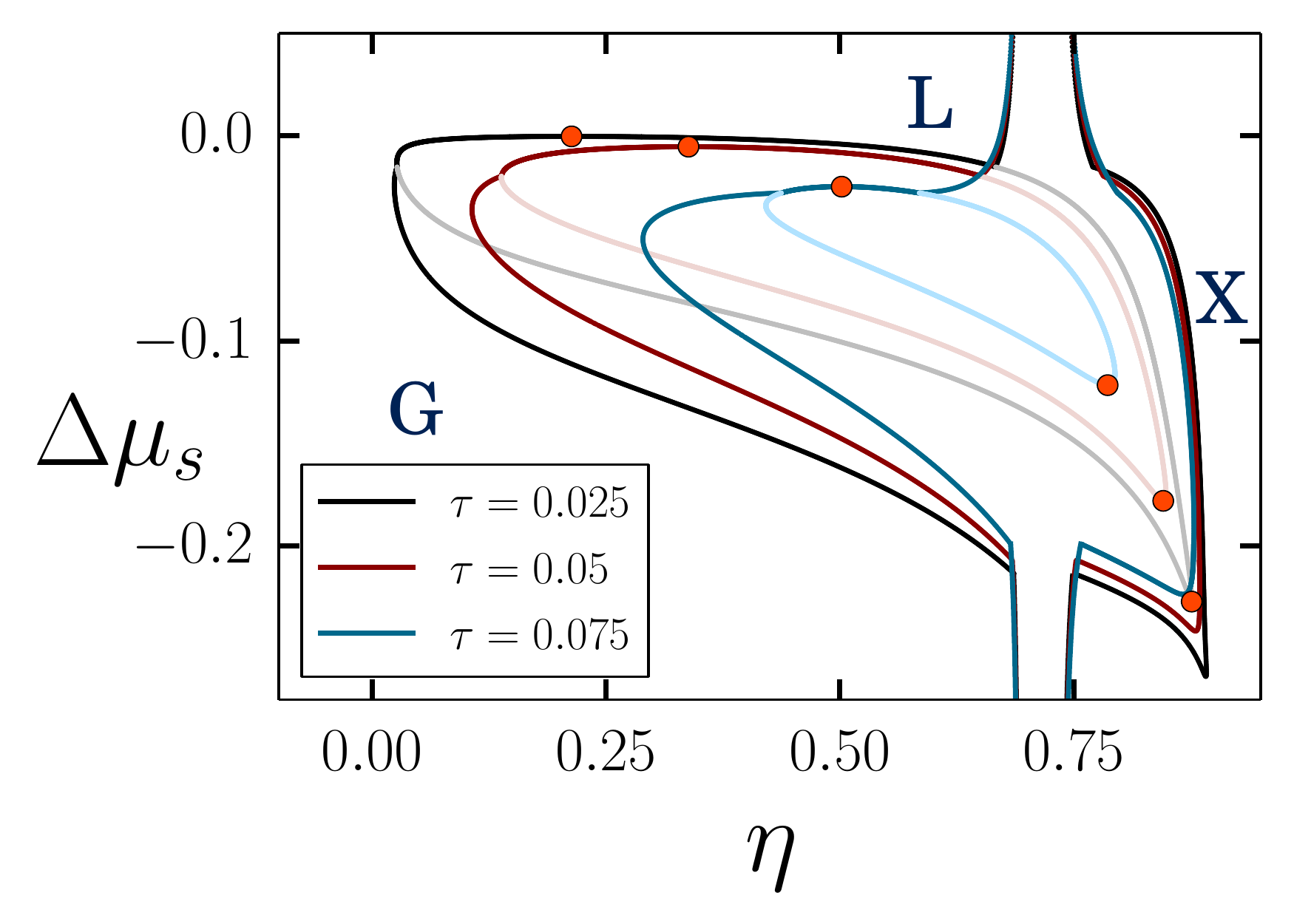}
         \hspace*{0.1cm}        
         \includegraphics[width=0.45\textwidth]{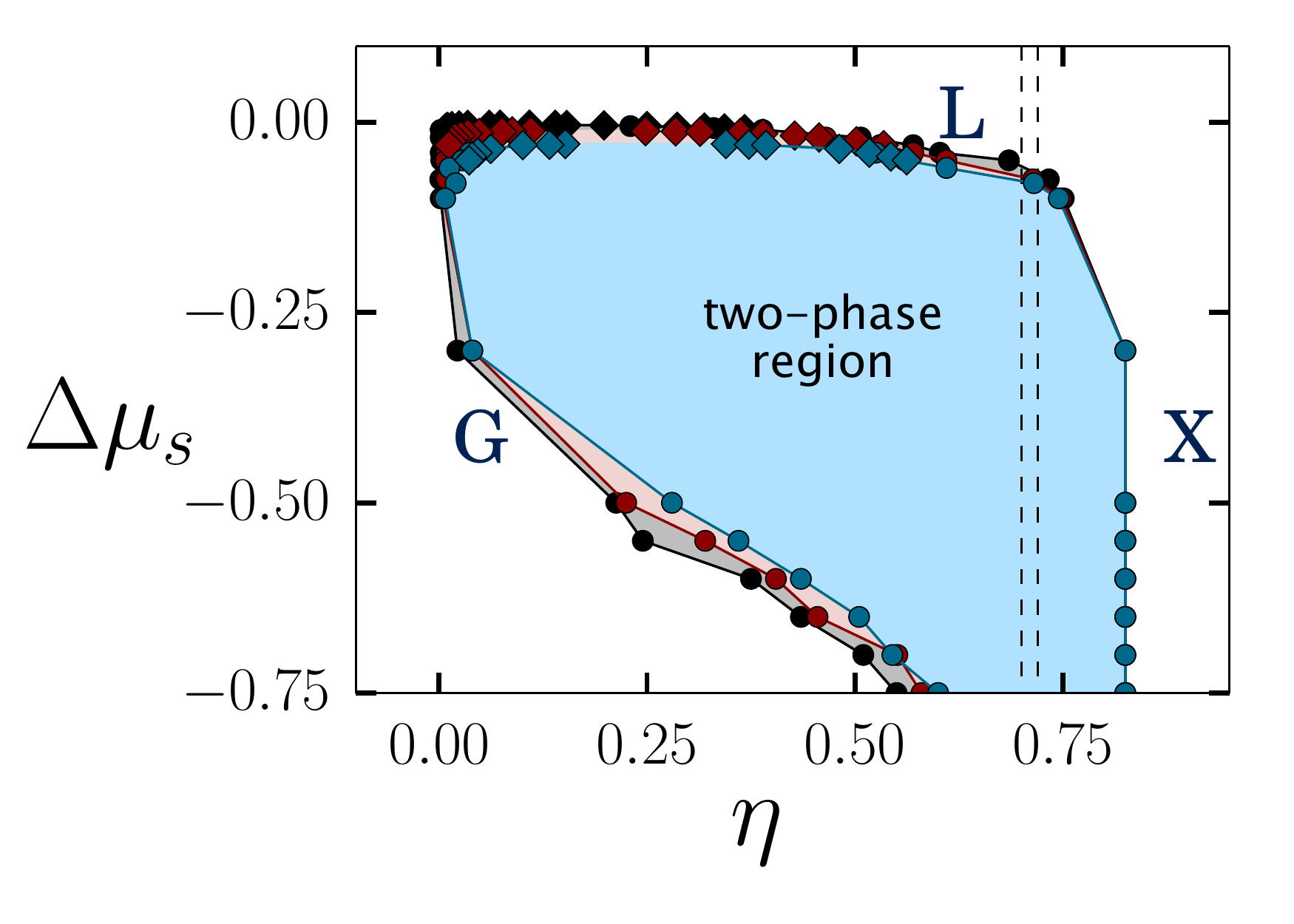}
        \caption{  Phase diagrams of the full  ternary colloid - solvent $a$ - solvent $b$
	$2d$ lattice model (Fig.~S2 in Ref.~\cite{bob-et:2014}) for three values of
	$t=(1-T/T_c^{(s)})$  (in Ref.~\cite{bob-et:2014} denoted by  $\tau$):  0.025 (black),  0.05 (dark red), and  0.075
	(blue)  in the $(\Delta\mu_s,\eta)$ plane of the solvent chemical potential difference and the colloidal packing fraction. 
	The quantity $\Delta\mu_s =\mu_a - \mu_b$  is the chemical potential difference between species $a$ and $b$ (in units of the solvent-solvent interaction strength). 
	The upper panel shows results obtained within mean field theory.  
	The grey, pale red, and pale  blue curves correspond to metastable
	colloidal gas-liquid (G-L) coexistence, which also terminates at the critical point. For each $\tau$ the upper (stable) and lower (metastable) gas-liquid critical points are
	indicated by red dots. X denotes the solid phase. (The various phases are inferred from monitoring their free energies.) 
	The lower panel shows the corresponding  phase diagrams as determined by Monte Carlo simulations.
       The diamonds and dots denote the phase boundaries   as obtained from grand canonical staged insertion MC simulations and $(\Delta\mu_s/(k_BT), \eta, \tau)$ ensemble MC simulations, respectively.
       Their color corresponds to the value of $\tau$ given in the upper panel. The blue area corresponds to the two-phase region for $\tau = 0.075$; for $\tau = 0.05$ and 0.025 the two-phase regions
       encompass the previous regions and have added (colored) slices. The vertical dashed lines denote fluid-solid coexistence for pure colloidal hard  discs.
        $T_c^{(s)}=T^{MFT}_c$  and $T_c^{(s)}=T^{MC}_c)$ is the  critical temperature of the binary ($ab$) solvent within mean field theory and  MC simulations, respectively.
      } 
    \label{fig:6}
\end{figure}

\begin{figure}
	\includegraphics[width=\textwidth]{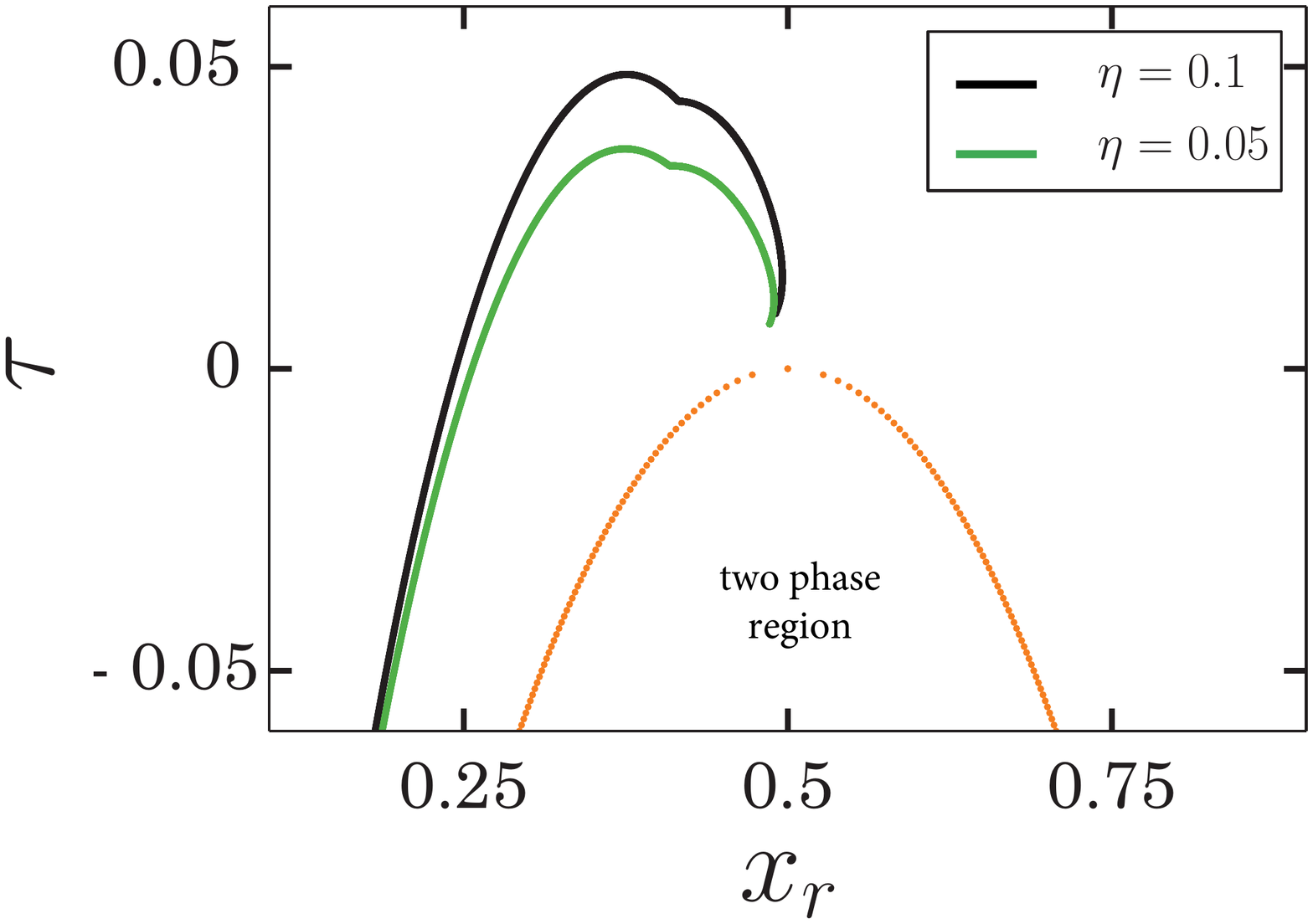}
	\caption{ Aggregation lines (black and green) for a ternary colloid - solvent $a$ - solvent $b$  mixture (interpreted as a colloidal condensation transition),
	as determined within   mean field theory for  the lattice model considered in Ref.~\cite{bob-et:2014} (Fig.~4 in Ref.~\cite{Edison-et:2015b}), plotted  in  the reduced temperature -  composition of the pure solvent in the {\it r}eservoir 
	$(t=(1-T/T_c^{(s)}), c_a)$
	representation (in Ref.~\cite{Edison-et:2015b} denoted by $(\tau=(T-T^{MFT}_c)/T^{MFT}_c,x_r))$.
	A point on the aggregation line at $x_r$ gives the temperature at which phase separation of colloids 
	is  observed first, upon cooling the suspension at a fixed packing fraction $\eta$ of colloids and for fixed solvent composition $x_r$.
	The $2d$ calculations have been performed  for a parameter $v_c$ (approximately equal to the area of the colloidal disc)  chosen to be equal to 1000${\it a}^2$
	(in order to reproduce most closely the MC simulations results of Ref.~\cite{bob-et:2014}), where ${\it a}$ is the lattice spacing, for the 
        coupling strength between colloid  and solvent species  $b$  equal to 32 (in units of  the solvent - solvent interaction strength), and for two fixed values $\eta = 0.1$ and $\eta = 0.05$ of the colloid packing fraction.
        The colloids interact with the members of solvent species $a$ and with each other via a hard core repulsion.  The dotted orange line is the binodal of the colloid-free $ab$ solvent. Each aggregation
        line ends at a critical point of the ternary mixture, which is removed from the binodal  of the solvent reservoir (see the main text). The origin of the break in  slopes of
        the aggregation curves is not discussed  in Ref.~\cite{Edison-et:2015b}.  The bent of the  aggregation lines near their critical points implies the occurrence
        of reentrant dissolution upon cooling.
 }
    \label{fig:7}
\end{figure}

The   mean field approximation of the model considered  in Ref.~\cite{Tasios-et:2016} was studied in more detail in Ref.~\cite{Edison-et:2015b}.
It is based on free-volume arguments, according to which  the Helmholtz free energy  is taken as the sum of three
contributions, describing the direct interactions among  the pure colloids,  the pure binary mixture of solvent $a$ and solvent $b$,  and  the colloid - solvent $b$ mixture. 
For the pure-colloid contribution the hard-disc free energy has been employed, with its distinct expressions  for
 the fluid \cite{Santos-et:1995} and  the solid \cite{Young-et:1979} phase. In the free space between the colloids, the mean field 
free energy for the  $a-b$ mixture has been taken; the solvent is excluded from the volume of the colloids. Concerning the colloid - solvent $b$ 
contribution, the mean adsorption energy of component $b$ at the colloid surface has been taken.
As can be inferred from Fig.~\ref{fig:6}, the topology of the resulting phase diagram and its temperature dependence agree  qualitatively
 with the one obtained from simulations (see Fig. S2 in the Supplemental Material
for Ref.~\cite{Tasios-et:2016} and Fig.~1 in Ref.~\cite{Edison-et:2015b}).
In particular, a  lower critical point has  been found.
However, as already expected on the basis of the different spatial dimensions, there is no  quantitative agreement; the parameter values
 used in order to obtain  phase diagrams, which are similar to  those from simulations,
were chosen empirically. It seems that in these studies the
underlying mean field approximation  severely underestimates the effect of the
hard-core repulsion of the colloids. 
In addition to the colloidal gas and colloidal liquid, this theory predicts the occurrence of two crystal phases which 
have  the same (hexagonal) structure
but with different lattice spacings. In the 3d thermodynamic state space of a ternary mixture spanned by  $\Delta\mu_s$, $\eta$, and $t$ one finds also upper  and  (metastable) lower colloidal gas-liquid
 critical lines and a colloidal solid-solid critical line. (The critical points in Fig.~\ref{fig:6}(a) lie in the  
planes  cutting  the
 full phase diagram at three constant  temperatures.)  The fluid-solid transition corresponds to a first-order freezing transition.
The  critical point of the
colloid-free solvent  ($t=0$, $\Delta\mu_s = 0$, $\eta = 0$)  is shifted considerably upon adding a
small amount of colloids (see the red dots in Fig.~\ref{fig:6}(a)). 
The authors of Ref.~\cite{Edison-et:2015b} have pointed out that experiments on colloidal aggregation
are  typically  performed by suspending a fixed number of colloids in a
solvent at a fixed (pure) solvent composition, which we denote by $c_a^{(s)}$, and then by adjusting the temperature
of the system  to reversibly induce aggregation.
They have identified the so-called aggregation line, i.e., the loci of the points $(t,c_a^{(s)})$ at fixed $\eta$ and $c_a^{(s)}$ at which
aggregation is observed first, as the 
 line which demarcates the one-phase region of the ternary mixture, which lies at the  outside of the line,
 and the region at its inside where colloidal phase separation
can be found.
Such a line (see Fig.~\ref{fig:7}) ends  at an $\eta$-dependent critical point of the ternary
 mixture, which is shifted from the critical point of the pure solvent towards a  higher ($\eta$-dependent) temperature.
 In the vicinity of their end points these lines  are slightly bent, which indicates  reentrant dissolution.
 Such lines  qualitatively resemble the experimental results  for strong aggregation reported
 by Beysens and Est\`eve  in Ref.~\cite{Beysens-et:1985}  (see the solid line in Fig. 3 therein).  
 This  tends to  support the   view on the aggregation lines 
 as the line of onset of  colloidal phase separation
in the full ternary mixture.
The above mean field calculation has been performed in $d=2$ \cite{bob-et:2014} and then extended to $d=3$ \cite{Edison-et:2015b} but no new features of the phase behavior
have been found. Although the uncertainties generated by the large differences
of length scales prevent a quantitative comparison with the available experimental data,  the  studies 
reported in Refs.~\cite{bob-et:2014,Edison-et:2015b} are very valuable because they reveal the possible
scenarios for the phase behavior, which can occur in such ternary mixture, highlighting the physical mechanisms 
behind it.

\section{Experiments}
\label{sec:exp}

\subsection{Phase transitions and aggregation in  bulk}
\label{subsec:ag_ph}

The  view, 
that  aggregation, as observed in  pioneering experiments by Beysens and Est\`eve \cite{Beysens-et:1985},
is in fact a reversible phase transition in a ternary  suspension, has been tested via early experimental
studies  by  Wegdam, Schall {\it et al.}~\cite{Guo-et:2008}. 
In these studies,   charge-stabilized
polystyrene spheres of radius $R=105$ nm suspended in a mixture of 3-methylpyridine (3MP), water, and heavy water near its lower critical point  have been considered.
The  mass fractions of the components of this latter mixture have been chosen such 
that the mass density of the solvent mixture closely matches  that of the colloidal particles in the  region of the  parameter space
where phase transitions have occurred. If the system is density-matched,
the   growth  of the nucleated liquid or solid phases is not perturbed by gravity at a very early stage and can be followed until macroscopically large coexisting phases are formed.
In order to characterize the phase behavior of the system the authors have used small angle $X$-ray scattering (SAXS). Transition temperatures
were also determined by measuring the sample turbidity. Moreover, the samples have been observed directly with
a  CCD camera. The only information, which has been provided by the authors about the phase diagram of the 3MP-water-heavy water mixture at the  mass fractions
actually used, is the {\it{}coe{\it{}x}istence} temperature  $T^{(s)}_{cx}\approx  65$\textcelsius \; at which this (chemically in fact binary) solvent mixture demixes.
The schematic phase diagram shown as Fig.~1(a) in Ref.~\cite{Guo-et:2008} suggests that the considered
mass fraction $\omega_{\mathrm{3MP}}$  of 3-methylpyridine investigated in this study (and therein denoted as $c_{\mathrm{3MP}}$) has been  smaller than   the critical one.   
Instead of $t=(1-T/T^{(s)}_c)$,  the authors of Ref.~\cite{Guo-et:2008} have used  the deviation $\Delta T = T-T^{(s)}_{cx}$ as the actual control parameter. 
Upon  increasing the temperature from a  value in the one-phase region of the
pure solvent towards  $T^{(s)}_{cx}$, the onset of colloid  aggregation, at which the system separates into a colloid-rich and  a colloid-poor phase as indicated by
regions of high and low turbidity and by the appearance of peaks in the structure factor, has been observed at a sharply defined
temperature $T_a$. Depending on the particle volume fraction   of the colloids in the suspension (which is  the ratio between the volume of all colloids in the suspension  and  the volume of the system),
two or three phases of the colloidal suspension have been found:
at low volume fractions this is  a  fluid phase in equilibrium with a fcc crystal, whereas at larger volume fractions 
there is a gas phase in equilibrium with a liquid phase and this liquid phase is coexisting with a fcc solid.
 Interestingly, upon increasing  temperature quickly  the measured structure factor
 indicated the formation of a glassy state as in a molecular system.
On the basis of the experimental results reported in Ref.~\cite{Guo-et:2008}, it is not possible to infer uniquely  the physical origin of the attractive
 potential which gives rise to the observed reversible aggregation.
 The authors concluded that all mechanisms  discussed  in Sec.~\ref{sec:eff} can play a role.
  Concerning the role of CCFs in this experiment, one finds that the
 temperature $T_a$ of the onset of aggregation is ca. $8-4$ K (depending on the volume fraction
of the colloids) below  $T^{(s)}_{cx}$, which corresponds to $(T^{(s)}_{cx}-T_a)/T^{(s)}_{cx} \simeq 0.02 -0.01$, which seems to lie  outside the critical region.
However, in Ref.~\cite{Guo-et:2008} the actual  critical temperature $T^{(s)}_c$ of the solvent is not given; if  $T^{(s)}_c$  is   lower, 
it is possible that the corresponding values of the reduced temperature  actually  lie within the critical region.
The  bulk correlation length of the solvent at $T=T_a$ has been estimated from light scattering 
to be ca. 8 nm, which is somewhat short. However, the experiments were performed at an off-critical composition
of 3MP such that along the corresponding thermodynamic path the phase is relatively  poor in the solvent component preferred by the colloids.
If $T_a < T_c^{(s)}$, this composition  precludes that  colloidal aggregation arises from complete wetting and
capillary condensation \cite{Evans:1990}. On the other hand, it is this thermodynamic region   where  the CCFs are expected to be  strongest 
although their range, governed by bulk correlation length, is  smaller than the one at the critical composition. Thus it
is rather plausible that  CCFs are playing a  crucial role in this experiment.

In a subsequent study by Wegdam, Schall {\it et al.}~\cite{Bonn-et:2009}, the aggregation of
colloidal particles  suspended 
in the same (quasi-) binary mixture of 3MP, water, and heavy water has been observed
directly by using confocal microscopy. The fluorescent fluorinated
latex colloids  of radius $R=200$ nm  used in this  study exhibit  affinity for water. 
In this experiment the refractive indices between the colloids and the solvent have been  matched closely.
Two different compositions of 3MP have been studied, one below (mass fraction $\omega_{3MP}^{(1)} = 0.24$)
and one above  (mass fraction $\omega_{3MP}^{(2)} = 0.37$)  the critical composition $\omega_{3MP,c} = 0.31$.
(In Ref.~\cite{Bonn-et:2009} the critical  mass  fraction of 3MP  is denoted as  $X_c$.)
The colloid volume fraction has been kept fixed at the rather small value 0.002 and the D$_2$O/H$_2$O mass ratio was chosen as $0.25$.
For the suspension poor in 3MP no aggregation has been observed upon increasing temperature within the homogeneous 
one-phase region of the solvent until phase separation of
the whole mixture has taken place. This is expected to occur for 3MP mass fractions so far off the critical value
that the CCFs are negligibly  small. Above $\omega_{3MP,c}$, within a rather
 wide temperature range of $T^{(s)}_{cx} - T_a = 8$ K,  reversible aggregation has been observed.
For the very  small volume fraction of colloids used, the formation of clusters rather than a colloid phase transition
has been observed and the kinetics of this aggregation process has been studied. The clusters have been a few particle diameters wide.
The sizes of the  clusters, as inferred from the maximum of the confocal image intensity $I(q)$ in  Fourier space,
have been found to to increase  linearly as function of time. The fractal dimension of these aggregates has indicated that the growth  process is diffusion limited.
Because the particles have desorbed from  the clusters at  a certain finite rate, the authors have been able to estimate
the energy scale of the attraction between the particles by comparing the observed escape frequency with
the corresponding attempt frequency, which is  the inverse of the Brownian time, i.e., the time in which a particle diffuses a distance equal to its radius, and by assuming that this whole process is thermally
activated; this renders   3$k_BT$ as the estimate of for the energy scale of attraction. 
The authors rightfully argued that the CCFs alone are sufficient to induce aggregation in charged-stabilized colloids. In order to show that this is indeed the case for their experiments, 
they analyzed their data using a simple expression for the effective pair potential between the colloidal
particles. As in  Eq.~(\ref{eq:17}), it consists of two competing contributions with 
the repulsive part decaying exponentially on the scale of  the Debye screening length $\kappa^{-1}$ (denoted as $\ell_D$ in Ref.~\cite{Bonn-et:2009}.) 
Concerning  the  CCP  contribution, a rather simple expression has been adopted. First, although the solvent in the actual experiment has not been at 
the critical composition, the occurrence of the corresponding relevant scaling variable  $\Lambda$ (Eq.~(\ref{eq:4})) or, equivalently, $\Sigma$ (Fig.~\ref{fig:1}) has been neglected. Second,
the Derjaguin  approximation (Eq.~(\ref{eq:13})) has been employed with $\mathcal{A}_+$ set to 2. As both contributions decay exponentially,
the hypothesis was put forward that  aggregation should occur when both decay lengths, $\xi_t$ and
$\kappa^{-1}$, become comparable. This hypothesis  has been  tested by adding salt in order to vary the range
of the repulsive Coulomb interactions and  by changing temperature in order to vary the critical Casimir attraction.
The Debye screening length $\kappa^{-1}$ follows from the formula $\kappa^{-1} =\sqrt{\epsilon\epsilon_0 k_BT/(e^2\sum_i\rho_i)}$  
(see below Eq.~(\ref{eq:1})) with
 $\sum_i\rho_i = 2N_A\ell$, where $\{\rho_i\}$ are the number densities of all ions (regardless of the sign of their charges), $N_A$ is the Avogadro number, and $\ell$ is the ionic strength (i.e., the molar density of positively charged ions)
of the solvent. The bulk correlation length $\xi_t$ has been determined independently using X-ray and light scattering. The data representing 
the onset of aggregation for various values of $\kappa^{-1}$ fit well to the anticipated relation  $\xi_t(T) = \kappa^{-1}$.

\begin{figure}
        \includegraphics[scale=0.8]{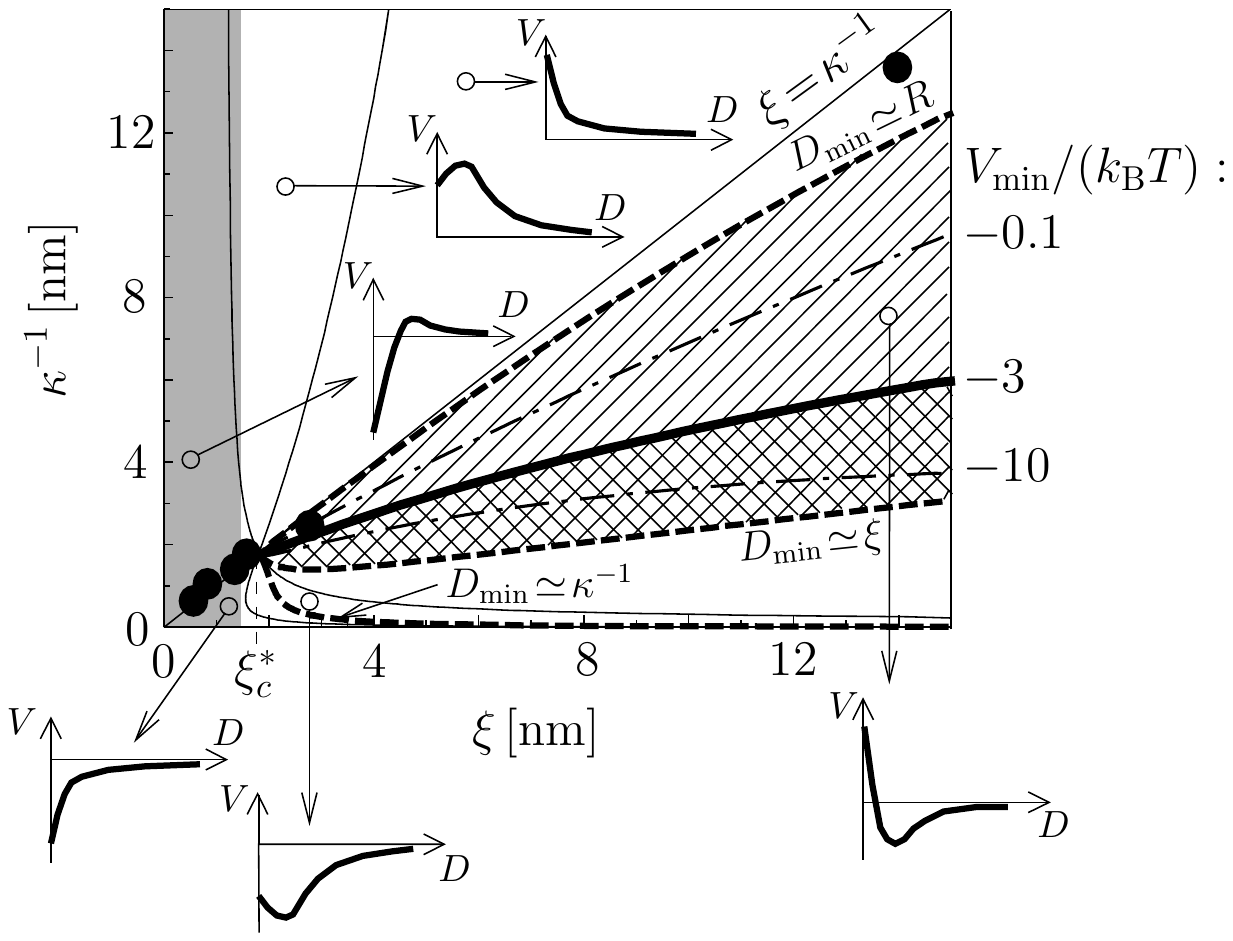}
        \vspace{-10cm}
	\caption{Shapes of the effective total  potential $V(D)= V_{el}(D) + V_{C}(D)$  of the force acting on two identical colloids of radius $R$ for various values of the bulk correlation length $\xi=\xi(t,h_b=0)=\xi_t$  and the Debye screening
	length $\kappa^{-1}$, where $D$ is the surface-to-surface distance  of the two colloids (Fig.~1 in Ref.~\cite{Gambassi-et:2010}). Six different regions of distinct  shapes  of 
	$V(D)$ are limited by the thin solid lines. These lines  meet at  $\kappa^{-1}=\xi_t \equiv \xi^*_c$.  $D_{min}$ is the position  $V_{min}$ of the minimum of $V(D)$.
	 Within the hatched area enclosed by the thick dashed lines the condition  $R\gg D_{min} \gtrsim \xi_t,\kappa^{-1}$ is satisfied. The additional requirement
	that $V_{min} \lesssim -3 k_BT$ is fulfilled 
	in the cross-hatched part of the hatched are.
	The black dots mark the experimentally determined \cite{Bonn-et:2009} aggregation line.  For further details see  Ref.~\cite{Gambassi-et:2010} and Ref.~1 therein.
   }  
    \label{fig:8}
\end{figure}

This data analysis has been  objected and redone in Ref.~\cite{Gambassi-et:2010} with the conclusion that, in fact, 
most of the   data reported in Ref.~\cite{Bonn-et:2009} for the onset of aggregation have been located within  ranges of values of  $\xi$ and $\kappa$ 
for which the proposed pair potential does not apply or, in other words, the model proposed in Ref.~\cite{Bonn-et:2009}
 does not predict aggregation to occur at $\xi_t = \kappa^{-1}$.
The  comment \cite{Gambassi-et:2010} linked to Ref.~\cite{Bonn-et:2009} pointed out the following weaknesses of the theoretical
analysis in Ref.~\cite{Bonn-et:2009}. The majority of the experimental data correspond to rather small  bulk correlation lengths,
i.e., $\xi_t \lesssim 5 \xi_{t,+}^{(0)}$, for which the adopted form of the CCP is not valid and should be complemented by 
corrections to scaling. Moreover, the effect of
screening of the surface charge upon  adding salt to the solution, which changes the amplitude $A$ of the repulsive
contribution to the pair potential (Eq.~(\ref{eq:17})), has been neglected. Therefore,  as a function of $\xi_t$ and
$\kappa^{-1}$ the total pair potential exhibits richer forms than the two forms (a purely repulsive potential above the line $\xi_t(T) = \kappa^{-1}$
and in addition an  attractive potential  well below the line $\xi_t(T) = \kappa^{-1}$) shown in Ref.~\cite{Bonn-et:2009}
 (see Fig.~\ref{fig:8}). The model pair potential actually used in Ref.~\cite{Bonn-et:2009} is valid in the range where the distances $D$  between the colloids (denoted as $\ell$ in Refs.~\cite{Bonn-et:2009} and \cite{Gambassi-et:2010}),
which are close to  the position $D_{min}$ of the minimum of the total potential, satisfy $R\gg D_{min} \gtrsim \xi_t, \kappa^{-1}$.
Together with the requirement that the potential depth should be at least $-3k_BT$, this leads to a range of applicability of the model which is much  more stringent than   $\xi_t(T) < \kappa^{-1}$ as indicated in Refs.~\cite{Bonn-et:2009}.
However, the reanalysis presented in Ref.~\cite{Gambassi-et:2010} does not exclude the possibility
that the observed aggregation is not solely due to the competing effects of repulsive electrostatic
and attractive critical Casimir interactions.

\subsection{Effective colloid-colloid pair potential}
\label{subsec:pp}

In Refs.~\cite{Nguyen-et:2013,Dang-et:2013}  
the effective pair potential between colloidal particles  for a dilute (volume fraction 2\%) suspension of
 poly-$n$-isopropyl acrylamid microgel (PNIPAM) particles with a radius of $R=250$ nm has been inferred from the pair correlation function $g(r)$. 
For  a sufficiently dilute suspension of solute, the potential of  mean force $V_{mf}(r) \sim - \ln g(r)$ can be identified with the effective pair potential
$V(r)$; $g(r)$ has been determined from the $2d$ images obtained by confocal microscopy. The spatial resolution has been estimated to be ca.~$0.03\mu m$ 
in the image plane  and ca.~$ 0.05\mu m$ in the direction perpendicular to it. The solvent was, as before, a (quasi) binary mixture of 3MP - water - heavy water
with  various  compositions, including the critical one  and compositions slightly off the critical one (towards the 3MP poor phase).
The measurements have been performed  for various temperatures upon approaching the solvent two-phase coexistence curve  from the homogeneous mixed phase.
The obtained pair potential displayed a very soft repulsion at small separations and developed  an increasingly deep minimum as 
$T$ approaches  the coexistence temperature $T^{(s)}_{cx}$ of the solvent. At low temperatures, at which $V(r)$ is purely repulsive, the authors have been able to fit $V(r)/(k_BT)$  to the screened electrostatic, 
exponentially decaying potential $U_{rep}$ (see Eq.~(\ref{eq:2})) with   plausible values of the parameters. From this the  CCP has been  then determined by
the best fit to the exponential form given in Eq.~(\ref{eq:13}), but with the amplitude and the  decay length treated as fit parameters, assuming
that the total $U(r)$ is the  sum of $U_C$ and $U_{rep}$  (as in Eq.~(\ref{eq:16})).
This assumption is justified for small salt concentrations
\cite{ions} which has been the case  for the samples studied there. (There is experimental \cite{Nellen-et:2011} and theoretical
\cite{ions}  evidence that at larger salt concentrations  the coupling between
the charge density and the order parameter  can  alter significantly the standard critical adsorption and
CCFs.) Moreover, it has been assumed that for the temperatures studied, within a range  $\Delta T < -1\mathrm{K}$,
the `background' repulsive  contribution is de facto  temperature independent.
This  data analysis has rendered the  length scale for the experimental decay of the
 CCP, which, however, differs from  the bulk correlation length. This discrepancy as well as the softness of the  repulsion at small separations have been  attributed to the fluffiness of the colloidal particles.
The choice of PNIPAM particles has been motivated by their special  properties when dissolved  in the  solvent described above.  They swell,
which prevents sedimentation, and in this swollen state their refractive index matches    
that of the solvent. 

\begin{figure}
\includegraphics{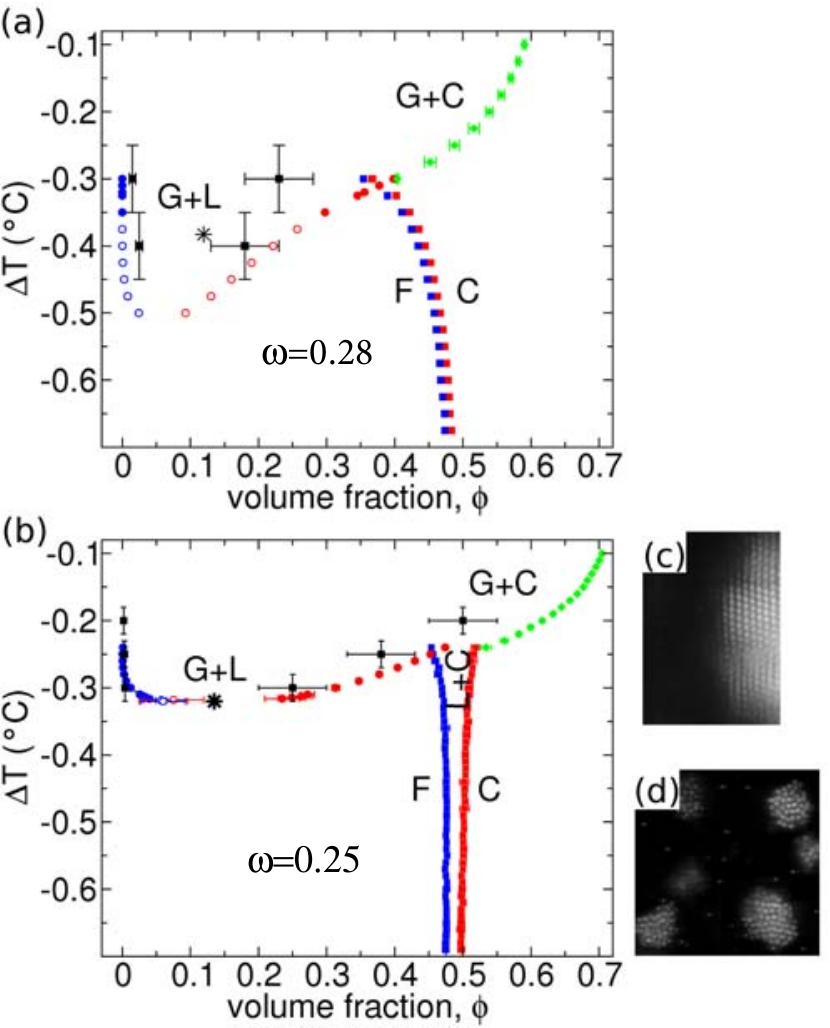}
	\caption{ Phase diagram in terms of $\Delta T = T- T^{(s)}_{cx}$  and volume fraction, in Ref.~\cite{Nguyen-et:2013} denoted as $\phi =V_s/V_{tot}$, where $V_s$ is the volume of colloids and $V_{tot}$ is the total volume of the sample. These data have been obtained from MC simulations (colored symbols)  for the effective one-component colloidal system governed by a pair potential
	which is the sum of repulsive and attractive, exponentially decaying, functions describing screened electrostatic and critical Casimir 
	interactions (Fig. 2 in Ref.~\cite{Nguyen-et:2013}). G, L, F, and C denote colloidal gas, liquid, fluid, and crystal phases, respectively. G+L, G+L, and L+C stand for the gas-liquid, gas-crystal, and the liquid-crystal coexistence regions. Black squares with error bars are experimental data 
	for PNIPAM particles suspended  (a) in  the 3MP - heavy water mixture at the critical 3MP mass fraction
	$ \omega_{3MP}= 0.28 \simeq \omega_{3MP,c}$  and (b) in the 3MP - water - heavy water mixture  at the off-critical 3MP mass fraction
	$\omega_{3MP} = 0.25$.   The authors of Ref.~\cite{Nguyen-et:2013} interpret the  snapshots of confocal microscopy images for the mixture in (b) as  the coexistence of colloidal gas and crystal 
	at $\Delta T = - 0.2$  (c) and colloidal gas and liquid at $\Delta T = - 0.3$ (d).  Stars indicate the experimental position of the colloidal gas-liquid
	critical point as estimated by using the law of rectilinear diameters. In (a) the star appears as being placed at a too small value of $|\Delta T|$.
	} 
    \label{fig:9}
\end{figure}

In a subsequent paper~\cite{Dang-et:2013},  the fitting procedure 
has been improved in the sense that the bulk correlation length $\xi_t$ has not been anymore an adjustable 
 parameter. Assuming   the standard
scaling law for $\xi_t$  as a function of temperature, the  amplitude $\xi_{t,+}^{(0)}$  has been used as  a fit parameter in addition to
the amplitude and the  decay length of $U_{rep}$. Using one set of these three fit parameters, Dang and coworkers   \cite{Dang-et:2013} have fitted
  all experimental data to the sum of two exponentially decaying functions. 
 The fitted analytic expression for the  potential has been further used for performing  Monte Carlo  (MC) simulations  of  the   colloidal sample. The colloidal  gas-liquid coexistence has been investigated  by 
using Gibbs Ensemble MC \cite{Frenkel-et:2001}
 whereas colloidal liquid-solid coexistence has been determined by using Kofke's Gibbs-Duhem integration technique \cite{Kofke:1993}. 
The conclusiveness  of these MC simulation results for the actual colloidal phase behavior may be questioned  because the  fits of the experimental data to the analytic expression for the potential are  not completely satisfactory. The strongest deviations have occurred at small separations, 
where  the potential has been finite even for center-to-center separations smaller than the diameter of the particles. Moreover, the  pair potentials obtained from fitting have not  followed the data around the minimum of
the total potential, especially close to $T_c$. Also these deviations have been attributed to the  softness of the PNIPAM particles
and to the fact that the scaling variable $\Lambda$ describing off-critical compositions has been neglected.
 It has been tested if all these deviations may  significantly affect
the observed phase behavior by calculating  the reduced second virial coefficient $B^*_2=B_2/B_2^{(HS)}$.
Empirically, gas-liquid phase transitions are   expected to  occur for $B^*_2 \lesssim -1.5$~\cite{Vliegenthart-et:2000,Noro-et:2000}. 
Applying this criterion for experimentally determined and fitted potentials 
has led to an estimate for an upper bound for the deviation  $\Delta T= T_{cx} - T^{(s)}_{cx}(c_a) < - 0.3$K  of the temperature $T_{cx}$, at which a colloidal gas-liquid transition takes place, from that of the pure  solvent 
phase separation (i.e., $T^{(s)}_{cx}(c_a)$), which is 
in agreement both with the experiments and the simulations reported in Ref.~\cite{Dang-et:2013}.
The comparison between computed and measured phase diagrams  at fixed composition of a pure solvent  in the plane spanned
by   $\Delta T = T_{cx} - T^{(s)}_{cx}(c_a)$
and by the colloidal volume fraction (in Ref.~\cite{Dang-et:2013} denoted  as $\phi$) is shown in Fig.~\ref{fig:9} (Fig. 2 in Ref.~\cite{Dang-et:2013}).
 For solutions with off-critical compositions of a (pure) solvent,  a reasonable agreement has been found, given the large uncertainties in 
the experimental determinations of the volume fraction and of $\Delta T$,  as well as given the simplified functional form taken for  $U_C$. On the other hand, for solutions with the (supposedly)
critical composition, the agreement is less good. Deviations occur concerning  the shape of the colloidal gas-liquid  coexistence curve, which in simulations is
shifted towards  values of $\Delta T$ lower than those for the experimental data. Such a shift suggests that the fitted potentials
underestimate the attractions between the colloids. According to the authors, this is due to 
 many-body interactions and   too small simulation boxes; both are particularly relevant for  systems at the critical composition.
  Another possibility is that the effective one-component approach  is not adequate to  describe the considered experimental system.
 Concerning the colloidal liquid-solid phase coexistence in this system, there are no experimental data.

The  experimental data for $g(r)$  at a 3MP mass fraction of $0.28$, 
which is close to the critical value,  have been   re-analyzed in Ref.~\cite{Mohry-et:2014}. The main improvement
over the earlier analyses has been due to a better treatment of  the CCP contribution  $U(r) = U_{bck} + U_C$
to the total pair potential, where $U_{bck}$ is a background term. Within this approach, also the dependence
of $U_C$ on the  solvent mass fraction has been taken into account. The  scaling function of  $U_C$ 
 as a function of the scaling variable $\Sigma =\mathrm{sgn}(th_b)\xi_t/\xi_h$  (see Eq.~(\ref{eq:17}) and Fig.~\ref{fig:1}) has been  calculated within   the Derjaguin approximation  by
using a local functional approach. In order to minimize the 
number of fit parameters, the amplitude $\xi_{t,+}^{(0)}$ has been  set to the value extracted from the experimental data presented in Ref.~\cite{Sorensen-et:1985}.
However, the  value of the  critical mass fraction $\omega_{3MP,c}$ of the 3MP-heavy water binary liquid mixture
 is not well established. The inaccuracy 
of the value for $\omega_{3MP,c}$ enters into the reduced  order parameter
$\widetilde\phi = (\omega_{3MP,c}-\omega_{3MP})/{\cal B}$, where ${\cal B}$ 
 is the non-universal amplitude of the bulk coexistence curve 
$\omega_{3MP,cx}(t<0) = \omega_{3MP,c} \pm {\cal B}{|t|}^{\beta}$. 
Via the equation of state, which close to the critical point attains its scaling form, this determines the 
scaling variable $\Sigma = \mathrm{sgn}(th_b)\xi_t/\xi_h = \mathrm{sgn}(th_b)\mathcal{E}_{\pm}(|t|^{\beta}\widetilde{\phi})$, where $\pm$ refers to the sign of $t$. 
(In Ref.~\cite{Mohry-et:2014} the scaling functions $\mathcal{E}_{\pm}$  have been  obtained by using the equation of state within a linear parametric model.)
Because  $U_C$ depends  sensitively on $\Sigma$, it has been natural to use  the reduced order parameter 
$\widetilde\phi$
 as a fitting parameter for achieving the weakest variation of the background potential $U_{bck}$ with 
temperature. 
A fair agreement with the experimental data has been obtained by allowing  the value of  $\omega_{3MP,c}$ to differ significantly from the value cited by the authors of
the experiment  (see Fig.~5 in Ref.~\cite{Mohry-et:2014}).
The fit has rendered  the background potential which  varies slightly with temperature and has an  attractive part. This might be
due to the  the  coupling between critical fluctuations and electrostatic interactions
and thus deserves further studies. 

\begin{figure}
        \includegraphics{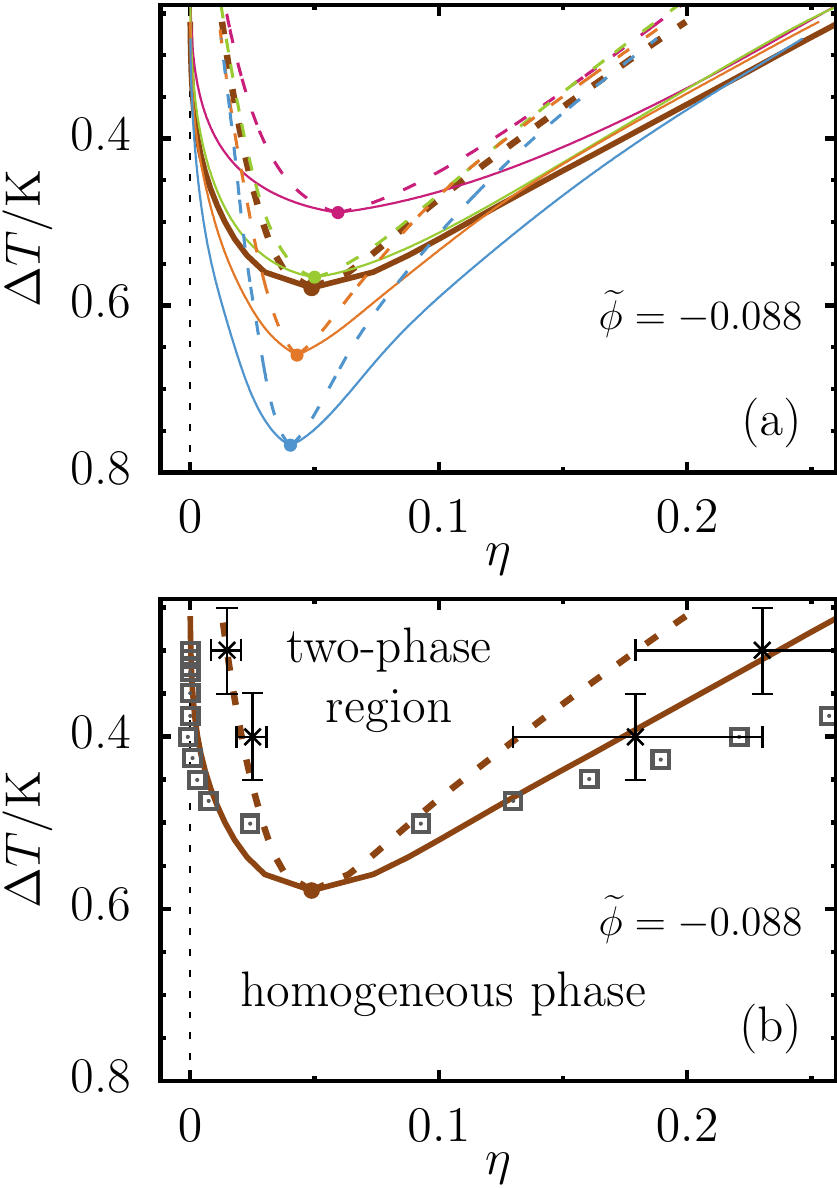}
        \caption{
	Segregation phase diagrams obtained from theory (RPA), experiment, and simulations (MC) (Fig. 6 in Ref.~\cite{Mohry-et:2014}). 
	(a)~The phase diagram obtained within RPA using the four available background 
	potentials $U_{bck}$  extracted from  effective potentials inferred from experimental data~\cite{Mohry-et:2014}
	at $\Delta T/\textrm{K}=0.6$ (magenta), 
	$0.5$ (green), $0.4$ (orange), and $0.3$ (blue), and their mean curve (thick dark red curve).
	The solid lines show the phase 
	boundaries in terms of the packing fraction $\eta$ of the colloids, whereas the dashed 
	lines correspond to the spinodals, and dots represent critical points. 
	(b)~Comparison of the theoretical predictions for the phase boundaries (based
	on the mean $U_{bck}$) with MC simulation data ($\boxdot$) and experimental data 
	($\boldsymbol{\times}$, with error bars) from Ref.~\cite{Dang-et:2013}. 
        On the temperature axis $\Delta{}T= T^{(s)}_c - T$ 
	increases from top to bottom in order to mimic the visual impression of a lower critical point $T^{(s)}_c$ 
	(of the solvent) as observed experimentally. $\widetilde\phi = (\omega_{3MP,c}-\omega_{3MP})/{\cal B}  \quad = -0.088$, where ${\cal B}$ 
         is the non-universal amplitude of the bulk coexistence curve (see the main text). }
        \label{fig:10}
 \end{figure}

This detailed knowledge of the CCFs as  function of $T$ and $h_b$ has  been  used by the authors of  Ref.~\cite{Mohry-et:2014}  to re-analyze the  experimental data for the phase segregation, obtained in Ref.~\cite{Dang-et:2013}
and shown in Fig.~\ref{fig:9}(a).
 The colloidal gas-liquid coexistence has been calculated within the effective one-component DFT
 approach using the pair potential
  $U(r) = U_{bck} + U_C$ fitted to the experimental data as described above, where   $U_C$ 
 has the scaling form as in  Eq.~(\ref{eq:17}). The random-phase (RPA) approximation 
has been  employed for the free energy together with the  Percus-Yevick expression  for the hard-sphere
reference contribution. For the off-critical composition of the solvent, which renders the best expression (i.e., the least temperature dependent one) for $U_{bck}$ extracted from the experimental
data of Ref.~\cite{Dang-et:2013} (with, however, $U_{bck}$ still varying slightly with temperature), the phase diagram has been  calculated by taking the mean   curve  of those $U_{bck}$ which correspond to the various 
temperatures  considered. 
The corresponding results together with the experimental and  MC simulation data
 from Ref.~\cite{Dang-et:2013} are shown in Fig.~\ref{fig:10}.  At high colloidal densities, 
the RPA is in surprisingly good agreement with the experimental data. On the 
other hand, at low densities the RPA agrees well with the Monte Carlo simulations, 
while for these densities  both approaches underestimate the experimental values which, in turn, agree well 
with the RPA spinodal. 
While this latter `agreement' might be accidental, it 
nevertheless raises the question whether the experimental system had actually been 
fully equilibrated  at the time of the measurements. 
As pointed out in Refs.~\cite{Nguyen-et:2013,Dang-et:2013}, the phase diagrams shown in Fig.~\ref{fig:9} are analogous to those of molecular fluids modeled, e.g.,
by Lennard-Jones or square-well fluids, but  with a lower critical point. The peculiarity of the CCP, i.e.,  the strong temperature
dependence of the shape of the potential and of its range, is mirrored by  the small temperature range over which  colloidal  gas-liquid coexistence extends
and in the shift of the critical point and of colloidal liquid-solid coexistence to lower volume fractions.

\begin{figure*}
\begin{minipage}{0.47\textwidth}
   \hspace{-19em}(a)\vspace{-1em}\\
   \includegraphics[width=\textwidth]{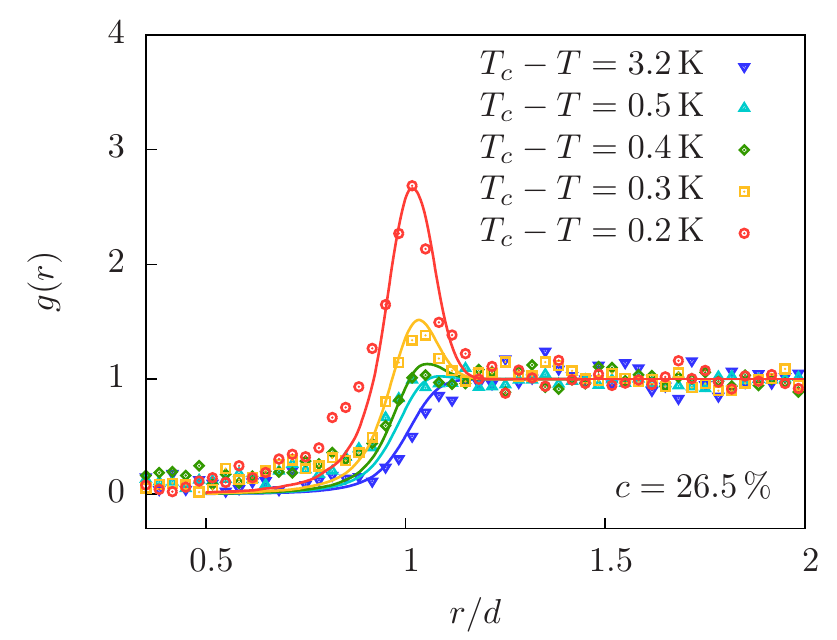}
  \end{minipage}
  \begin{minipage}{0.47\textwidth}
   \hspace{-19em}(b)\vspace{-1em}\\
   \includegraphics[width=\textwidth]{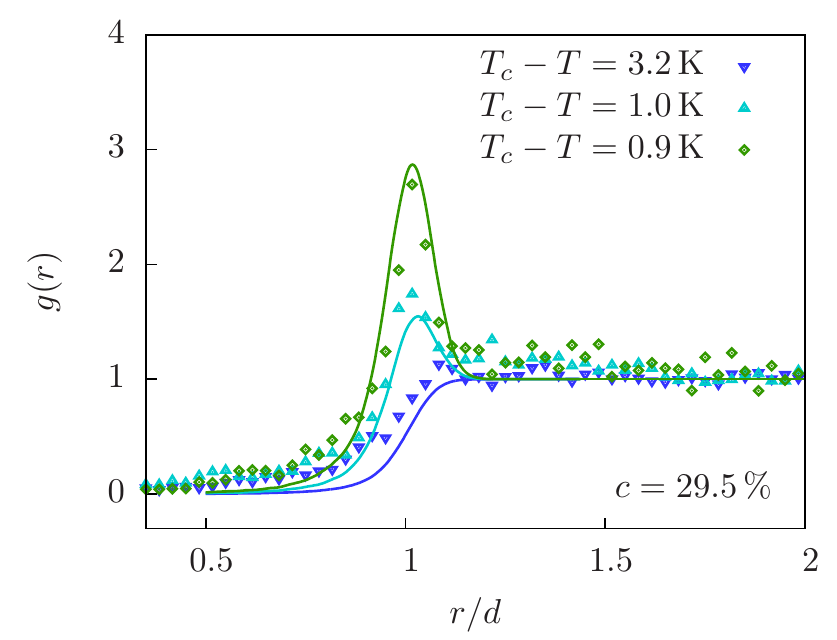}
  \end{minipage}
  \begin{minipage}{0.47\textwidth}
   \hspace{-19em}(c)\vspace{-1em}\\
   \includegraphics[width=\textwidth]{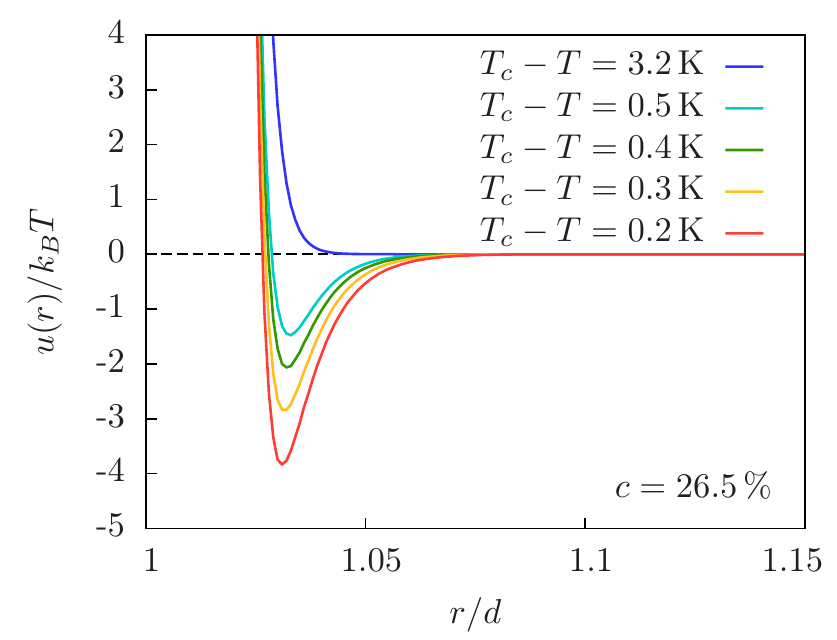}
  \end{minipage}
  \begin{minipage}{0.47\textwidth}
   \hspace{-19em}(d)\vspace{-1em}\\
   \includegraphics[width=\textwidth]{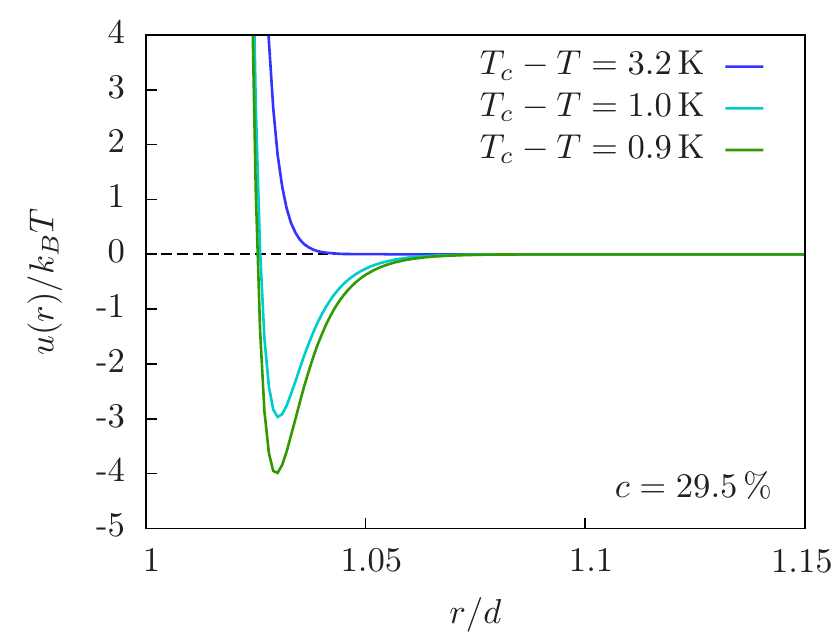}
  \end{minipage}
  \caption{(a) and (b): Radial distribution function $g(r)$ of the PNIPAM particles suspended in the 3MP - heavy water liquid mixture  for the off-critical mass fractions $\omega_{3MP}$ (in the figures denoted as $c$ and given in unites of weight percentage wt\% 3MP $ = 100 \times \omega_{3MP}$) 
$c=26.5\,\%$ (with $\Delta T_\text{off}=0.7\,\mathrm{K}$) and $c=29.5\,\%$ (with $\Delta T_\text{off}=0.18\,\mathrm{K}$), where the effective temperature offset $\Delta T_{off}$ defined via $(T_c-T+\Delta T_{off})/T_c \quad \equiv\quad t+t_{off}$ is a fit parameter. The experimental data (symbols), obtained in Ref.~\cite{Marcel-et} by using confocal microscopy,  are compared with the theoretical predictions (solid lines) based on the model given by Eq.~(\ref{eq:18}) with the Derjaguin approximation for the CCP. (c) and (d): Theoretically predicted pair potentials $u(r)$ 
for the same compositions (Fig. 7 in Ref.~\cite{Marcel-et}).}
  \label{fig:11}
\end{figure*}

\begin{figure}
\centering
 \hspace{-19em}(a)\hspace{6.7cm}(b)\vspace{-1em}\\
  \includegraphics{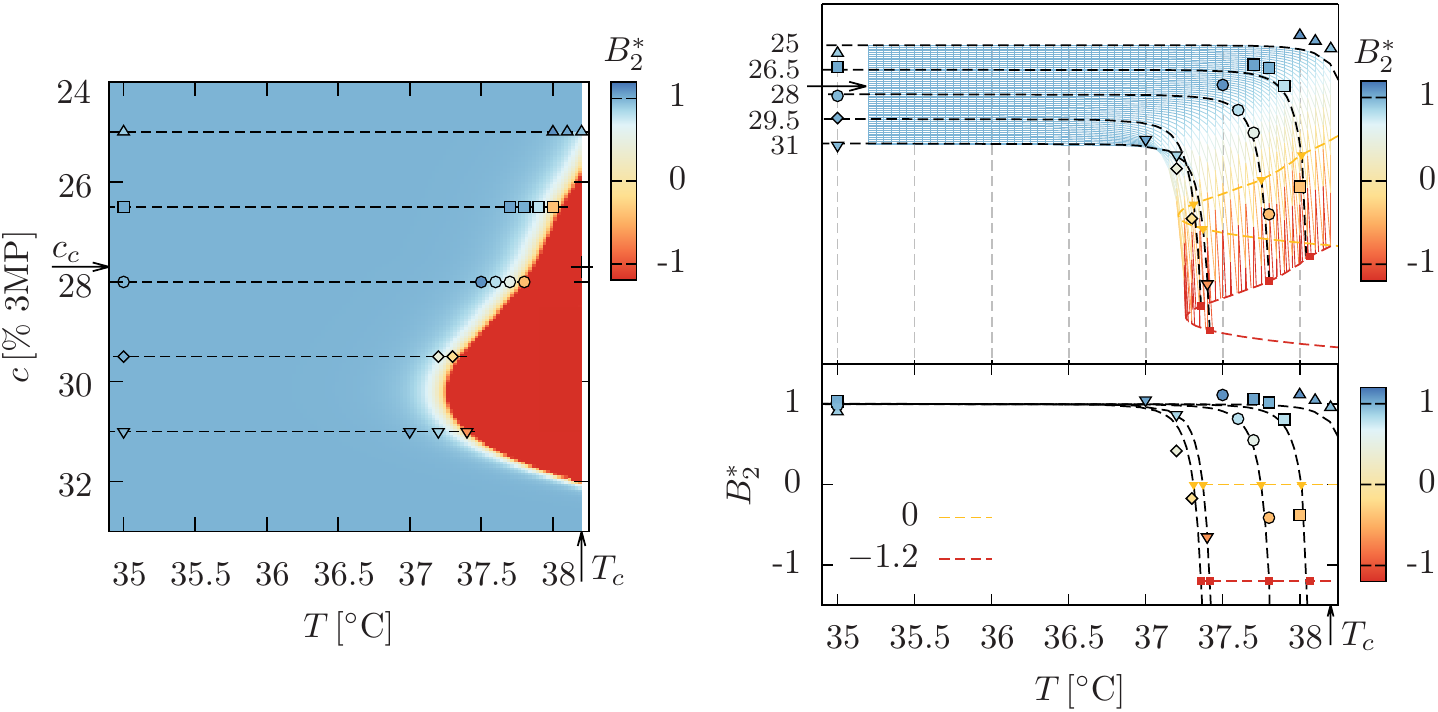}\\
 \vspace{-4cm}(c)\vspace{4cm}
 \linespread{1.2}
  \caption{
  Reduced second virial coefficient $B_2^*=B_2/B_2^{(HS)}$ as function of $T$ and $c$ for the same system as in Fig.~\ref{fig:11}  (see Fig. 9 in Ref.~\cite{Marcel-et}). 
  The color of the shading provides  the theoretically predicted values of $B_2^*(T,c)$, while the colored symbols provide 
  the experimental value of $B_2^*(T,c)$ 
  as obtained by numerically integrating the measured radial distribution function $g(r)$ (Eq.~(\ref{eq:18})).
 The weak color contrast between the colors of the symbol and the corresponding underlying shading indicates agreement between the experimental and theoretical data. 
The  values $B_2^*  \approx  1$ depicted in blue indicate a significant repulsion, while the values $B_2^* \approx -1$ depicted in red indicate strong attraction.  Yellow marks the crossover. 
The critical temperature $T^{(s)}_c$ (denoted as $T_c$) and the critical mass fraction $c_{c}$ are indicated by arrows and the critical point is indicated by $+$.
  In (a), via the color, $B_2^*$ is shown in the entire  $(T,c)$ plane.  The black dashed lines correspond to those five concentrations  for which there are experimental data; each dashed line corresponds to a certain symbol type. 
  In (c) the values of $B_2^*$  are shown as  functions of temperature for the five  values of $c$ as introduced  in (a) via  the corresponding symbol type.
The  horizontal dashed yellow and red lines indicate the isolines for  $B_2^*=0$ and $B_2^*=-1.2$, marking the crossover from repulsion to attraction and 
the critical value of the sticky spheres model, respectively. The value of $B_2^*$ can be read off both from the vertical axis and from the color of the symbols.
  Panel  (b) shows the same data but the curves corresponding to distinct values of $c$  are  shifted up vertically in order to gain visual clarity.
  The shifts are chosen such that the shifted  values of $B_2^*$ at $T=35$ \textcelsius \; are  separated equally. This generates a perspective view of the $B_2^*$ values above the $(T,c)$ plane, 
  revealing  a folded curtain-like shape
  of the surface $B_2^*(T,c)$. The color shading is discretized due to the shifts of the curves.  The yellow and red dashed lines emerge from the corresponding ones in (c) by the shifts.
  The vertical axis in (b) is $B_2^*$, but shifted. In the upper left corner of (b), the five dashed lines are labeled both by the symbol types and the corresponding numerical values of the concentration. 
 }
  \label{fig:12}
\end{figure}

Experience gained from the  studies summarized above tells that  a meaningful comparison of experimental data and theoretical predictions for the CCP requires  accurate
experimental knowledge of the solvent bulk phase diagram and  of the solvent bulk correlation length. Such corresponding dedicated measurements
 have been reported in  Ref.~\cite{Marcel-et} for  PNIPAM particles  suspended in a binary liquid mixture of 3MP and heavy water with addition of 1mM KCl salt. The solvents have been prepared with 3MP  mass fractions $\omega_{3MP}$ (and therein denoted as $c$ and given in  units of weight percentage wt\% 3MP$ = 100 \omega_{3MP}$) ranging from 23.5\% to 33\% around the critical value $\omega_{3MP,c}^{(s)}\approx 28.0$\%,  and for each of them  the temperature 
has been varied from the homogeneous, mixed phase towards the solvent two-phase coexistence.
In order to achieve intrinsic consistency of the experimental data and to link them to theoretical predictions, the measurements of  the  pair
potentials of the particles in terms of radial distribution functions  $g(r)$ have been complemented 
with dynamic light scattering measurements of the solvent phase diagram and of the bulk correlation length.
 Below the lower critical temperature $T_c^{(s)}$, the solvent
correlation lengths have been inferred from the temperature-dependent correlation functions
determining the scattered intensity, which are well-described by single-exponential decays with a characteristic
time scale $\mathsf{t}_{d}$ related to the effective diffusion coefficient $D=(q^2\mathsf{t}_{d})^{-1}$, where $q$ is the wave vector of the incident
wave (equal to 19$\mu m^{-1}$ in these specific measurements). This effective diffusion coefficient depends on the linear extent $\xi$  of the
correlated regions  via a relation analogous to the Stokes-Einstein relation for Brownian particles. However, close to the critical point 
one has to take into account that $D=D_\text{c} + D_\text{bg}$  decomposes into a {\it c}ritical and a {\it b}ack{\it g}round part \cite{Sorensen-et:1985}, where \cite{Sorensen-et:1985, Burstyn-et}
$
 D_\text{c} = \frac{\mathcal{R}\,k_B T}{6\pi\,\eta\,\xi}\,K(q\,\xi)\left(1+b^2(q\,\xi)^2\right)^{z/2}$ and  $D_\text{bg} = \frac{k_B T}{16\,\eta_\text{bg}\,\xi}\,\frac{1+(q\,\xi)^2}{q_c\,\xi}.
$
Here, $\mathcal{R}\approx 1.05$ is a universal dynamic amplitude ratio \cite{Das-et:2007, Burstyn-et}, $K(x)=3/(4 x^2)[1+x^2+(x^3-x^{-1})\arctan x]$ is the Kawasaki function \cite{Kawasaki},   $b=0.55$ ~\cite{Burstyn-et} is  a correction to scaling amplitude and
$\eta_\text{bg}$ is the background viscosity. The  latter has been obtained as a function of $T$ and $\omega_{3MP}$ by extrapolating the available viscosity data \cite{Oleinikova-et:1999} to the critical region.
This detailed analysis has allowed one to determine  the correlation length for various compositions and to extract 
the amplitude $\xi^{(0)}_{t,+}=0.44$nm. The phase separation temperatures have been defined as those which correspond the minimum of the diffusion constant for for various values of  $\omega_{3MP}$. They are  fitted to the relation $\omega_{3MP}-\omega_{3MP,c}=\mathcal{B}|t|^{\beta}$ with the fixed critical exponent $\beta=0.3265$  leading to the estimate $\mathcal{B}=0.6$ for the OP amplitude.
The diameter $2R=2.12\mu$m  of the particles  has been deduced by using confocal microscopy whereas their surface charge density $\Upsilon \simeq -0.17$e nm$^{-2}$ has been obtained from  electrophoresis.
The experimental radial distribution function $g(r)$, determined by  particle tracking, reveals a  very soft repulsion, according to which the measured effective potential fulfills $U(r<2R) > 0$ and is very large but not infinite as one would expect for hard core repulsion.
As already mentioned in the present section, in the previous studies \cite{Nguyen-et:2013} this softness has been attributed   to the fluffiness of the colloidal particles.  A  plausible alternative  explanation for this softness
 consists of the inaccuracy associated with  determining a three-dimensional $g(r)$ from two-dimensional images and, in addition, a certain  polydispersity. In Ref.~\cite{Marcel-et}  this inaccuracy
of  the experimental radial distribution  function has been compared  
with the projected theoretical function $g_\text{proj}(r'=\sqrt{x'^2+y'^2})=\int_{-\infty}^\infty\mathrm{d}z\int_{-\infty}^\infty\mathrm{d}y\int_{-\infty}^\infty\mathrm{d}x\, f_{\{0,\sigma_z\}}(z)f_{\{y',\sigma\}}(y)f_{\{x',\sigma\}}(x) g(\sqrt{x^2+y^2+z^2})$, where
 $f_{\{x',\sigma\}}(x)$, $f_{\{y',\sigma\}}(y)$, and $f_{\{0,\sigma_z\}}(z)$ are the probability distributions, which  account for the uncertainty in the two horizontal directions and the vertical direction, respectively, with the in-plane spreads $\sigma=\sigma_x=\sigma_y$ being equal. 
In Ref.~\cite{Marcel-et} these distributions  have been taken to be the normal ones. 
Building upon the consistent description of the bulk properties of the binary solvent, the comparison between the measured 
pair potentials and the theoretical model, given by Eq.~(\ref{eq:17}) within the Derjaguin approximation for the CCP,  would have not  required any fitting parameter if the surface of the colloids had exhibited strong preferential adsorption.
However, the PNIMP particles are only weakly hydrophilic. In such a case the CCP depends also on the surface field $h_s$  via
the  scaling variable $\hat h_s = h_s |t|^{-\Delta_1}$, where $\Delta _1$ is the surface counterpart of the bulk gap exponent \cite{Diehl:1986} (see  Secs.~\ref{subsection:CCFpot}   and \ref{subsubsec:DA}).
 For the film geometry, within mean field theory the dependence on $\hat h_s$ at the critical concentration ($\Sigma=0$) can be reduced to a re-mapping
 $\tilde\vartheta^{(d)}_{||}(\mathcal{Y}, \mathsf{\Sigma}; \hat h_s) = s^{d}\,\vartheta^{(d)}_{||}(s^{-1}\mathcal{Y}, \mathsf{\Sigma})$
with a rescaling parameter $s$, which depends on $\hat h_s$ \cite{Mohry-et:2010}. The authors of Ref.~\cite{Marcel-et} have found that within the experimentally accessible range of the
CCP scaling function, which usually consists of only its exponential tail, one may mimic such a rescaling by using an effective temperature offset $t_{off}$ which  shifts the reduced temperature according to
$t'=(T_c-T+\Delta T_{off})/T_c=t+t_{off}$. Assuming that such a rescaling is valid also beyond mean-field theory, the authors of Ref.~\cite{Marcel-et} have been able to fit all pair correlation functions, for a variety of  temperatures and even of off-critical compositions, by taking the single parameter $t_{off}$ 
to vary smoothly  with composition. Based on these fits the predictions for the pair potential are also given (see Fig.~\ref{fig:11}).
 Remarkably, any  distorting influence on  $g(r)$, which can be described by  probability distribution functions which are symmetric about their argument, leaves the second virial coefficient unchanged. This holds because $B_2$ is determined by the integrated pair potential (see Eq.~(\ref{eq:18})). Taking advantage of this property, the authors of Ref.~\cite{Marcel-et} could  compare virial coefficients computed from the raw data of $g(r)$ directly with theoretical predictions, without any need to account for experimental inaccuracies  and particle polydispersity. 
The experimental and the theoretical values of $B_2^*$ show very good agreement in the entire temperature-composition plane (see  Fig.~\ref{fig:12}). The comparison based directly on the raw data provides clear  evidence that  in the investigated solvent composition range it is indeed the critical Casimir interaction which underlies the colloidal attraction. Hence, this direct comparison suggests that not only at the critical composition, but also at these off-critical compositions, the attraction is described in terms of a critical Casimir force rather than by wetting effects. Yet, at even larger off-critical compositions, wetting effects are expected to eventually take over and
dominate the attraction as  observed clearly  in Ref.~\cite{Hertlein-et:2008}.

Using the same  experimental imaging and particle tracking techniques, in Ref.~\cite{Newton-et:2017} the effective  interaction potentials for non-spherical dumbbell particles 
have been extracted from observed radial and angular distributions. The colloidal patchy dumbbell particles
have been suspended in  heavy water and 3MP with a mass  fraction $\omega_{3MP} = 0.25$ (therein denoted as $c_{3MP}$). At this subcritical composition,  the hydrophobic spherical ends
prefer 3MP, while the  neck joining these two spherical ends (called by the authors ``shell'')  prefers water.  While the one-to-one mapping between
radial distribution function  and the effective, angularly averaged pair potential
still holds for the anisotropic particles,  the simple procedure of inferring the effective potential from  the radial distribution
function is not valid anymore. In order to find an optimal effective potential 
the authors of Ref.~\cite{Newton-et:2017} have assumed that dumbbells are the rigid construction of two isotropic spheres   each of which interacts via an isotropic
pair potential. They  have used the following three distinct  distribution functions which facilitate the comparison
of  simulations, theory,  and experiment \texttwelveudash this  way  determining that set of  potential parameters which renders 
the best match: (1) the minimum distance
radial distribution, for which only the minimum center-to-center distance $r_{i\alpha,j\gamma}$ between two spheres $\alpha$ and $\gamma$ belonging to dumbbell $i$ and $j$, respectively, contributes, (2) the, ``site-site'' radial distribution  between sphere ( called site) $\alpha =1,2$ on dumbbell $i$ 
 and a  sphere (site) $\gamma =1,2$ on a different dumbbell $j$, and (3) the bond-angle distribution, which measures
the mutual orientation of the particles forming  dumbbells. 
As a first approximation, the observed site-site radial distribution function has been matched to an underlying potential between the dumbbell particles  using the so-called reference interaction site model (RISM) in the context of integral equation theory for molecular fluids \cite{Chandler-et:1972}.
Within the latter theory,  a molecule is taken to contain several interaction sites so that the total interaction between two molecules is the sum of the site-site interactions. The RISM equation is a generalization of the Ornstein-Zernike equation \cite{OZ}
for hard-core site-site interactions. The authors of Ref.~\cite{Newton-et:2017} have used the hypernetted chain approximation as a closure.
All three measured distributions have been compared with the ones computed using   MC simulations  of the dumbbell system in which every pair of spheres, apart from the one on the same dumbbell interacts via an effective potential as given by Eq.~(\ref{eq:17}). Because the corresponding experimental  system has not been  density matched, the particles  sedimented, which was taken 
into account by including the gravitational potential  and a  surface field in order to describe the interaction with the bottom wall.
Concerning  the CCP, the authors of Ref.~\cite{Newton-et:2017} adopted an   oversimplified expression as the one given by  Eq.~(\ref{eq:13}), in line with  Refs.~\cite{Bonn-et:2009,Nguyen-et:2013,Dang-et:2013}. As  discussed  above, such a form 
is valid only in the  limit $D/\xi \to \infty$ and at the critical composition.
Both conditions are not met in the experiment under consideration. Moreover, the amplitude of the bulk correlation length of the solvent and the surface charge, which determines the strength and the decay length of the electrostatic repulsion, respectively, have not been measured. As a consequence, the numerous adjustable parameters have been used
to fit the theoretically proposed effective potentials to the measured ones. Following Ref.~\cite{Marcel-et},
the authors of Ref.~\cite{Newton-et:2017} have used the projected theoretical distribution functions in order to mimic
the  experimental uncertainties. Nevertheless,
the extent of agreement between the measured and the theoretically proposed effective potentials depends on temperature and is not satisfactory.
The  results for the dumbbell effective pair potentials, based on so many crude approximations and numerous fit parameters, are not conclusive and cannot be predictive. In particular, the quantitative modeling carried out in Ref.~\cite{Newton-et:2017} is not able to provide
the explanation for the observed two ranges of temperatures featuring distinct aggregation behaviors  (see Sec.~\ref{subsec:applic} below).

 \subsection{Applications of CCF for self-assembly and aggregation of colloids}
 \label{subsec:applic}

The possibility of controlled aggregation by exploiting  CCFs has been used, near the lower critical point of the suspension, for the self-assembly of cadium telluride quantum dots in  water - 3MP liquid mixtures with  NaCl salt  \cite{Marino:2016}.
By measuring the intensity of scattered  light as a function of time and 
by following the time evolution of the intensity correlation  function, it has been found that  1 K below 
the  phase separation temperature of the suspension the hydrophilic quantum dots with a  size of ca. 2.6 nm form aggregates with an average radius of ca. 700 nm.
As expected, for a composition with the 3MP volume fraction being larger than its critical value, i.e., corresponding to the phase poor in the component preferred by the quantum dots, aggregation takes place on 
shorter time scales and within a larger temperature interval than for a mixture with  a 3MP volume fraction being smaller than the critical one.

\begin{figure}
        \includegraphics[width=0.7\textwidth]{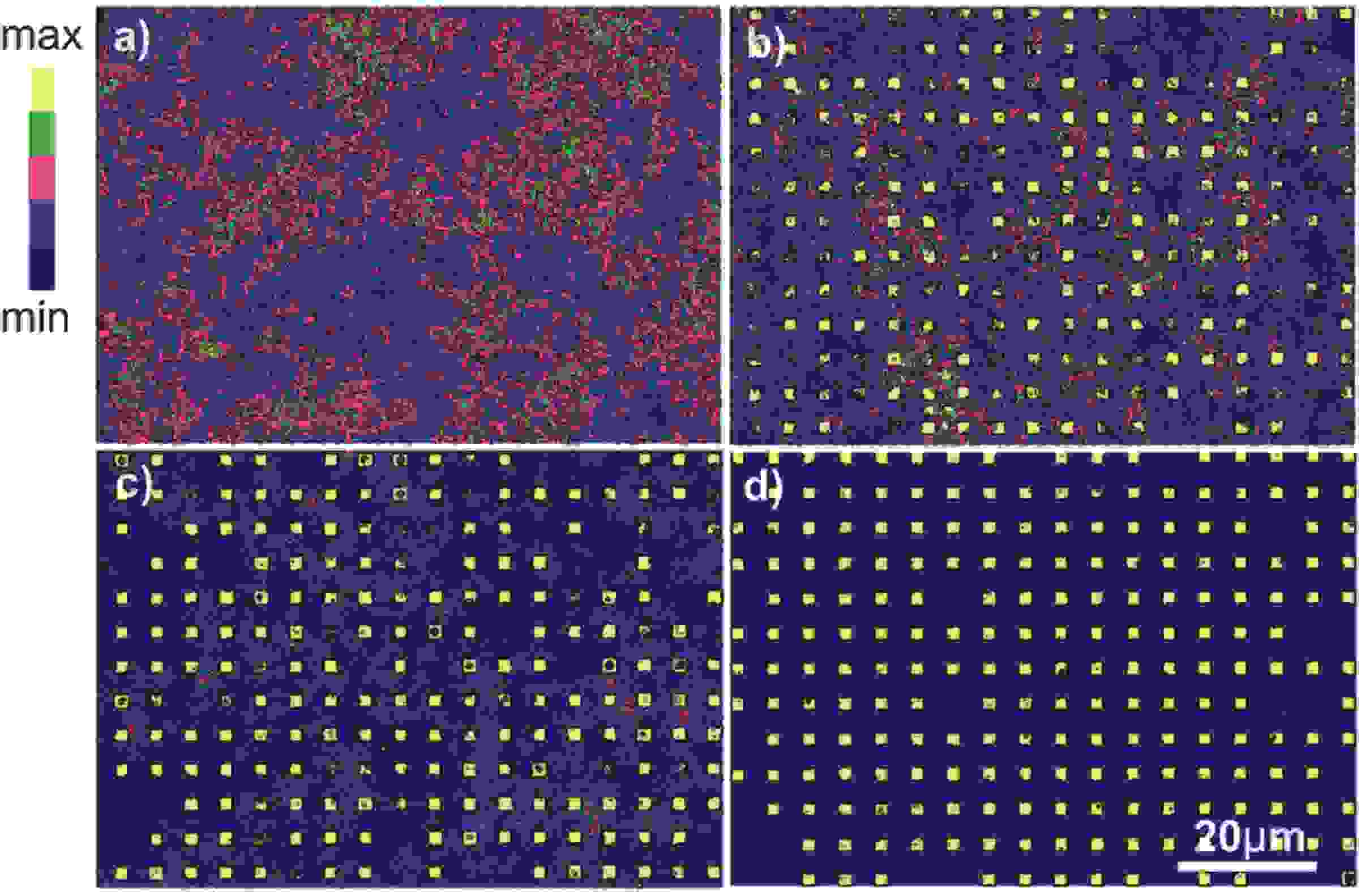}
        \caption{Mean particle density distribution (represented by different colors ranging from black to yellow for minimal to maximal density, respectively) of a dilute colloidal
  suspension of spherical particles with $R= 2.4 \mu m$  dissolved in a critical WL
  mixture in the presence of a chemically patterned substrate (Fig.~1 in Ref.~\cite{Soyka-et:2008}). Particle positions
were determined by digital video microscopy with a spatial
resolution of ca. $50\mathrm{nm}$. 
  $T_c^{(s)}-T=0.72\mathrm{K}$ (a), $0.25\mathrm{K}$ (b), $0.23\mathrm{K}$ (c), and $0.14\mathrm{K}$ (d).
}
\label{fig:13}
 \end{figure}

Critical Casimir interactions can be utilized  for aggregation of colloids being induced by a  substrate  
such that the colloids follow the chemical  pattern designed on the surface of a substrate which characterizes the boundary conditions.
This is because  CCFs respond
sensitively to the chemical properties of the confining surfaces.
As already discussed  in Sec.~\ref{sec:eff},  depending on whether the surfaces of the colloid and of the substrate  have the same or opposite preferences for the species of the solvent
 (symmetric or asymmetric boundary conditions), attractive or repulsive CCFs  arise.
In Refs.~\cite{Soyka-et:2008,Troendle-et:2011},
 charged polystyrene (PS) spheres with $R=1.2\mathrm{\mu m}$ have been suspended in a water-2,6-lutidine (WL) mixture at
its critical composition, i.e., a lutidine mass fraction of $\omega_{L,c} \cong 0.286$.
The suspension has been exposed to a chemically patterned substrate with well defined, spatially varying
adsorption preferences. This has been achieved by  first coating the glass surface
with a monolayer of hexamethyldisilazane (HMDS) which  rendered the glass surface
hydrophobic with a preferential adsorption of lutidine, corresponding
to a ($+$) boundary condition.
Spatial patterning of the boundary conditions has been obtained  by using  a focused ion beam (FIB) of
positively charged gallium (Ga) ions which  created
well-defined hydrophilic ($-$) areas with a
lateral resolution on the order of several tens of nanometers
extending over an area of approximately $400 \times 400$ $\mathrm{\mu m}^2$.
 Close to the lower critical demixing  point  
 $T_c \approx 307\mathrm{K}$ of WL \cite{Beysens-et:1985}, normal and   lateral
CCFs lead  to a strongly temperature dependent
attraction between the hydrophilic ($-$) PS particles and the hydrophilic squares
(forming a $2d$ square lattice of locally symmetric boundary conditions) and to a repulsion from the
hydrophobic regions ($+$) (locally asymmetric boundary conditions). This gives rise   to the formation of highly ordered
colloidal self-assemblies, the structure of which is controlled by the
underlying chemical pattern (see Fig.~\ref{fig:13}).
 At higher particle
concentrations, additional CCFs between neighboring
particles  arise and eventually lead to the formation of
three-dimensional, facetted colloidal islands on the substrate.
In order to quantify lateral CCFs,  substrates with a  periodic one-dimensional chemical pattern have been created
 forming  
hydrophilic ($-$) and hydrophobic ($+$) stripes with widths of 24.6 $\mathrm{\mu m}$ and 5.2 $\mathrm{\mu m}$, respectively.
For these  one-dimensional surface patterns the particle distribution has been
measured by  digital video microscopy and from the two-dimensional projection of this distribution the effective
one-dimensional CCP has been determined by resorting to the Boltzmann factor. 
In Ref.~\cite{Troendle-et:2011} the experiment has been repeated for substrates with  
the chemical pattern  created  by micro-contact printing, which provided sharper, chemical steplike, onedimensional 
interfaces between alternating regions of antagonistic adsorption preferences than the FIB technique does.  
The reason for the disadvantage of FIB is that the ion beam charges the substrate surface and thus deflects the incoming beam which gives rise to fuzzy chemical steps.
It turns out that the agreement between  theory and the experimental data is so sensitive (Fig.~\ref{fig:13}), that the CCP can be used to probe the geometry  of the chemical structures,
which at present cannot be achieved by other experimental techniques. Accordingly CCFs  can be used as a novel surface sensitive probe.

Reversible aggregation of  spherical Janus particles  has been studied experimentally 
for micro-sized silica particles half covered by a layer of gold and suspended in the WL mixture
at the critical concentration \cite{Iwashita-et:2013}.  The gold caps of the particles have been  modified chemically by sulfonic groups,
which have bestowed a large charge density on their  surfaces, leading  to  an  adsorption property  which differs from  that of the bare silica surfaces.  
Due to this anisotropy, the particles assembled to  clusters with a very specific structure, followed by a  hierarchical growth of the clusters.
These structures depend on the ``valence'' of the Janus particle, i.e., the maximum possible number of bonded nearest neighbors.
In two spatial dimensions, which applies to the dispersion of the sedimented particles considered in the experiment, the valence is equal to six. 
Direct visual observation has revealed that  particles,  randomly dispersed at low temperatures,
have started to form micellar structures upon increasing $T$ towards  the lower critical point of the WL mixture.
Most of the clusters have been  trimers and tetramers, with  gold-coated  hemispheres inside the micelle 
and silica hemispheres always facing outwards. No ``inverted'' micelles have been  observed, which
indicates that the effective attraction between golden patches of colloids has been much  stronger than the attraction between the silica parts of the particles.  At higher temperatures small clusters, mostly tetramers,
have assembled into chain-like structures, finally forming a percolating network. The reversibility of the aggregation, the value of the onset temperature, and the apparent
increase of the strength of attraction upon approaching  $T_c^{(s)}$, have supported { \color{red}the} expectation  that this aggregation
is due to CCFs. The structures of the clusters observed in the experiments are similar to those obtained within
MC simulations of hard discs with a pairwise square-well  potential acting between the semi-circular patches of the particles. 
However, the strength of attraction could not be determined from the experimental data.
The analysis of the cluster structures corresponding to the lowest internal energy
and of the hierarchical clustering suggested the possibility that 
the self-assembly of Janus particles is governed by the valence structure of the clusters and not by that of a single particle.

Recent progress in synthesizing  colloidal building blocks allowed the authors of Ref.~\cite{Nguyen-et:2017,Nguyen-et:2017a}
to fabricate particles with complex shapes and surface patch properties which act as  analogues of
molecular valence. Multivalent particles, such as dimers, trimers, and tetramers have been  produced by swelling and polymerizing
clusters of PMMA spheres with a methylmethacrylate/methacrylic acid shell, resulting in geometrically well-defined patches. 
The specific solvent affinity of the particle patches has been achieved by grafting a polyhydroxy stearic acid-copolymer (PHSA) onto the surface patch, which renders it hydrophobic. The central part
of the patchy particles is made hydrophilic by using the ionic initiator potassium persulfate. The particles have been suspended in the  homogeneous phase of a binary solvent of heavy
water and 3-methylpyridine (3MP)  at temperatures  below the lower critical temperature $T_c=38.55$ \textcelsius, which has been determined by light scattering and microscopy from solvent phase separation at the
critical composition. The solvents were prepared with 3MP mass 
fractions $\omega_{3MP} = 0.25$ and 0.31 (therein denoted as $c_{3MP}$), i.e., slightly below and above 
the critical composition $\omega_{3MP,c} = 0.28$, respectively. The hydrophobic patches  have a strong affinity for the non-aqueous
component 3MP of the binary solvent, while the hydrophilic central part has an affinity for water. 
It has been observed  that, in solvents poor in the component preferred by the particle patches, 
the patches approach each other at temperatures close to $T^{(s)}_c$, and that dimer particles assemble into directed,
chain-like structures. The bending stiffness of the chains has been measured directly
by monitoring  thermally activated bending fluctuations. In contrast, in
3MP-rich solvents, the particles approach each other sideways
resulting in distinct parallel structures. For trimers, the patch-to-patch binding in 3MP-poor solvents
leads to staggered chains, while the side-by-side
binding in 3MP-rich solvents leads to bent filaments associated
with the dense alternating stacking of trimers. In
all cases, the assembly is fully reversible as confirmed by the
break-up of aggregates upon lowering the temperature several
degrees below $T^{(s)}_c$. Interestingly,  in the case that the chain structure is formed by dimers, upon further approaching
$T^{(s)}_c$, the authors of Ref.~\cite{Nguyen-et:2017} have observed that the chain spontaneously collapses into a
compact state, the dimer particles
approach each other sideways, and eventually form a close-packed arrangement.
It has been argued that in this close-packed state, a particle has
more bonding neighbors, and hence a more negative bond energy. The authors of Ref.~\cite{Nguyen-et:2017}
regard this as being similar to the  collapse transition of a polymer, which  occurs if the   solvent conditions go from good to poor.
For polymers, too, the reduction of the  conformational entropy 
of the chain is
offset by the stronger interparticle interaction energy.  
Using MC simulations with the effective pair potential determined in Ref.~\cite{Newton-et:2017} (see the preceding Sec.~\ref{subsec:pp}), 
it has been argued in Ref.~\cite{Nguyen-et:2017},  that the colloidal chain collapse results from the enlarged interaction range due
 to the increase of the solvent correlation length upon approaching  the solvent
 critical point. In Ref.~\cite{Nguyen-et:2017a}, the effect of patch width on the topology of colloidal
 aggregates has been experimentally investigated.

\subsection{Colloidal mixtures}
\label{subsec:cm}

The way the unique properties of  CCFs can be harnessed to manipulate colloidal suspensions
has been demonstrated by experiments on colloidal mixtures \cite{Zvyagolskaya-et:2011}. 
This experimental system has been composed of  a binary mixture of micro-sized latex particles with slightly different diameters 
which were suspended in the WL mixture. For this small size difference of the colloids the inherent depletion interaction
cannot induce demixing. However, the CCFs  can accomplished this, if the two types of particles carry  opposite adsorption preferences.
 The $a$-type particles were functionalized with  silane
rendering them  hydrophobic, i.e., $(+)$ BC, whereas the $b$-type particles had a strong adsorption preference 
for water, i.e., $(-)$ BC. For such BCs the CCFs among the  $a$-type  particles and among the $b$-type particles are attractive, whereas between $a$-particles and $b$-particles they are repulsive.
At the same temperature and the same distance, the repulsive CCF is stronger than the attractive CCF.
In the system under consideration, the van der Waals forces have been eliminated by index matching so that besides the CCFs the only remaining 
forces have been the screened electrostatic interactions.
In the suspension, this mixture of colloids sedimented at the bottom of the cell  forming  a dense monolayer.
One species has been labeled with a fluorescent dye and traced by using video microscopy.
As  expected, upon approaching the (lower) critical point of the binary solvent at its critical composition,
 large structural changes in the colloidal mixture have been observed,  signaling the demixing process.
It was found that the morphology of this process depends strongly  on the mixing ratio $x_{a,b}= \rho_{a,b}/(\rho_a+\rho_b)$ of $a$- and $b$-particles, 
where $\rho_{a,b}$ are  the number  density of $a$ and $b$ particles  (see Fig.~\ref{fig:16}, which has been  taken from Ref.~\cite{Zvyagolskaya-et:2011}).
For  $x_a = 0.54$    and at low temperatures an initially  random distribution of particles transforms
into a bicontinuous network, which is  coarsening further  upon increasing the temperature. 
For $x_a = 0.28$, no bicontinuous structure has been  observed; demixing has proceeded via the growth of small clusters of the minority phase
(here rich in $a$-particles). 

In the theoretical part of this study  \cite{Zvyagolskaya-et:2011}, the effective approach has been employed
in order to construct an approximation for the Helmholtz free energy of the colloid mixture. It has been  assumed that the effective pair potential $U_{i,j}, i,j = a, b$,  can be split into 
a  hard-disc part and into a tail, which for identical particles   is  attractive and for the distinct ones repulsive.
For the hard-disc part of the free energy the scaled particle approximation has been used, whereas
the attractive or repulsive tail has been treated within the mean field van der Waals approximation. The predictions of this simple theory 
for the locations of the colloid demixing transitions  and their critical point agree fairly well with the experimental data.

\subsection{Effects of depletants on colloidal phase separations }
\label{subsec:eff_dep}

Piazza and collaborators \cite{Buzzaccaro-et:2010,Piazza-et:2011}  have
studied  experimentally the colloidal  phase separation in  suspensions with a depletion  agent, which occurred near the critical point 
of the depletant enriched solvent. They  have used  the so-called Hyflon$^{TM}$ MFA latex particles (a copolymer of tetrafluoroethylene (TFE) and perfluoromethylvinylether (PF-MVE); the producer does not provide the meaning of the acronym MFA) 
of average size $R \sim 90$ nm  suspended  in water with  the  nonionic surfactant C$_{12}$E$_8$. Salt (NaCl)  has  been added  in order to  screen electrostatic interactions between the colloids, caused by their surfaces carrying  a negative charge.
In addition,  this  surfactant  has provided   also  steric stabilization of  the suspension due to its spontaneous adsorption on  the  colloid  surfaces.
What makes this system interesting is, that  at concentrations above the critical micellar concentration,  C$_{12}$E$_8$ in water forms   globular micelles with a radius $\sigma = 3.5$ nm. These micells  act as  a depletant for the MFA particles.
Moreover, the surfactant-water mixture  exhibits a liquid-liquid phase transition terminating at a  lower critical point $(T_c^{(s)}, c^{(c)}_s)$
with a very small value of the {\it c}ritical concentration $c^{(c)}_s$  of the {\it s}urfactant (1.8\% mass fraction $c_s=m_s/m_{tot}$, where $m_s$ is the mass of the surfactant and $m_{tot}$ is the total mass of the sample). (We  
 use the superscript $s$   as an acronym for  the solvent and the subscript $s$ as the one for the surfactant.) Above this critical point two liquid phases exist, one rich and the other poor in micelles.
The authors have determined, as a function of temperature, the {\it m}inimum  concentration $c^{(m)}_s$  of surfactants  required to induce 
  colloidal gas-liquid phase separation. The onset of this phase separation
has been assumed to manifest itself via  the sudden increase of turbidity followed by fast sedimentation of the colloidal particles.
The results of these measurements  (see Fig.~\ref{fig:15}, taken from  Ref.~\cite{Piazza-et:2011}) show  a drastic decrease of $c^{(m)}_s$  upon increasing temperature towards
 the consolute point $(T^{(s)}_c,c^{(c)}_s)$  of the surfactant-water mixture, such that $c^{(m)}_s$ approaches $c^{(c)}_s$ as $T \to T^{(s)}_c$.
Far below  the  surfactant-water miscibility gap, the colloidal phase separation has been obtained due to the action of depletion forces, provided that a sufficient amount 
of surfactant has been  added. The range of the depletion interaction is  set by the ratio $\approx 0.03$ between 
the micelle and the particle size and is very short, but the strength depends on the  concentration of the micelles.
In Fig.~\ref{fig:17}, the full dots $(T^{(m)}_s,c^{(m)}_s)$   show the minimum amount $c^{(m)}_s$  of surfactant  required to induce colloidal phase separation at the temperature $T^{(m)}_s$.
These points are expected to correspond to the thermodynamic states of the surfactant-water solvent for which the strength of the effective attractive potential between the colloidal particles
is roughly constant. The authors have interpreted  the reduction in $c^{(m)}_s$ observed at higher temperatures as an increase of depletant ``efficiency''
suggesting that  the emerging critical Casimir interactions near the critical point of the surfactant-water mixture can be considered as an extreme case of depletion forces
in which the correlated regions of micelles act as a  depletant of ``renormalized'' size, i.e.,  bigger than that of a micelle, which acts as a bare depletant. However, this  interpretation is overstreched because 
depletion and Casimir forces originate from  very different physical mechanisms. In particular, upon approaching  the critical point the correlated regions of the emerging  phases do not form bubbles but self-similar, bicontinuous structures.  

Inspired by these experiments, Sciortino and co-workers  \cite{Gnan-et:2012a} have performed  a numerical study of the phase separation of hard spheres 
dispersed in an implicit solvent  (i.e., there are no direct interactions, beyond the hard-core ones, between  the solvent molecules and these spheres) in the presence of  interacting depletant particles. 
The particles have been taken  to interact via the corresponding effective potential $V_{eff}$ as determined by 
MC simulations for SW and 3P models of depletants (see Eq.~(\ref{eq:7}) as well as   Secs.~\ref{sec:eff} and \ref{subsec:eff_dep}).
The   phase separation of the colloids as a function of the depletant  concentration for 
various depletant-colloid size ratios has been determined 
within grand canonical MC simulations. The numerically determined  loci of the onset of colloidal gas-liquid phase separation
has been displayed within  the phase diagram  of the depletant  which exhibits an upper critical point. This revealed
 that almost for all  parameters studied in the MC simulations, the colloidal phase separation induced by CCFs
is preempted by the one driven by standard depletion forces. The important message of this study is that
only by weakening the attractive depletion interactions,
either by lowering the critical depletant concentration $c^{(c)}_s$ 
or  by introducing a repulsive interaction between the colloids, 
it is possible to exploit  CCFs for the fine tuning of the self-assembly of colloids in solvents with interacting depletant agents.

\subsection{Aggregation kinetics and  structures of aggregates}
\label{subsec:kin_aggreg}

Aggregation of colloids in a near-critical binary solvent
has been studied  experimentally on the ground and under microgravity conditions  
~\cite{Bonn-et:2009,Veen-et:2012,Shelke-et:2013,Potenza-et:2014}.
The advantage of the critical Casimir effect of providing the ability to tune the effective interactions between the colloids by varying the temperature 
has been used in order to study systematically the  internal structure of the aggregates
as a function of their interparticle attraction.
Direct visual observations via confocal microscopy of a dilute suspension
of latex particles of radius $R$ in a $\mathrm{D_2O/H_2O - 3MP}$ mixture have  revealed the formation of fractal
clusters of colloidal particles near the critical point of the solvent~\cite{Bonn-et:2009}. 
The analysis of the time-dependent  intensity $I(q,\mathsf{t})$  revealed that  $I(q,\mathsf{t})$ exhibits a maximum at  $q=q^*(\mathsf{t})$, which corresponds to the inverse 
of the mean cluster size.  It has been found that in the course of time $q^*$ decreases as  $q^*(\mathsf{t}) \propto \mathsf{t}^{-1}$ and that this linear  growth of the size of the clusters with time $\mathsf{t}$ is 
 self-similar. The latter property has been inferred  from the  data collapse of $I(q,\mathsf{t})$  plotted as a function 
 of the  rescaled wave vector $q' = q/q_0 = (\mathsf{t}/\mathsf{t}_0)qR$ with $\mathsf{t}_0 = 33$ min\footnote{In the original paper~\cite{Bonn-et:2009} $q'$ is given as $q'=  (\mathsf{t}_0/\mathsf{t})qR$, which we consider as a typo.}
 The fractal dimensions $d_f$ determined from $2d$ projection images have indicated that this system exhibits diffusion-limited cluster aggregation 
rather than  diffusion-limited particle aggregation, for which  $d_f$ is larger \cite{Dutcher-et:2004}.
The fractal dimension reflects the internal structure of an aggregate and relates its radius $R_g$  of gyration  to the number of particles 
$N$ according to  $R_g \propto N^{1/d_f}$; $d_f \lesssim 3$ corresponds to a close packed structure.  $R_g$   measures  the mean  size of the fractal aggregate. It is defined in terms of the mass density $\rho(r)$ of the particles at a distance $r$ from the center of mass of the aggregate as
$R_g^2 = \int_{0}^{\infty} r^4 \rho(r) dr/\int_{0}^{\infty} r^2 \rho(r) dr$ or, alternatively,  based on the pair correlation function $g(r)$ of colloidal particles  $R_g^2 = (1/2) \int_{0}^{\infty} r^4 g(r) dr/\int_{0}^{\infty} r^2 g(r) dr$ (see, e.g., Ref.~\cite{vanSaarloos:1987}). For fractal structures consisting of spheres
with spherically symmetric mass density distribution $\rho(r)$, one has $R_g = R_c$, where $R_c$ is the largest radius beyond  which the aggregate has zero mass.  The images from confocal microscopy have shown that the width of the branches of the clusters is  a
 few  particle sizes  and that the particles escape from the aggregate at a finite rate. The latter observation and the assumption 
that this process is thermally activated, allows one to estimate
the energy scale of attraction between the particles as ca.  $\approx  3\times k_BT$. 
A more detailed investigation of the  structure of the clusters and its evolution  
was impeded by sedimentation.

In order to avoid this complication,  using near field light scattering (NFS), new measurements for the same type of particles have been performed
under microgravity conditions on board of the International Space Station  and, simultaneously, on the ground ~\cite{Veen-et:2012,Potenza-et:2014}. 
(For these  experiments, the composition of the solvent and the volume fraction
of the colloidal particles have been taken to be different from those in Ref.~\cite{Bonn-et:2009}. Moreover, different amounts of  salt (NaCl) have been added. Thus the results of both studies 
cannot be compared  directly.)
The  light scattering intensity    $I(q,\mathsf{t})$  has  provided information about the time evolution of the  structure and of the mean size of the   aggregates.
Its normalized variance, defined as  $\langle I^2(q,\mathsf{t}) \rangle/\langle I(q,\mathsf{t}) \rangle^2 -1 $, with the angular brackets $\langle \cdots \rangle$ denoting the
time average, has revealed   the onset  and  the  time scale of the aggregation process.   The   time dependence  of the normalized variance 
  results from variations  of the number density of the scatters. Accordingly, the start of {\it a}ggregation 
 has been marked by the time $\mathsf{t}_a$ at which   this quantity starts to increase  from zero.
From the slope  in the log-log plot of $I(q,\mathsf{t})$  at large momentum transfer $q$, which has been temporally constant during the growth process (which is characteristic of  scattering off fractal structures),
a fractal dimension has been  inferred. It has been found that, upon varying the temperature towards $T_c^{(s)}$, $d_f$ has decreased from  a value, which is close to the theoretical one of 2.5 for diffusion-limited aggregation, to about 1.8. 
This  indicates that more open structures of clusters
are formed if - as expected - the strength of the attraction increases. This has been  interpreted to the effect that  for an  attractive interaction potential with a well depth of ca. $1 \times k_BT$ the restructuring of the aggregates into more compact objects can proceed, whereas for deeper attraction wells restructuring
gradually stops and the resulting structure is more open.
Such more open structures have been observed  in the presence of gravity with $d_f \simeq 1.6 - 1.8$ for all temperatures studied. In both cases, i.e., in the presence and absence of gravity,  only a slight variation with the salt concentration has been found. 
Moreover, it has been observed that independent of temperature, and thus the strength of the attraction, the aggregates grow similarly.
Specifically, the scattered intensity  of the growing aggregates as obtained  at different times  has  been {\it red}uced to a scaling function  $I(q,\mathsf{t}) \approx \left(q_{red}\right)^{-d_f}F(q/q_{red}(\mathsf{t}))$,
where the time dependent characteristic momentum transfer $q_{red}(\mathsf{t})$ has been determined from the best fit  in the sense of  $I(q,\mathsf{t})\left(q_{red}\right)^{d_f}$ yielding data collapse to a function $F$ of a single variable $q/q_{red}(\mathsf{t})$.
Based on this apparent scaling behavior, an  analogy has been drawn between  the observed aggregation process  and  spinodal decomposition processes. For the latter,
 a similar scaling holds with $d_f$ replaced by the spatial dimension $d$ \cite{Carpineti:1992}.
As for  spinodal decomposition,  $I(q)$ exhibits a maximum at  $q\simeq q_{red}$ corresponding  to the inverse of the characteristic length in the system. In the course of time the momentum
$q_{red} $ shifts to  smaller values  indicating the growth of this characteristic length.
The decrease of $I(q)$ for small values of $q$ reflects the fact that a region around the cluster is depleted of colloids \cite{Carpineti:1992}.
The growth rate of the characteristic length scale  $q_{red}^{-1} \propto R_g\propto N^{1/d_f}$  of the aggregates has been
 found to be  a power law  with  exponent $1/d_f$, regardless of the salt concentration. This  is characteristic of
diffusion-limited models of aggregation \cite{WittenMeakin}.
Under gravity, the growth rate has been observed to be significantly faster and nonmonotonic (see Fig.~\ref{fig:18}, which is 
taken from Ref.~\cite{Veen-et:2012}). At early stages it has followed the purely diffusive behavior. After that,  aggregation has been  influenced strongly by convection of the solvent 
and by sedimentation. The  aggregation among  clusters has been  found to set in rather fast, resulting 
in an exponential growth rate, which is characteristic of reaction limited aggregation with a strong dependence on  temperature and thus on the strength of the attraction.
At still later times, a sudden drop in $q_{red}^{-1}$ has been observed, which corresponds to the sedimentation of the largest clusters.

 Whereas in Ref.~\cite{Veen-et:2012} the static properties of the aggregates, i.e., their structure factor $S(q)$, have been measured as  aggregation proceeds,  in the subsequent  NFS  study of this  system  \cite{Potenza-et:2014}  
the  dynamical counterpart of $S(q)$, the intermediate scattering function  $\mathfrak{S}(q,\mathsf{t})$   has been measured under microgravity conditions. The aim has been to  determine
  the ratio $\beta  \equiv  R_h/R_g$ of the hydrodynamic radius $R_h$ of the aggregate and its gyration radius $R_g$. This ratio provides information about  the density distribution within an aggregate as function of its fractal dimension, which in turn depends on the strength 
  of the particle attraction.
  The hydrodynamic radius $R_h$ of the aggregate is defined through the translational diffusion coefficient $D = k_BT/(6\pi\eta R_h)$, where $\eta $ is viscosity of the solvent.
  For  a densely packed spherical aggregate of radius $R_c$ and with $d_f = 3$ one expects $R_h$ to be very close to $R_c$.
The  intermediate scattering function $\mathfrak{S}(q,\mathsf{t})$  is the spatial (three-dimensional) Fourier transform of the van Hove distribution function  $G(r,\mathsf{t})$, which is the dynamical counterpart of the  radial distribution function $g(r)$~\cite{Hansen-et:1976}.  
The NSF technique enables one to measure $\mathfrak{S}(q,\mathsf{t})$ instantaneously, on the time scale of the much slower aggregation and diffusion processes. This  provides the concurrent  measurement of 
 $R_h$ from the dynamic and $R_g$ from the static structure factor, simultaneously for all accessible 
wave vectors. $R_g$ follows from the Fisher-Burford expression \cite{FB} $S(q; R_g) = (1 + (2/3d_f)q^2R^2_g)^{-d_f/2}$, which is valid for monodisperse fractal aggregates.  $R_h$ has been  determined   via the effective diffusion constant $D_{eff}$  obtained from the measured  decay  time $\mathsf{t}_d$ of $\mathfrak{S}(q,\mathsf{t})\propto e^{-\mathsf{t}/\mathsf{t}_d}$  via $\mathsf{t}_d = 1/(D_{eff}q^2)$ followed by  relating $D_{eff}$ to
$D$ by using $S(q)$ \cite{Lin}.
  It has been  concluded that in order to obtained reliable data for $R_h$, one has to consider the late stages of aggregation, in which the clusters reach sizes large enough to exhibit 
a well-defined fractal dimension and internal structure and in which   the actual polydispersity of the aggregates does not  play a significant role.
The analysis of the data has shown, that upon approaching the  critical point of demixing (i.e., for stronger attraction)
 the ratio $\beta$ has varied  between 0.76 for $d_f = 1.8$ 
and 0.98 for $d_f = 2.5$ at the onset of aggregation (i.e., for  weaker attraction). The authors of Ref.~\cite{Potenza-et:2014} have assumed that the fractals are spherically symmetric. Accordingly they have defined the radial density distribution of the fractal objects  as
$\rho(r) = r^{d_f-3}f_{cut}(r)$, where $r$ is the distance from the  center of mass of  the cluster  and $f_{cut}$ is a cutoff function which  accounts for the finite size of the  aggregates. (The exponent follows from the fact that $\rho(r) = \mathcal{N}(r)/V(r) \sim r^{d_f-3}$, where 
$\mathcal{N}(r)\sim r^{d_f}$ is the number of particles  within a sphere of radius $r$ from the center of the cluster  and $V(r)\propto r^{-3}$ is the volume of such sphere.) 
The expected behavior of the ratio for  various forms of a cutoff function has been then compared with the experimental data.
It has been observed (see Fig.~\ref{fig:19}, which is taken from Ref.~\cite{Potenza-et:2014}) that the data for the ratio $\beta$ have been  closest to those  values which  correspond to the assumption of fully compact objects with a  Heaviside step function for $f_{cut}$,
independent of  the strength of the attraction and of the fractal dimension.

The  way to avoid sedimentation for experiments performed on the ground \cite{Shelke-et:2013} has been to use  poly-n-isopropyl acrylanid particles (PNIPAM), which swell in  solution.  This swelling adjusts their buoyancy, preventing particle sedimentation.
This  also has  allowed the authors to observe directly  individual particles (after labeling them  with a fluorescent dye)  even  deep in the bulk of the suspension.
Using confocal microscopy, the compactness of aggregates of PNIPAM particles, suspended at small volume fractions in the 3MP - D$_2$O/H$_2$O mixture, has been  studied by determining the number of particles $\mathcal{N}(r)$ within a sphere of radius $r$ from  the center of the cluster
for   quenches with various temperature deviations  $\Delta T$ from and  below the lower critical point $T_c^{(s)}$ 
of the  demixing transition  of the solvent. These data have confirmed the fractal character of the structures, i.e., the 
relation $\mathcal{N}(r) \sim r^{d_f}$, with $d_f$ decreasing  continuously upon increasing temperature towards $T_c^{(s)}$
from $d_f \approx 3$ to $d_f\approx 2.1$. This indicates the formation of more compact structures for weaker attraction
(see Fig.~\ref{fig:19}(h)
which is taken from Ref.~\cite{Shelke-et:2013}),  in agreement with the behavior observed for the latex particles under microgravity (see the above paragraph).
Monte Carlo simulations of diffusion-limited particle aggregation in $d=2$ and $d=3$ have been used 
in order to relate $d_f$ to the attractive effective interaction potential of the particles. Experimentally, the pair potential has been  determined from the radial distribution function of the colloids in a way similar to that described in Ref.~\cite{Nguyen-et:2013}.
The depth $V_0 < 0$ of the attractive well  of this  pair potential has been related to the  
 quantity $\alpha \propto \exp (V_0/(k_BT))$ used in the simulations for the probability with which the particles can detach from the growing cluster. 
The  experimental variations  of $d-d_f$ as a function of $V_0$  and of $\alpha$ in simulations exhibit a  
common curve after suitable rescaling. The authors have interpreted this result as a manifestation of a certain type of universality
which  occurs although the corresponding simulations have been carried out for different spatial dimensions ($d=2$ and $d=3$)
 and  although the mechanisms of aggregation differ, i.e., diffusion-limited cluster aggregation in the actual experiments, in which clusters aggregate to form a fractal, and 
diffusion-limited particle aggregation  in simulations, in which  a fractal grows by aggregation of  single particles.

\section{Perspectives}
\label{sec:persp}

It remains as a  challenge to accurately determine the  critical  Casimir pair  potential 
and its scaling function within the
 range of the relevant parameters and boundary conditions. This  applies to chemically homogeneous or Janus particles of   spherical  or
 anisotropic  shapes such as cylinders, ellipsoids, cubes, or even  more complex shapes such as  dummbells or $L$-like ones.
Quantitative reliability    demands to go  beyond   mean field theory and the  Derjaguin approximation. This requires the development of new theoretical approaches and simulation algorithms.
Janus particles    have  already been used in experiments on aggregation 
in  near-critical binary solvents \cite{Iwashita-et:2013}.
Colloids with chemically homogeneous or inhomogeneous surfaces, forming patchy colloids~\cite{Schall_review,Garcia-et:2017}, with various shapes as well as  their mixtures are attractive building blocks for generating a large variety of self-assembled
structures by using CCFs. 
In order to make  further progress in understanding the interplay between CCFs and other forces, such as  electrostatic
ones, for aggregation phenomena, more elaborate and combined experimental and theoretical investigations are 
required.
Concerning the thermodynamic properties of colloidal suspensions with near-critical solvents, a multi-component theory beyond the effective approach
and beyond mean field theory is needed. One of the challenges along this line is to construct a suitable theory, which accounts for all
relevant degrees of freedom of  the  solvent and of the solute particles and masters  the 
size differences of the various species. 
An interesting line of research  would be to study aggregation of colloids in lipid membranes {\texttwelveudash}
close to the corresponding lipid phase segregation {\texttwelveudash}  or in 
a liquid analogue of such quasi two-dimensional  systems.

\pagebreak

\pagebreak

\pagebreak
\pagebreak

\pagebreak
\pagebreak

\pagebreak

\pagebreak
\vspace{-5cm}
\pagebreak

 \pagebreak
\pagebreak
\pagebreak
\pagebreak
\begin{figure}
  \includegraphics[width=0.9\textwidth]{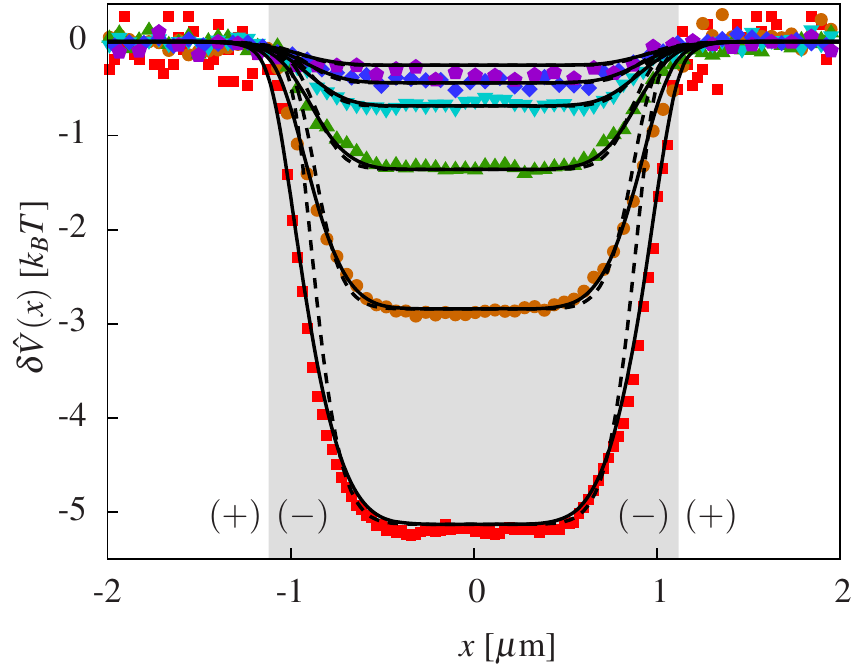}
  \caption{Total effective potential $\delta\hat{V}(x)=\hat{V}(x)-\hat{V}(x=P/2)$ of the forces acting on a hydrophilic  polystyrene spheres of radius $R=1.2\mathrm{\mu m}$ above
  a chemically striped pattern of periodicity $P$ with alternating  ($-$) and ($+$) boundary conditions  and immersed in a
  water - lutidine   mixture at its critical concentration, 
     as  function of its lateral position $x$ for various temperatures $T^{(s)}_c - \Delta T$ below the experimental value $T^{(s)}_c$ of the lower critical temperature of the solvent (Fig.~5 in Ref.~\cite{Troendle-et:2011}).
    The potential is given by  $\hat{V}(x)=-k_BT\ln(\hat{\rho}(x))$, where $\hat{\rho}(x)$ is the effective number density of the colloids at $x$, obtained by projecting the actual number density onto the $x$ axis.
  The width of ($-$) and ($+$) stripes  is  $2.25\mathrm{\mu m}$.  Symbols indicate experimental data, whereas the
lines are the corresponding theoretical predictions for sharp
 (dashed lines) and fuzzy  (solid lines) chemical steps. From top to bottom the measured temperature deviations
 $\Delta T$ are  0.175 (0.165), 0.16 (0.152), 0.145
(0.143), 0.13, 0.115, 0.10 K.   If indicated, the values in paranthesis are
corrected values of temperature (but compatible within the experimental inaccuracy) which
have been used for evaluating the theoretical predictions.
}
\label{fig:14}
\end{figure}
\pagebreak

\begin{figure}
  \includegraphics[width=\textwidth]{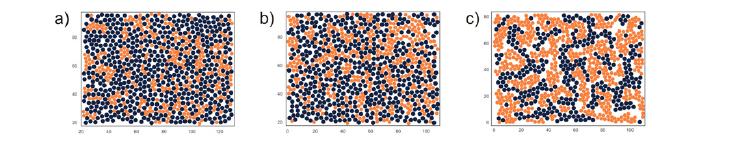}
  \caption{Snapshots of the twodimensional configurations of binary colloidal systems (black and orange discs, Fig.~1 in Ref.~\cite{Zvyagolskaya-et:2011})
  for three different compositions, expressed in terms  of the concentration $x_a=\rho_a/(\rho_a+\rho_b)$, where $ \rho_a$ and $\rho_b$ are (areal) number densities 
  of the particles of type $a$ and $b$, respectively: (a) $x_a=0.28$,  (b)  $x_a = 0.32$, and (c) $x_a =0.54$, with $a=$ orange,
  exhibiting distinct structures formed after 1 hour at a temperature deviation  $\Delta T  = T_c^{(s)} - T = 0.01$K from the lower critical point of the solvent.
  The horizontal  scales range from 20 to 120 in (a) and from 0 to 100 in (b) and (c), whereas the vertical scales
  range from 20 to 80 in (a) and (b) and from 0 to 80 in (c); the  authors have not provided the units. 
  }
  \label{fig:15}
\end{figure}
\pagebreak
\begin{figure}
  \includegraphics[width=0.6\textwidth]{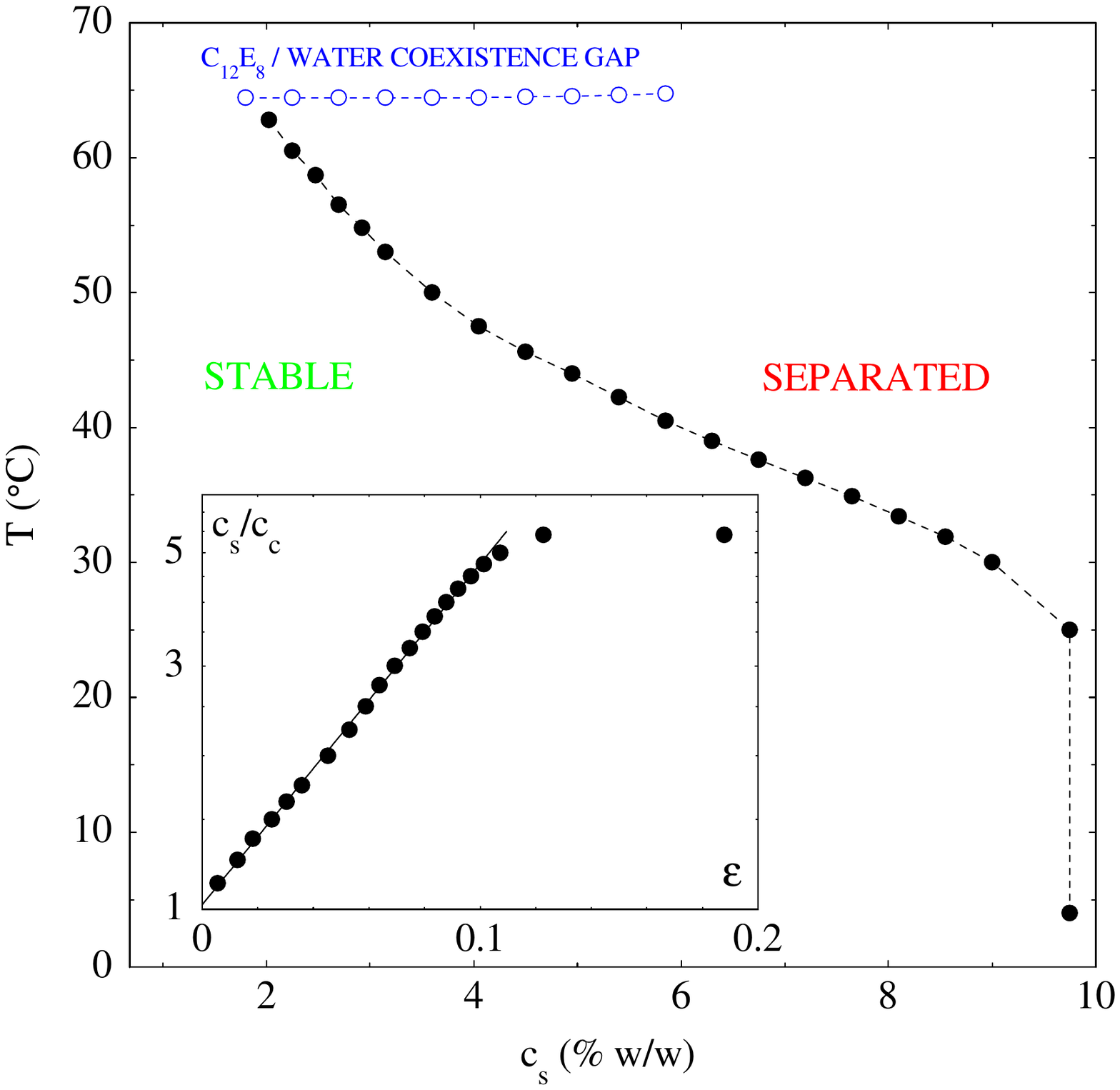} \vspace{-2cm}
  \caption{Experimental phase diagram of aqueous suspensions of  MFA latex particles in the presence of a nonionic surfactant of
  mass fraction $c_s=m_s/m_{tot}$  (in units of mass percentage, sometimes called weight percentage,  \% w/w $ = 100\times c_s$), where $m_s$ is the mass of all surfactant particles and $m_{tot}$ is the total mass of the sample  (Fig. 8 in Ref.~\cite{Piazza-et:2011}). 
  For state  points of the surfactant - water mixtures to the left of the line of dots (denoted as ``stable'') the dissolved colloidal particles with a volume fraction $V_c/V_{tot} = 0.03$ form a homogeneous colloidal phase whereas for state
  points to the right of the line of dots (denoted as ``separated'') the colloids phase separate into a colloidal gas and a colloidal liquid.
   ($V_c$ is the volume taken by  colloidal particles and $V_{tot}$ is the total volume of the sample). At a given temperature, the full black dots represent the {\it m}inimum amount $c^{(m)}_s$ of {\it s}urfactant required to induce colloidal gas-liquid phase separation  at a given temperature.
  The rather shallow coexistence curve  of the surfactant-water mixture with  a lower critical point (located around 1.8 \% of $c_s$ - not marked in the plot) is shown by open blue dots.
  The data correspond  to 250 mM added NaCl salt.
  The inset shows (in our present notation) $c^{(m)}_s$ in units of  its critical value $c^{(c)}_s$ (i.e., $c_s$ at the critical point) as a function of reduced temperature $t =(T^{(s)}_c-T)/T^{(s)}_c$ (note that - as in Ref.~\cite{Piazza-et:2011} - in this figure $c^{(m)}_s/c^{(c)}_s$ and $t$ are denoted by $c_s/c_c$ and $\varepsilon$, respectively.) 
  }
  \label{fig:16}
\end{figure}
\pagebreak
\begin{figure}
  \includegraphics[width=0.8\textwidth]{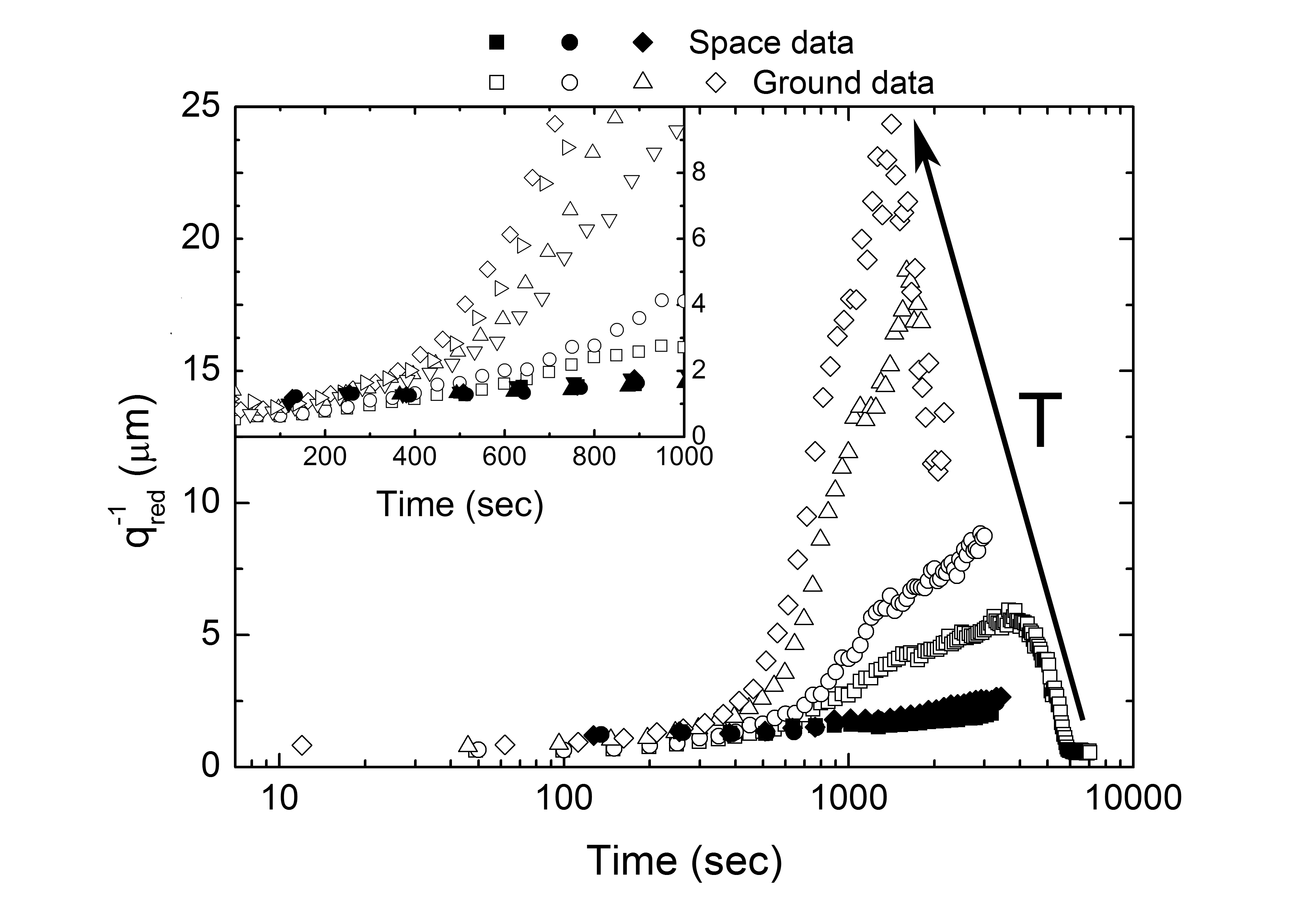}
  \caption{The evolution of the characteristic size scale of the aggregating colloidal suspension studied in Ref.~\cite{Veen-et:2012} (see Fig. 3 therein)
    as given by the quantity  $q_{red}^{-1}(\mathsf{t})$,  which renders a {\it red}uced description of the light scattering intensity  $I(q,\mathsf{t})$ 
    of the growing aggregates in terms of a scaling function $F(q/q_{red}(\mathsf{t}))$  of a single variable (see the main text).
    Results  both from microgravity (full symbols) and from ground experiments (open symbols) are shown.  (The  sample
  contains 1.5 mmol/liter NaCl.) The curves correspond to various temperatures $T$ ramped up  beyond the aggregation temperature $T_{agg}$,  
   up to  $T_{agg} + 0.4°$C
 (from bottom to top as indicated by the arrow;  the  caption to Fig. 3  in Ref.~\cite{Veen-et:2012} does not provide the 
 temperature values corresponding to the various symbols). The temperature $T_{agg}$ at which the aggregation
 starts  is identified (somewhat loosely) as  the onset of the rapid increase   of the normalized variance of $I(q)$ (see the main text). The sudden drop of $q^{-1}_{red}$ at late times  is due to the massive sedimentation of the aggregates
  so that the suspension becomes poor in large aggregates and rich in small ones which shifts down $q^{-1}_{red}$. 
  The inset provides an enlarged view of the data concerning the  onset of the aggregation process,
   at which the  characteristic length scale $q_{red}^{-1}$
    starts to increase.}
  \label{fig:17}
\end{figure}
\clearpage
\pagebreak
\begin{figure}
  \includegraphics[width=0.8\textwidth]{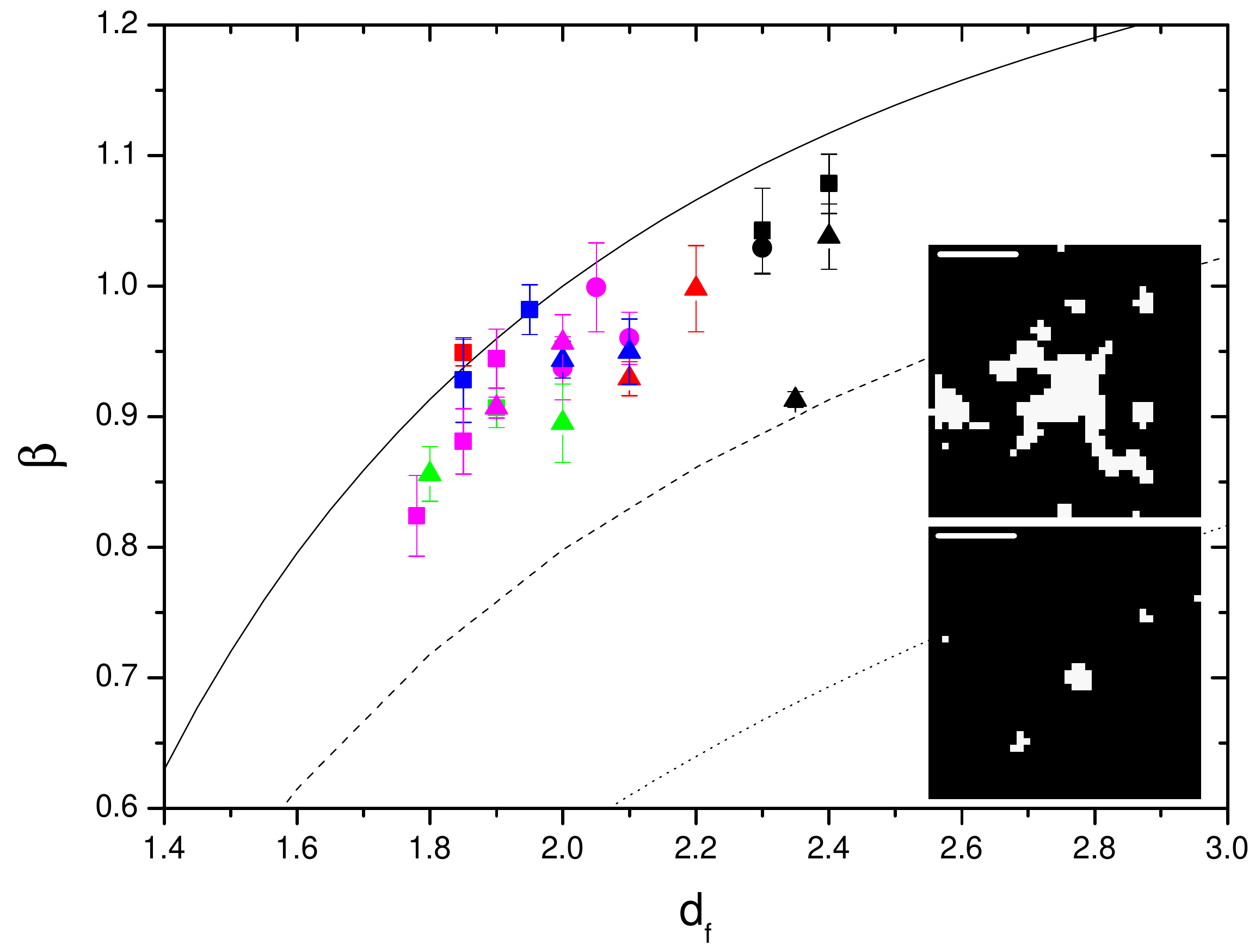}
  \caption{Ratio $\beta = R_h/R_g$ of the hydrodynamic to the gyration radius of aggregates as a function of their fractal dimension 
  $d_f$ for various temperature deviations $\Delta T$  above the aggregation temperature 
  $T_{agg}$  (Fig.~4  in Ref.~\cite{Potenza-et:2014})  characterized by the colors of the symbols: $\Delta T =0$ (black), 0.1 K (red), 0.2 K (blue), 0.3 K (green), and 0.4 K (violet).
  $T_{agg}$   is determined as in Fig.~\ref{fig:18}, i.e., from the behavior of the normalized variance of the scattered intensity $I(q)$.
The types of symbols indicate the salt concentrations of 0.31 mmol/liter (squares), 1.5 (circles), and 2.7 (triangles),
corresponding to the Debye screening length $\kappa^{-1}$ (denoted as $\lambda_D$ in Ref.~\cite{Potenza-et:2014})
of 14, 6.4, and 4.8 nm, respectively. Concerning  the values of $\beta$, from top to bottom the lines indicate the  dependence on $d_f$  as expected for steplike, Gaussian, and exponentially decaying cut functions of the  density distribution
of the aggregates, respectively (see  the main text).
The  insets show holographic reconstructions of the the real-space shape of the aggregates (white regions) grown at $T=T_{agg} +0.4$K (closest to $T^{(s)}_c$ and hence strongest attraction between the colloids, top) and $T=T_{agg}$ (furthest from $T_c^{(s)}$ and hence weakest attraction between the colloids, bottom).
The length of the scale bar is 25 $\mu$m.}
  \label{fig:18}
\end{figure}
\pagebreak
\begin{figure}
 \includegraphics[width=\textwidth]{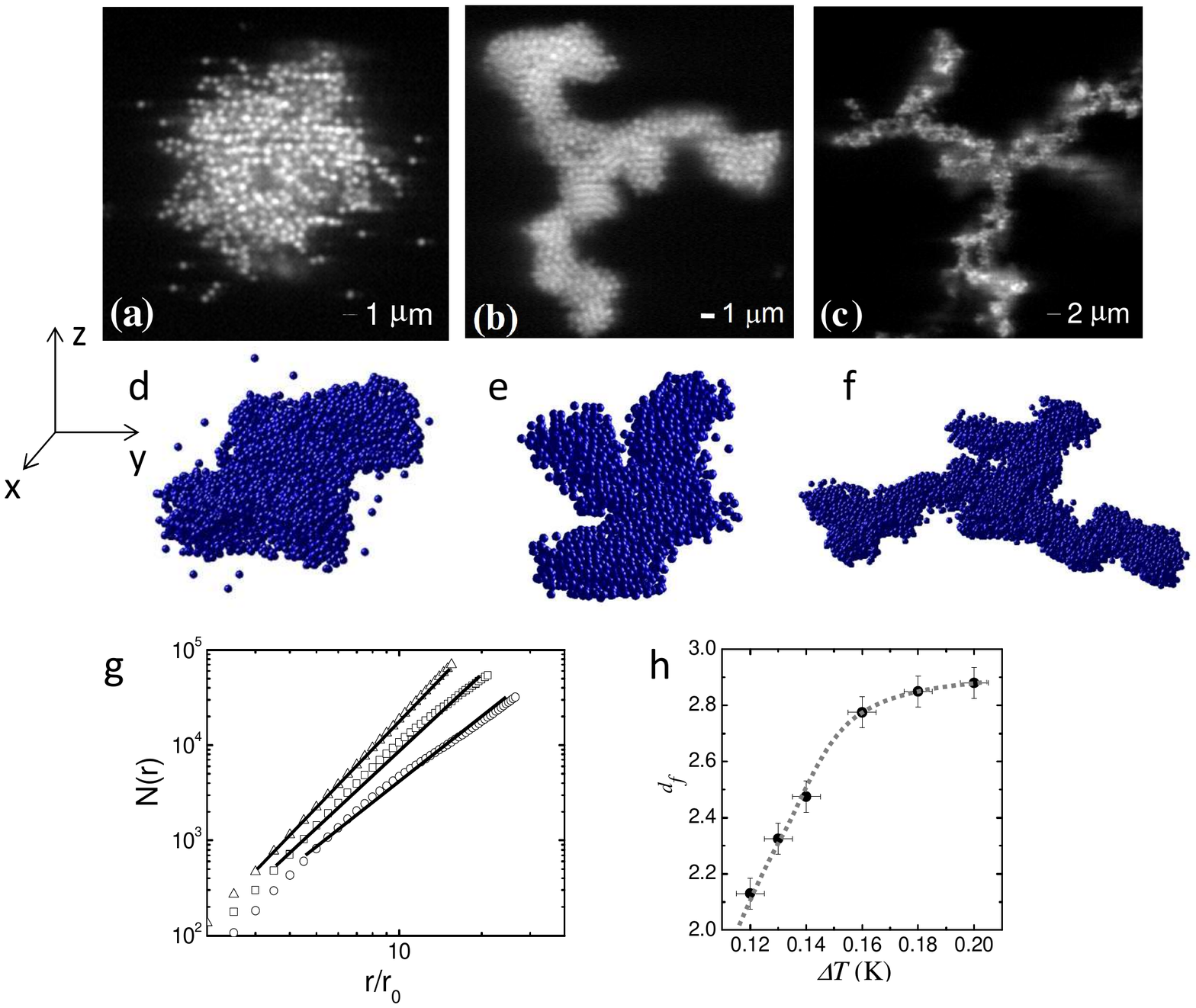}
 \caption{Tuning the morphology of colloidal aggregates by critical Casimir forces (Fig.~1 in Ref.~\cite{Shelke-et:2013}). 
 Confocal microscopy images and the threedimensional reconstructions of the aggregates formed upon  temperature quenches
to $\Delta T = T_c^{(s)}-T = 0.2$K (a,d), 0.14K (b,e), and 0.12K (c,f) below the lower critical phase separation temperature $T_c^{(s)}$ of the binary solvent.
(g) Scaling of the number of particles $\mathcal{N}(r)$ within a sphere of radius $r$ from the center of the  cluster  at $\Delta T = 0.2$K (triangles),
$\Delta T = 0.14$K (squares), and $\Delta T = 0.12$K  (circles); $r_0$ is the radius of the particles. (h) Fractal dimension as a function
of $\Delta T$ determined from $\mathcal{N}(r) \sim r^{d_f}f_{cut}(r)$ and thus from  the slopes in (g). The dashed line is a guide to the eye. Upon increasing $\Delta T$
a continuous increase of the fractal dimension to the space-filling limit $d_f=3$ is observed (see the main text).
 }
 \label{fig:19}
\end{figure}

\end{document}